\documentclass[aps,prb,floatfix,twocolumn]{revtex4}%
\usepackage{graphicx}
\usepackage{amsfonts}
\usepackage{amssymb}
\usepackage{amsmath}
\providecommand{\U}[1]{\protect\rule{.1in}{.1in}}
\providecommand{\U}[1]{\protect\rule{.1in}{.1in}}

\begin{document}
\title{Cavity QED with hybrid nanocircuits: from atomic-like physics to condensed
matter phenomena}
\author{Audrey Cottet$^{1}$, Matthieu C Dartiailh$^{1}$, Matthieu M. Desjardins$^{1}$,
Tino Cubaynes$^{1}$, Lauriane C Contamin$^{1}$, Matthieu Delbecq$^{1}$,
J\'{e}r\'{e}mie J. Viennot$^{1}$, Laure E. Bruhat$^{1}$, Benoit
Dou\c{c}ot$^{2}$ and Takis Kontos$^{1}$}
\date{\today}
\affiliation{$^{1}$Laboratoire Pierre Aigrain, Ecole Normale Sup\'{e}rieure, CNRS UMR 8551,
Laboratoire associ\'{e} aux universit\'{e}s Pierre et Marie Curie et Denis
Diderot, 24, rue Lhomond, 75231 Paris Cedex 05, France}
\affiliation{$^{2}$Sorbonne Universit\'{e}s, Universit\'{e} Pierre et Marie Curie, CNRS,
LPTHE, UMR 7589, 4 place Jussieu, 75252 Paris Cedex 05}

\begin{abstract}
Circuit QED techniques have been instrumental to manipulate and probe with
exquisite sensitivity the quantum state of superconducting quantum bits
coupled to microwave cavities. Recently, it has become possible to fabricate
new devices where the superconducting quantum bits are replaced by hybrid
mesoscopic circuits combining nanoconductors and metallic reservoirs. This
mesoscopic QED provides a new experimental playground to study the
light-matter interaction in electronic circuits. Here, we present the
experimental state of the art of Mesoscopic QED and its theoretical
description. A first class of experiments focuses on the artificial atom
limit, where some quasiparticles are trapped in nanocircuit bound states. In
this limit, the Circuit QED techniques can be used to manipulate and probe
electronic degrees of freedom such as confined charges, spins, or Andreev
pairs. A second class of experiments consists in using cavity photons to
reveal the dynamics of electron tunneling between a nanoconductor and
fermionic reservoirs. For instance, the Kondo effect, the charge relaxation
caused by grounded metallic contacts, and the photo-emission caused by
voltage-biased reservoirs have been studied. The tunnel coupling between
nanoconductors and fermionic reservoirs also enable one to obtain split Cooper
pairs, or Majorana bound states. Cavity photons represent a \ qualitatively
new tool to study these exotic condensed matter states.

\end{abstract}
\maketitle

\section{Introduction}

Since the 1980's, the continuous progress of nanofabrication techniques has
enabled the fabrication of a wide diversity of nanoelectronics devices which
reveal the oddities of quantum mechanics when placed at low temperatures. The
strong confinement of electrons in narrow conductors leads to a quantization
of transport into a few transverse channels. This phenomenon has been observed
for instance in quantum point contacts made in a two-dimensional electron gas
in a semiconductor\cite{van Wees: 1988}, or in break-junctions between
metals\cite{Krans:1993}, where the electric current can be carried by a very
low number of transverse channels. Another major ingredient of nanoelectronics
is the longitudinal confinement of electrons between two potential barriers
along the transport path\cite{Kouwenhoven:1997}. This leads to the formation
of quantum dots with a discrete energy spectrum, which are often seen as
artificial atoms. The fabrication of quantum dot circuits has reached a very
high level of control in two-dimensional electron gas structures, where small
quasi-zero dimensional dots are contacted to large two-dimensional reservoirs
through quantum point contacts\cite{Shulman:2012}. Interesting alternatives
are offered by self-assembled quantum dots\cite{Klein:1997}, carbon
nanotubes\cite{Tans:1997} and semiconducting nanowires\cite{De
Franceschi:2003}. To form a nanocircuit with these nanoconductors, one must
contact them with metallic electrodes which can be made out of normal metals,
ferromagnets or superconductors. Due to the versatility of nanofabrication
techniques, many circuit configurations can be used, with for instance
multiple metallic contacts, or superconducting flux loops. This leads to a
large variety of configurations to study quantum transport and obtain new
electronic functionalities. For instance, a quantum dot coupled to
ferromagnetic contacts can show a ferromagnetic proximity effect which is
interesting for the control of spin transport\cite{Cottet:2006a} or for local
spin manipulations\cite{Cottet:2010}. Other intriguing example, the Cooper
pair splitter enables the spatial separation of the two spin-entangled
electrons from a Cooper pair into two different quantum dots, which could be
an interesting resource for quantum information\cite{Recher:2001}. Finally,
semiconducting nanowires coupled to superconductors raise a strong interest in
the context of the search for topological matter and Majorana bound
states\cite{Leijnse:2012}. In these three examples, the coupling between the
nanoconductors and the superconducting or ferromagnetic reservoirs deeply
modifies the electronic properties of the nanoconductors. The hybrid nature of
nanoelectronics devices therefore appears as an essential feature.

The first nanoelectronics experiments where naturally based on dc transport
measurements. However, it soon appeared that studying the response of
nanocircuits to a microwave excitation was also very interesting. For
instance, photo-assisted tunneling was observed in quantum dot circuits,
either between a dot and a reservoir\cite{Kouwenhoven} or inside a double
quantum dot\cite{Oosterkamp}. A microwave irradiation was used to investigate
the Kondo physics in a quantum dot with normal metal reservoirs\cite{Elzerman}%
. Radio-frequency single electron transistors\cite{Schoelkopf:1998} were used
for the electrometry of quantum dots
\cite{Fujisawa:2000,Cassidy:2007,Reilly:2007}. Wide-band\cite{Ota:2010} and
resonant\cite{Petersson:2010,Chorley:2012} techniques were developed to
measure the impedance of quantum dot circuits. Microwaves were finally used to
perform coherent manipulations of single charges\cite{Hayashi:2003} or
spins\cite{Koppens:2006,Nowack:2007}.

A new impulse on the microwave operation of electric nanocircuits is now
starting under the influence of Cavity and Circuit Quantum Electrodynamics
(QED). These experiments study respectively atoms strongly coupled to high
finesse superconducting mirror cavities\cite{Raimond: 2001}, or
superconducting Josephson circuits strongly coupled to microwave
resonators\cite{Wallraff:2004,Paik:2011}. In both cases, one can study the
interaction of light and matter at the most elementary level because the
cavity can trap a controlled low number of photons with a high spectral
purity. Furthermore, the atoms and Josephson circuits behave as effective
two-level systems. By analogy with these experiments, the idea of combining
quantum dot circuits and microwave cavities came out theoretically mainly with
the motivation of using quantum dots as quantum bits for quantum information
science\cite{Imamoglu,Burkard:2006}. In that context, quantum dot circuits are
operated in a well confined regime, i.e. the tunnel junctions between the dots
and reservoirs are very opaque in order to minimize possible decoherence due
to these reservoirs, and the metallic reservoirs are grounded to prevent dc
transport. Several strategies are possible to reach the strong coupling regime
between a nanocircuit and cavity photons. In particular, one can use the
charge\cite{bruhat:2017,Mi:2017,Stockklauser:2017} or spin degree of
freedom\cite{Viennot:2015} of a double quantum dot, or Andreev bound states on
a narrow superconducting contact\cite{Janvier:2015}. This could offer new
means to encode and manipulate quantum information in the context of the
development of quantum computing and quantum communication.

Nevertheless, using nanocircuits coupled to microwave cavities to mimic atomic
cavity QED or Circuit QED experiments is a bit restrictive since it evades
quantum transport effects which occur in out-of equilibrium conditions, as
well as strong correlations effects caused by highly transparent dot-metal
contacts. Along this direction, it appears that the use of microwave cavities
combined with hybrid nanocircuits enables experiments with no analogue in
atomic Cavity QED or metallic Circuit QED. Indeed, cavity photons provide
means to study quantum transport under a new perspective and with a very high
sensitivity. For instance, they can give a direct access to the
out-of-equilibrium state occupation of a double quantum
dot\cite{Viennot:2014a}, reservoir-induced quantum charge relaxation in a
single dot\cite{Bruhat:2016a}, or photo-assisted tunneling
processes\cite{Bruhat:2016a,Stockklauser:2015,Liu:2014,Liu:2015a,Liu:2015}.
They can also represent a powerful tool to characterize exotic condensed
matter states caused by the existence of the dot/reservoir interfaces, such as
Kondo clouds\cite{Desjardins:2017}, split Cooper
pairs\cite{Cottet:2012,Cottet:2014}, or Majorana bound
states\cite{Trif:2012,Schmidt:2013a,Schmidt:2013b,Cottet:2013,Dmytruk:2015,Dartiailh:2016}%
.

Following Ref.\cite{Childress:2004}, we will refer to experiments combining
microwave cavities and electronic nanocircuits as Mesoscopic QED experiments
because the metallic reservoirs in a nanocircuit are typically separated by a
micronic distance (see for instance Fig.\ref{Figure0}c). The purpose of this
short review is to introduce this recent but fast growing new field of
research and give possible directions for its future developments. In section
\ref{building}, we describe the Mesoscopic QED architecture. In section
\ref{theory}, we present a theoretical description of Mesoscopic QED devices,
from the mesoscopic QED Hamiltonian to the semiclassical description of the
cavity signals. In section \ref{Closed}, we review experiments performed so
far in the artificial atom limit. In section \ref{Open}, we discuss Mesoscopic
QED experiments beyond the artificial atom limit, i.e. when the tunneling
dynamics between the nanoconductor and metallic reservoir leads to significant
effects. Section VI presents conclusions and perspectives. In this review we
use $\hbar=1$ in most equations, and we will thus define many parameters as
pulsations. We note e the absolute value of the electron charge ($e>0$).

\section{Building a Mesoscopic QED experiment\label{building}}

\subsection{Embedding a hybrid nanocircuit in a microwave cavity}

\begin{figure}[th]
\includegraphics[width=1\linewidth]{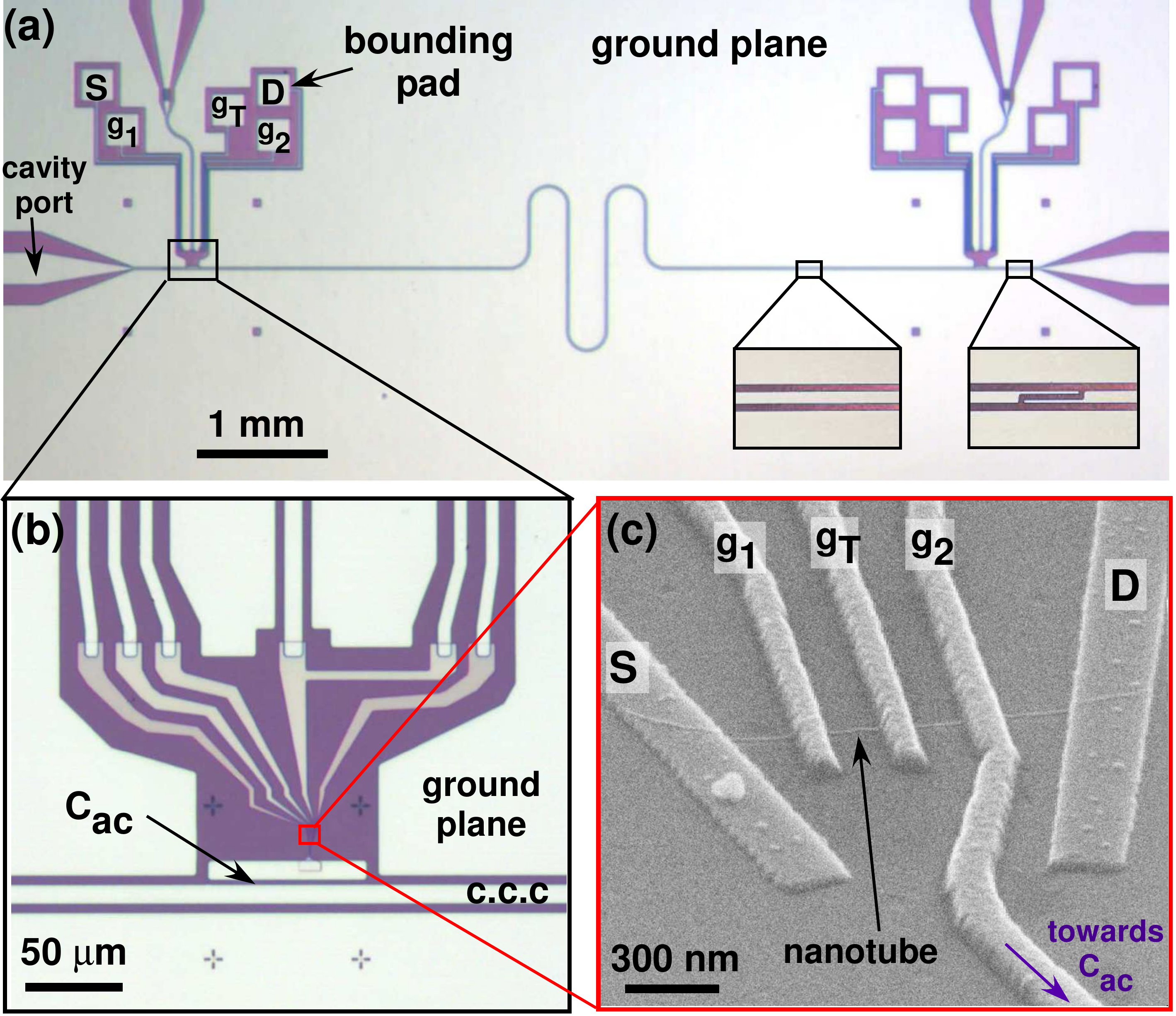}\caption{Example of Mesoscopic
QED device (a) Optical micrograph of a coplanar wave-guide microwave resonator
coupled to a hybrid nanocircuit. The squares are bonding pads for the
nanocircuit, which are isolated from the cavity ground plane and carry DC
voltage or current. The left inset shows a zoom on the coplanar waveguide,
which is a central conductor surrounded by two ground planes. The right inset
shows one of the capacitances which interrupts the waveguide to form a
microwave cavity. (b) Closeup on the connection between the cavity and the
nanocircuit. In this particular sample, an extra superconducting pad was
placed next to the resonator line, providing a large coupling capacitance
$C_{ac}$ between the cavity central conductor (c.c.c.) and one of the sample
gates (c) Scanning electron micrograph of the nanocircuit coupled to the
cavity, here a single wall carbon nanotube (SWNT) connected to source and
drains (S and D) reservoirs as well as three top gates $g_{1}$, $g_{2}$ and
$g_{T}$. Adapted from Ref.[\cite{Viennot:2014a}].}%
\label{Figure0}%
\end{figure}

The combination of hybrid nanocircuits with coplanar microwave cavities pushes
further the on-chip design initially introduced in the context of Circuit QED
experiments to control and readout the state of a superconducting quantum
bit\cite{Wallraff:2004}. Many different types of nanoconductors have already
been embedded in coplanar cavities, such as lateral quantum dots defined on a
GaAs/AlGaAs heterostructures\cite{Frey:2011,Toida:2012} or Si/SiGe
heterostructures\cite{Schmidt:2014,Mi:2016}, quasi-one dimensional conductors
such as carbon nanotubes\cite{Delbecq:2011,Ranjan:2015}, InAs
nanowires\cite{Petersson:2012,Larsen:2015,Lange:2015}, or InSb
nanowires\cite{Wang:2016}, but also graphene quantum dots\cite{Zhang:2014} and
atomic contacts\cite{Janvier:2015}. Different types of metallic contacts can
be used, such as normal metals, superconductors \cite{Bruhat:2016a} and
ferromagnets with collinear\cite{Cottet:2006a} or non-collinear
magnetizations\cite{Crisan: 2016,Viennot:2015}. Therefore, a large variety of
geometries and situations can be studied. Figure \ref{Figure0} shows an
example of Mesoscopic QED sample. Here, the hybrid nanocircuit is a double
quantum dot fabricated out of a carbon nanotube on top of which source (S),
drain (D), and top dc gates ($g_{1}$, $g_{2}$ and $g_{T}$) have been
evaporated (Fig.\ref{Figure0}c). The double dot is coupled capacitively to the
cavity central conductor (c.c.c.), through the capacity $C_{ac}$, near a
cavity electric field antinode (Fig.\ref{Figure0}b). The cavity central
conductor is interrupted by on-chip capacitances such as the one visible in
the right inset of Fig.\ref{Figure0}a. Openings are fabricated across the
cavity ground plane to allow for an electric connection of the source, drain
and gate electrodes of the double dot at bonding pads visible as squares in
Fig.\ref{Figure0}a. These openings must be designed in order to preserve the
cavity quality factor. To avoid spurious photon dissipation, it is also
important to introduce as little conductors as possible close to the cavity.
In experiments realized with semiconducting nanowires and first experiments
realized with carbon nanotubes, numerous nanoconductors have been dispersed on
the substrate during the fabrication process. Stamping
techniques\cite{Waissman:2013} are now used to deposit few carbon nanotube
inside the cavity\cite{Viennot::2014b}, which leads to cavity quality factors
$Q_{0}>10000$ \cite{Viennot:2015,Bruhat:2016a}. In the case of nanostructures
based on two-dimensional electron gases, the coupling to the whole electronic
substrate seems more difficult to avoid. Among other technical progresses, one
can mention the measurement of a double quantum dot in a cavity by using a
Josephson parametric amplifier which considerably speeds up data
acquisition\cite{Stehlik:2015}. Microwave-frequency resonators based on NbTiN
nanowires\cite{Samkharadze:2016} and SQUID arrays\cite{Stockklauser:2017} have
been recently developed in order to increase by a factor of 10 the cavity
electric field in comparison with standard coplanar cavities based on Al or Nb
metallic stripes. This can be used to increase the light/matter coupling.
Other alternative cavity technologies compatible with nanocircuit
architectures are being investigated\cite{Kopke:2015,Gotz:2016,Blien:2016}.

\subsection{Tailoring the spectrum of a hybrid nanocircuit with fermionic
reservoirs}

One important specificity of circuit QED experiments performed with
superconducting quantum bits, in comparison with atomic cavity QED
experiments, is that the spectrum of a superconducting quantum bit is not set
by nature like the spectrum of an atom, but it can be designed at the
nanolithography stage by choosing the circuit geometry and the value of the
capacitive and Josephson elements. This spectrum can also be tuned during the
experiment by using gate voltages or magnetic fluxes. This represents a
significant advantage for performing various tasks such as the selective
microwave control of different quantum bits in an experiment. In Mesoscopic
QED, the use of fermionic reservoirs offers other resources to tailor the
spectrum of a nanocircuit. One must find configurations where the nanocircuit
displays energy scales comparable with the cavity frequency. In this section,
we discuss different possibilities to do so.\begin{figure}[th]
\includegraphics[width=1\linewidth]{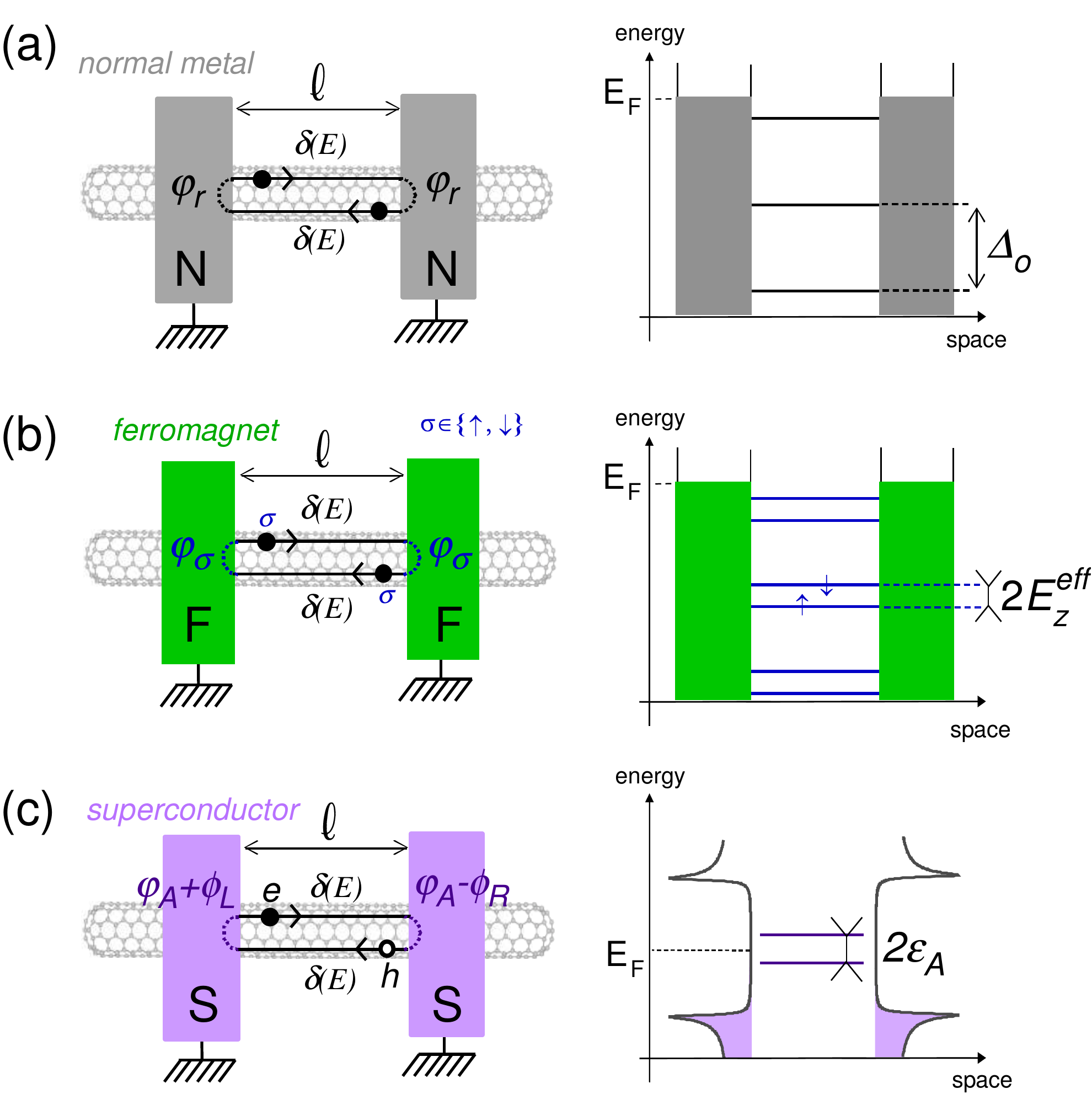}\caption{Hybrid quantum dot
circuits (left panels) and the corresponding energy spectra (right panels) (a)
Case of a quantum dot formed by a portion with length $\ell$ of nanoconductor
(here a carbon nanotube) delimited by normal metal reservoirs (in gray). The
separation between the orbital levels in the dot is $\Delta_{o}=\pi\hbar
v_{F}/\ell$ (b) Case of ferromagnetic reservoirs. The dot levels have an
effective Zeeman spin splitting $2E_{z}^{eff}=\left(  \varphi_{\uparrow
}-\varphi_{\downarrow}\right)  \hbar v_{F}/\ell$ because electrons with spin
$\sigma$ have a spin dependent reflection phase $\varphi_{\sigma}$ on the
ferromagnetic reservoirs (c) Case of superconducting reservoirs with
superconducting phases $\phi_{L}$ and $\phi_{R}$. Electrons and holes are
reflected on the left(right) superconductor with phases\ $\varphi_{A}%
+\phi_{L(R)}$ and $\varphi_{A}-\phi_{L(R)}$ respectively, with $\varphi_{A}$
the Andreev phase. This leads to the appearance of a twofold degenerate low
energy Andreev doublet with a splitting $2\varepsilon_{A}=2\Delta\cos\left(
(\phi_{R}-\phi_{L})/2\right)  $ in the DOS of the nanoconductor.}%
\label{Figure1}%
\end{figure}

For simplicity, we first consider the case where two normal metal reservoirs
delimit a single quantum dot with length $\ell$ inside a single channel
nanoconductor (Fig.\ref{Figure1}a). In a non-interacting scattering picture,
the phase shift acquired by an electron which crosses once the dot is
$\delta(E)=\ell(k_{F}+(E-E_{F})/v_{F})$, where $E_{F}$, $k_{F}$\ and $v_{F}$
are the Fermi energy, wavevector and velocity inside the nanoconductor and $E$
is the electron energy treated at first order. The electron is reflected on
the normal metal contacts with a spin-independent reflection phase
$\varphi_{r,\sigma}=\varphi_{r}$, so that the dot orbital energies are given
by the resonant condition $2\delta(E)+2\varphi_{r}=2\pi n$, with
$n\in\mathbb{N}$ (see Fig.\ref{Figure1}a). This corresponds to an orbital
level spacing
\begin{equation}
\Delta_{o}=\pi\hbar v_{F}/\ell\label{sep}%
\end{equation}
which is typically in the \textrm{THz} range, whereas the cavity frequency
$\nu_{0}$ is typically of the order of $10~\mathrm{GHz}$. Resonant effects
between microwave cavity photons and this local orbital degree of freedom are
therefore impossible. However, there exists other configurations more
favorable to reach the resonant regime as we will see below.

Ferromagnetic materials are widely used to control spin transport in
industrial spintronics devices\cite{Fert:2008}. They also represent a
promising resource for coherent nanocircuits. If a quantum dot is delimited by
ferromagnetic contacts, the reflection phases on its boundaries can become
spin-dependent, because the Stoner exchange fields inside the ferromagnets
provide a spin dependent confinement potential for electrons in the
dot\cite{Cottet:2006b} (see Fig.\ref{Figure1}b). Hence, an effective Zeeman
splitting
\begin{equation}
2E_{z}^{eff}=\left(  \varphi_{\uparrow}-\varphi_{\downarrow}\right)  \hbar
v_{F}/\ell\label{ZeemanEff}%
\end{equation}
occurs inside the dot. The factor $\ell^{-1}$ in the above equation occurs
because $E_{z}^{eff}$ is an interference effect between the two contacts,
given by the resonance condition $2\delta(E)+2\varphi_{r,\sigma}=2\pi n$, for
spin $\sigma\in\{\uparrow,\downarrow\}$. Due to this factor, $E_{z}^{eff}$ can
reach values of the order of $2~\mathrm{Tesla\sim}56~\mathrm{GHz}$ for small
dots with $g\mathrm{\sim}2$ ($\ell\sim500~\mathrm{nm}$)\cite{Cottet:2006a}. In
principle, this value is larger than stray fields from standard ferromagnets,
which are independent of $\ell$ and reach typically a few $100~\mathrm{mT}$.
Furthermore, the effective field of Eq.(\ref{ZeemanEff}) presents the
advantage of being local, which can be useful for building complex devices.

Another interesting possibility to modify the spectrum of a nanocircuit is to
use superconducting contacts which produce the Andreev reflection of an
electron quasiparticle into a hole quasiparticle and vice versa (see
Fig.\ref{Figure1}c). For simplicity, we will consider the case where only the
value of the superconducting gap changes at the superconductor/nanoconductor
interface, from $\Delta$ to 0, in a single channel model. In this case, the
resonant condition between the two contacts is $2\delta(E)+2\varphi_{A}%
\pm(\phi_{R}-\phi_{L})=2\pi n$, with $\varphi_{A}=-\arccos\left(
E/\Delta\right)  $ the Andreev phase and $\phi_{L(R)}$ the phase of the
superconducting order parameter in the left(right) contact. Therefore, in the
limit $\Delta\ll\Delta_{o}$, the interferences between electron and holes lead
to the creation of an Andreev doublet at energies $\pm\varepsilon_{A}$ with
\begin{equation}
\varepsilon_{A}=\Delta\cos\left(  \frac{\phi_{R}-\phi_{L}}{2}\right)
\end{equation}
One has for instance $\Delta\mathrm{\sim}42~\mathrm{GHz}$ for aluminium
electrodes. Hence, the scale $\varepsilon_{A}$ can become very close to the
cavity frequency $\omega_{0}$ if the superconductor/quantum dot/superconductor
junction is inserted inside a flux biased superconducting loop in order to
obtain $\phi_{R}-\phi_{L}<\pi$. Note that the Andreev doublet discussed above
has a two-fold degeneracy, since it can be produced by spin $\uparrow$
electrons and spin $\downarrow$ holes as well as spin $\downarrow$ electrons
and spin $\uparrow$ holes. Such a degeneracy has to be taken into account for
predicting the microwave response of superconducting nanostructures, as we
will see in section \ref{ABS}. In summary, ferromagnetic and superconducting
contacts offer interesting possibilities to make the spin or the electron/hole
energy scales of a single quantum dot comparable to the cavity frequency. In
contrast, the local charge degree of freedom associated to the scale
$\Delta_{o}$ is expected to be off resonant. Note that there can be a local
orbital degeneracy related to the atomic structure of the nanoconductor, such
as the K/K' degree of freedom in a carbon nanotube. Effects related to this
type of degree of freedom will be evoked in section 4.3.3.

Above, we have discussed exclusively the spectrum of a single quantum dot
delimited by fermionic contacts, in order to find intradot degrees of freedom
which could be coupled resonantly to the cavity. However, we will see in the
next sections that non-local charge degrees of freedom associated to tunneling
processes also play a major role in Mesoscopic QED. First, there can be
tunneling between two dots separated by a tunnel barrier with a hopping
constant $t$ (see Fig.\ref{Figure10}). This strongly affects the spectrum of a
double quantum dot, where bonding and antibonding states appear. In practice,
$t\sim\omega_{0}$ can be obtained with many different types of nanoconductors.
Second, electrons can tunnel between a quantum dot and a metallic reservoir
with a tunnel rate $\Gamma$ which can also be of the order of $\omega_{0}$.
These two types of resonances can lead to interesting effects, as we will see below.

\section{Theory of light-matter interaction in mesoscopic QED
devices\label{theory}}

\subsection{The Mesoscopic QED Hamiltonian\label{hamQED}}

\begin{figure*}[t]
\includegraphics[width=1\linewidth]{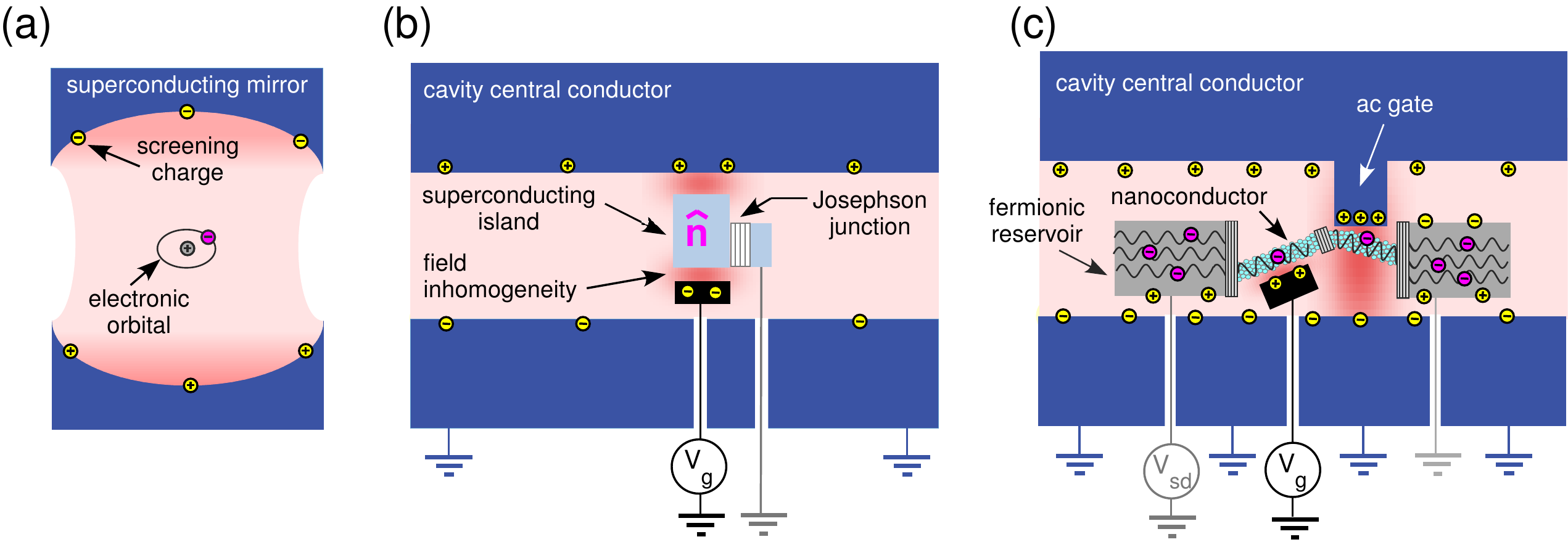}\caption{Schematic
representation of the different types of cavity QED experiments (a) Cavity QED
experiment with an atom (b) Circuit QED experiment with a charge
superconducting quantum bit (c) Mesoscopic QED experiment with a hybrid
nanocircuit. Cavity conductors are represented in blue and dc electrostatic
gates in black. The photonic field is represented in pink, with
inhomogeneities in dark pink. Electronic charges occupying orbitals of the
flying atom in panel a, or tunneling between quasi-localized orbitals of the
nanocircuit in panels c, are represented in fuchsia. Plasmonic screening
charges on the metallic elements are represented in yellow.}%
\label{Figure3}%
\end{figure*}

\subsubsection{Comparison between the different types of cavity QED
experiments}

It is instructive to make a comparison of the physical ingredients involved in
Cavity, Circuit and Mesoscopic QED to identify the specificities of the
light-nanocircuit interaction. Cavity QED focuses on the interactions between
electrons in the atomic orbitals of a flying atom and the photons trapped
inside a superconducting mirror cavity (see Fig.\ref{Figure3}a). In these
experiments, the effect of the cavity magnetic field on the atom can be
disregarded for weak microwave amplitudes\cite{Cohen-Tannoudji:book1}. In most
situations, one can consider that the cavity electric field is constant on the
scale of the atom because the atom is very small in comparison with the
cavity. In this case, the light-matter interaction can be expressed quantum
mechanically as $\hat{H}_{d}=\overrightarrow{\hat{E}}_{0}.\overrightarrow{\hat
{d}}$ with $\overrightarrow{\hat{E}}_{0}$ the quantized cavity electric field
and $\overrightarrow{\hat{d}}=%
{\textstyle\sum\nolimits_{i}}
q_{i}\overrightarrow{\hat{r}}_{i}$ the dipole associated to the atomic charges
$q_{i}$ at position $\overrightarrow{\hat{r}}_{i}$. Note that this charge
distribution includes electrons but also the ions of the atom nucleus.
However, the explicit description of these ions essentially grants the
electroneutrality condition $%
{\textstyle\sum\nolimits_{i}}
q_{i}=0$ which simplifies calculations. Cavity QED mainly focuses on
electronic transitions between the atomic orbitals, induced by the cavity
electric field.

In circuit QED, the concept of orbital degree of freedom is not relevant
anymore because only macroscopic collective degrees of freedom matter, due to
the rigidity of the superconducting phase. For instance, in a Cooper pair box,
which is a small superconducting island coupled to a superconducting reservoir
through a Josephson junction, only the total excess number $\hat{n}$ of
electrons on the island matters (see Fig.\ref{Figure3}b). A second important
difference with atomic Cavity QED is that the cavity field cannot be
considered as homogeneous on the scale of the superconducting quantum bit.
Indeed, its spatial profile is strongly modified by the presence of the
superconducting elements which tend to expel it. Figure \ref{Figure3}b
illustrates this situation in the case of a Cooper pair box embedded in a
coplanar microwave cavity. The cavity electric field concentrates in
capacitive areas between neighboring metallic elements, as represented by the
darker pink areas. This capacitive coupling scheme is often described with a
lumped element circuit model which discretizes the device into nodes with
uniform photonic potential and superconducting phase, connected by capacitors,
inductors or Josephson junctions. In the simplest picture, the cavity is
modeled as a distributed (L,C) line, and the superconducting island in
Fig.\ref{Figure3}b corresponds to a single node contacted through capacitors
and a Josephson junction to the rest of the circuit (see Figure 2 of
Ref.\cite{Blais:2004}). The cavity electric field shifts the island potential
due to the presence of the capacitive coupling between the dot island and the
cavity central conductor. Recently, more sophisticated lumped element circuit
models have been introduced for a more realistic description of circuit QED
devices\cite{Nigg:2012,Solgun:2014}. Note that Josephson circuits with
superconducting loops may equally couple to the cavity magnetic field (not
represented in Fig.\ref{Figure3}b).

Mesoscopic QED represents an intermediate situation between Cavity and Circuit
QED (see Fig.\ref{Figure3}c). Indeed, due to the existence of small confined
nanoconductor areas (like for instance quantum dots), there exists discrete
electronic orbital levels which recall the atomic orbitals of Cavity QED.
However, the cavity field is strongly inhomogeneous on the scale of the
nanocircuit, which rather recalls circuit QED. For instance, one can use ac
gates often connected directly (see Fig.\ref{Figure3}c) or sometimes
capacitively (see Fig.\ref{Figure0}) to the cavity central conductor to
reinforce locally the coupling between the cavity electric field and the
electrons in one small part of the nanocircuit. The area between the ac gate
and the nanoconductor in Fig.\ref{Figure3}c concentrates the electric field,
as represented by the darker pink shade. This provides a capacitive coupling
between the cavity central conductor and the nanoconductor. Field screening
effects represent another source of field inhomogeneity. First, the cavity
fields are confined between the superconducting cavity conductors, represented
in blue in Fig.\ref{Figure3}c. This effect naturally goes together with a
screening of the fields inside the cavity conductors. Second, the fermionic
reservoirs in the nanocircuit can screen at least partially the cavity fields.
These screening effects are due to electronic plasmonic modes, which are only
implicitly taken into account in the usual descriptions of Circuit QED,
through current conservation. In mesoscopic QED, it is not a priori obvious to
take into account plasmonic modes because one must take into account that
fermionic reservoirs host\ simultaneously plasmonic modes \textit{and}
fermionic quasiparticle modes which cause quantum transport effects in the
nanocircuit. These quasiparticle modes are coupled to the localized discrete
electronic orbitals inside the nanoconductors through tunnel junctions.
Tunneling is also at the heart of the Josephson coupling in superconducting
circuits. However, tunneling from a normal metal reservoir involves the
numerous quasiparticle modes in a reservoir on top of the nanoconductor
levels. Therefore, the study of Mesoscopic QED devices requires a description
which combines physical ingredients from both Cavity and Circuit QED. In the
following, we will follow the approach proposed by Ref.\cite{Cottet:2015}.

\subsubsection{Effective decomposition of a Mesoscopic QED device}

\begin{figure}[th]
\includegraphics[width=1.\linewidth]{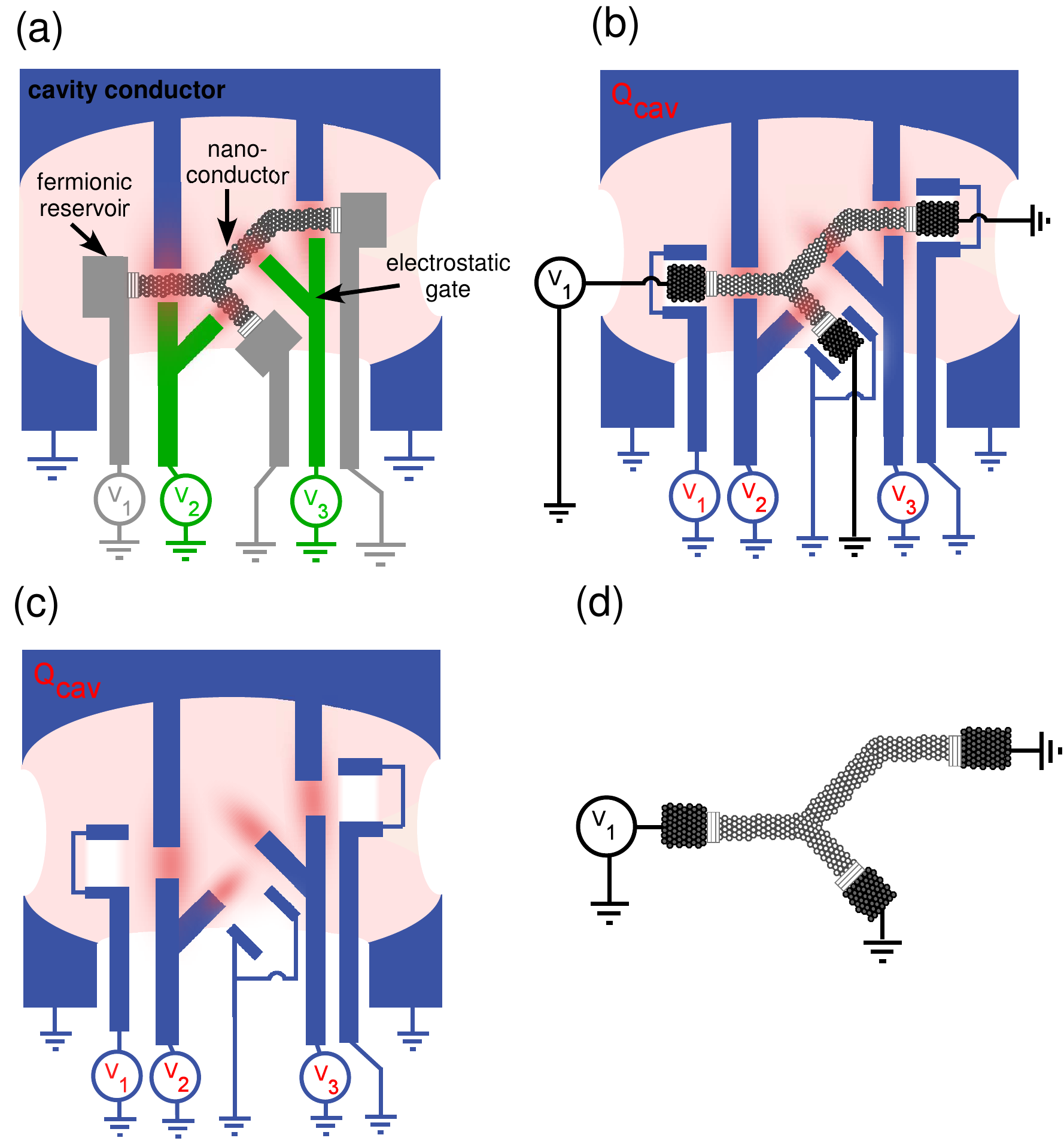}\caption{(a) Schematic
representation of a hybrid circuit QED device which includes a nanoconductor,
fermionic reservoirs and electrostatic gates (b) Heuristic decomposition of
the device into a an effective plasmonic circuit in blue, and an effective
quasiparticle nanocircuit in black (c) Separate representation of the
effective plasmonic circuit. The blue perfect conductors host the plasmons
which propagate together with the photonic modes. The photonic field outside
the blue conductors (in pink) is inhomogeneous, as represented by darker pink
areas near the ac gates, and white areas between the small blue conductors (d)
Separate representation of the effective orbital nanocircuit which hosts
electronic quasiparticle modes. The quasiparticles can tunnel through the
barriers represented by the striped rectangles. Adapted from
Ref.[\cite{Cottet:2015}].}%
\label{Figure4}%
\end{figure}

In order to take into account both quasiparticle tunneling and plasmonic
screening in a minimal way, one can assume that the plasmonic screening
charges on the cavity conductors or fermionic reservoirs have a frequency
which is much higher than all the other relevant frequencies in the device. In
particular, we assume that the plasmonic frequency is much higher than the
tunnel rate between a reservoir and a nanoconductor, or than the tunnel
hopping constant between two dots. Under this assumption, one can decompose
heuristically the nanocircuit of Fig.\ref{Figure4}a into two parts: an
effective orbital nanocircuit represented in black in Figs.\ref{Figure4}b and
d, in which tunneling physics prevails, and an effective plasmonic circuit
made out of perfect conductors, represented in blue in Fig.\ref{Figure4}b and
d. Below, we discuss this decomposition in more details.

In the black circuit of \ref{Figure4}d, the electronic orbital levels in the
different circuit elements are connected through tunnel junctions (striped
rectangles). In order to grant current conservation, the \textquotedblleft
orbital\textquotedblright\ reservoirs have to be connected to the black
voltage source and the ground, through wirings which necessarily host
plasmonic modes. However, one can assume that these plasmonic modes do not
have a significant influence on the value of the cavity field near the
nanocircuit. The role of the black voltage source in Fig.\ref{Figure4}b is
essentially to ensure that the electronic levels in the nearby reservoir are
filled up to the Fermi level plus a shift caused by the applied bias voltage.

The blue circuit of Fig.\ref{Figure4}c is electrically disconnected from the
black circuit. Its represents the physical host of the screening charges which
propagate together with the cavity photons. Some of these blue conductors
directly correspond to the cavity central conductor and ground planes, or to
the nanocircuit dc gates. The other conductors correspond to the nanocircuit
reservoirs, and account at least qualitatively for the local screening of the
cavity field in these reservoirs. This produces a renormalization of the
cavity field which can affect the coupling between the cavity photons and the
quasiparticles in the black orbital circuit.\ Importantly, the blue conductors
are connected to dc sources, drain, and gate voltage sources, similarly to the
initial circuit of Fig.\ref{Figure4}a. This enables one to make a complete
description of the cavity fields, including dc field contributions (this
description will be implemented mathematically in section \ref{Hodge}). Note
that when a tunneling event occurs in the nanocircuit, displacement currents
occur in order to sustain the reorganization of the screening charges in the
whole Mesoscopic QED device. In the model of Fig.\ref{Figure4}b, these
displacement currents are also carried by the blue plasmonic circuit.
Importantly, the model of Fig.\ref{Figure4}b which separates physically the
plasmonic and fermionic modes of the nanocircuit is only an effective model
which we will use in next section to justify the form of the Mesoscopic QED
Hamiltonian. In practice, the plasmonic modes and tunneling quasiparticles are
of course not spatially separated. To calculate in a realistic way how the
spatial profile of the cavity field is renormalized by the screening charges
of the nanocircuit, one should use a microwave simulation software (which
disregards tunneling physics). On the basis of the heuristic model discussed
above, we will introduce in next section a description of Mesoscopic QED where
plasmons are not described explicitly. This approach is allowed by the large
separation in the characteristic timescales associated to plasmons and tunneling.

\subsubsection{Hodge decomposition of the electromagnetic field\label{Hodge}}

In order to exploit the effective model of Fig.\ref{Figure4}b, we will first
quantize the electromagnetic field outside the blue perfect conductors.
Ultrafast plasmonic modes on these conductors will not be treated explicitly
but included through boundary conditions on the blue conductors. These
boundary conditions, which are disregarded in most descriptions of Cavity QED,
make the quantization of the electromagnetic field non-trivial. Some of the
blue conductors are biased with a voltage $V_{i}$, like for instance the
electrostatic gates which are used to tune the positions of the energy levels
in the nanocircuit. Some other are left floating with a constant charge
$Q_{i}$, like the central conductor of a coplanar waveguide cavity. To take
this into account, we will decompose the total electric field $\vec
{E}(\overrightarrow{r},t)$ outside the blue conductors into a longitudinal
component $\vec{E}_{\parallel}(\overrightarrow{r},t)$, which has a finite
gradient but no rotational, a transverse component $\vec{E}_{\perp
}(\overrightarrow{r},t)$, which has a finite rotational but no gradient, and a
harmonic component $\vec{E}_{harm}(\overrightarrow{r})$, which has none, such that:%

\begin{equation}
\vec{E}(\overrightarrow{r},t)=\vec{E}_{\perp}(\overrightarrow{r},t)+\vec
{E}_{\parallel}(\overrightarrow{r},t)+\vec{E}_{harm}(\overrightarrow{r})
\label{y0}%
\end{equation}
Then, it is very convenient to use the Coulomb gauge, defined by
$\overrightarrow{\nabla}_{\overrightarrow{r}}.\vec{A}(\overrightarrow{r}%
,t)=0$. In this case, the magnetic field in the system but also the transverse
electric field can be expressed in terms of the vector potential as
\begin{equation}
\overrightarrow{E}_{\perp}(\overrightarrow{r},t)=-\partial\vec{A}%
(\overrightarrow{r},t)/\partial t \label{y1}%
\end{equation}
and
\begin{equation}
\vec{B}=\overrightarrow{\nabla}_{\overrightarrow{r}}\wedge\vec{A}%
(\overrightarrow{r},t) \label{y2}%
\end{equation}
We now define scalar potentials $U_{\parallel}$ and $\Phi_{harm}$ from the
Eqs.
\begin{equation}
\vec{E}_{\parallel}(\overrightarrow{r},t)=-\overrightarrow{\nabla
}_{\overrightarrow{r}}U_{\parallel}(\overrightarrow{r},t) \label{y3}%
\end{equation}
and
\begin{equation}
\vec{E}_{harm}(\vec{r})=-\vec{\nabla}.\Phi_{harm}(\vec{r}) \label{y4}%
\end{equation}
By combining the Maxwell equations with Eqs.(\ref{y0}-\ref{y4}), one finds
that $\Phi_{harm}(\vec{r})$, $U_{\parallel}(\overrightarrow{r},t)$ and
$\vec{A}(\overrightarrow{r},t)$ are set by separate equations. First,
$\Phi_{harm}$ is a static field set by the boundary conditions on the blue
perfect conductors. More precisely, it fulfills the Poisson equation
\begin{equation}
\Delta_{\vec{r}}\Phi_{harm}(\vec{r})=0
\end{equation}
with boundary conditions corresponding to having charges $Q_{i}$ or voltages
$V_{i}$ on the blue conductors. Second, the potential $U_{\parallel}$ is
instantaneously set by the charge distribution $\{e_{\alpha}%
,\overrightarrow{q}_{\alpha}\}$ in the black nanocircuit, i.e.
\begin{equation}
U_{\parallel}(\overrightarrow{r},t)=%
{\textstyle\sum\nolimits_{\alpha}}
e_{\alpha}G(\vec{r},\overrightarrow{q}_{\alpha}) \label{pooisson}%
\end{equation}
Above, the function $G$ is the solution of the Poisson equation $\Delta
_{\vec{r}}G(\vec{r},\overrightarrow{r}^{\prime})=-\delta(\vec{r}%
-\overrightarrow{r}^{\prime})/\varepsilon_{0}$ with boundary conditions
$Q_{i}=0$ and $V_{i}=0$ on the blue conductors (see Ref.\cite{Cottet:2015} for
details). In other terms, $G(\vec{r},\overrightarrow{r}^{\prime})$ describes
the electrostatic interaction between two charges at points $\vec{r}$ and
$\overrightarrow{r}^{\prime}$, renormalized by the screening charges on the
blue conductors. Finally, $\vec{A}$ follows the propagation equation
\begin{equation}
\left(  \Delta_{\overrightarrow{r}}-\frac{1}{c^{2}}\frac{\partial^{2}%
}{\partial t^{2}}\right)  \vec{A}(\overrightarrow{r},t)=-\mu_{0}%
\overrightarrow{j}_{\perp}(\overrightarrow{r},t) \label{field}%
\end{equation}
with the same boundary conditions as Eq.(\ref{pooisson}). One can conclude
that $\vec{A}$ corresponds to the photonic field which can be quantized.
Interestingly, this picture is similar to the picture used for Cavity QED, up
to two differences. First, in cavity QED, the harmonic potential is generally
omitted because the boundary conditions on the cavity conductors are not
explicitly treated\cite{Vukics:2014}. Second, in the simplest descriptions of
cavity QED, the longitudinal potential $U_{\parallel}$ corresponds to the
potential caused by the atom in vacuum\cite{Cohen-Tannoudji:book1}, i.e.
$U_{\parallel}(\overrightarrow{r},t)=%
{\textstyle\sum\nolimits_{\alpha}}
e_{\alpha}/4\pi\varepsilon_{0}\left\vert \vec{r}-\vec{q}_{\alpha}\right\vert
$. In the case of Mesoscopic QED, this potential is dressed by the screening
charges on the blue conductors. This effect is taken into account by the $G$
function in Eq.(\ref{pooisson}).

\subsubsection{Minimal coupling Hamiltonian}

From the above section, the independent degrees of freedom in the Mesoscopic
QED device are the value of the cavity potential $\vec{A}$, and the position
$\overrightarrow{q}_{\alpha}$ of the charges $e_{\alpha}$ in the black
nanocircuit. In principle, the charge distribution $\{e_{\alpha}%
,\overrightarrow{q}_{\alpha}\}$ includes the crystalline background
$\mathbf{C}$ and electrons from the valence bands\ $\mathbf{V}$\textbf{ }of
the black nanocircuit. However, assuming that these charges are off resonant
with the cavity, one can describe them with the mean field approximation. This
requires to take into account that conduction electrons feel a confinement
potential $V_{conf}(\overrightarrow{r})=\left\langle
{\textstyle\sum\nolimits_{\alpha\in\mathbf{C,V}}}
e_{\alpha_{i}}G(\vec{r}_{\alpha_{i}},\overrightarrow{q}_{\alpha_{i}%
})\right\rangle $, with $\left\langle {}\right\rangle $ a statistical average
on the state of the charges of $\mathbf{C}$ and $\mathbf{V}$. This potential
accounts for the transverse confinement of the conduction electrons inside the
nanocircuit, and the tunnel barriers between the different nanocircuit
elements. Then, by using a quantization procedure analogous to the one used in
cavity QED, one can express the Mesoscopic QED Hamiltonian as (see
Ref.\cite{Cottet:2015} for details):
\begin{align}
\hat{H}_{tot}  &  =\int d^{3}r\hat{\psi}^{\dagger}(\vec{r})\hat{h}_{\rho}%
(\vec{r})\hat{\psi}(\vec{r})+\hat{H}_{Coul}+\omega_{0}\hat{a}^{\dag}\hat
{a}\label{1}\\
&  +\int d^{3}r\left(  \Delta(\vec{r})e^{2\Phi(\vec{r})}\hat{\psi}_{\uparrow
}^{\dagger}(\vec{r})\hat{\psi}_{\downarrow}^{\dagger}(\vec{r})+H.c.\right)
\nonumber
\end{align}
with%
\begin{equation}
\hat{h}_{\rho}(\vec{r})=\frac{1}{2m}\left(  \frac{\overrightarrow{\nabla
}_{\vec{r}}}{i}+e\overrightarrow{\hat{A}}(\vec{r})\right)  ^{2}-e\Phi
_{harm}(\overrightarrow{r})-eV_{conf}(\overrightarrow{r})~\text{,} \label{202}%
\end{equation}%
\begin{equation}
\hat{H}_{Coul}=\frac{e^{2}}{2}\int d^{3}rd^{3}r^{\prime}\hat{\psi}^{\dagger
}(\vec{r})\hat{\psi}^{\dagger}(\vec{r}^{\text{ }\prime})G(\vec{r},\vec
{r}^{\text{ }\prime})\hat{\psi}(\vec{r}^{\text{ }\prime})\hat{\psi}(\vec
{r})~\text{,} \label{coul}%
\end{equation}
and%
\begin{equation}
\overrightarrow{\hat{A}}(\vec{r})=\mathcal{\vec{A}}(\vec{r})i(\hat{a}-\hat
{a}^{\dag})~\text{,} \label{6}%
\end{equation}
We have introduced above the field operator $\hat{\psi}^{\dagger}(\vec
{r})=(\hat{\psi}_{\uparrow}^{\dagger}(\vec{r}),\hat{\psi}_{\downarrow
}^{\dagger}(\vec{r}))$ associated to the creation of conduction electrons in
the black nanocircuit. The potential $V_{conf}(\overrightarrow{r})$ can be
treated on the same footing as the harmonic potential $\Phi_{harm}%
(\overrightarrow{r})$. The term $\hat{H}_{Coul}$ describes Coulomb
interactions between electrons. This term depends on the function $G$ because
Coulomb interactions between the charges $\alpha$ of the black nanocircuit are
renormalized by the screening charges on the blue conductors. In
Eq.(\ref{202}), the vector potential $\overrightarrow{\hat{A}}(\vec{r})$
ensures the gauge invariance of the single electron term. For simplicity,
Eq.(\ref{6}) expresses $\overrightarrow{\hat{A}}(\vec{r})$ by using only one
cavity mode corresponding to the creation operator $\hat{a}$, but
Eqs.(\ref{1})-(\ref{6}) can be generalized straightforwardly to the multimode
case, either to take into account several cavity modes or to describe the
cavity bare linewidth with a bosonic bath. The second line of Eq.(\ref{1}) is
a pairing term which describes superconducting correlations in the
nanocircuit. This term must include a phase factor $\Phi(\vec{r})$ which
depends on the photonic operators, in order to ensure the gauge invariance of
the Hamiltonian (see next section for details).

\subsubsection{Photonic pseudo-potential picture}

Different types of light-matter interactions appear in Hamiltonian (\ref{1}).
Indeed, Eq.(\ref{202}) contains a linear term in $\overrightarrow{\nabla
}_{\vec{r}}.\overrightarrow{\hat{A}}(\vec{r})$ and a non linear term in
$\hat{A}^{2}$. It also contains the exponential of the phase factor $\Phi
(\vec{r})$ which is non-linear. The effect of the non-linear terms is not
negligible, in principle (see Appendix B of Ref.\cite{Cottet:2015} for
details). Therefore, in this section, we introduce a unitary transformation of
the Hamiltonian $\hat{H}_{tot}$ which simplifies the form of the light-matter
interaction. For simplicity, we consider nanocircuits with standard dimensions
and without loops, so that one can disregard magnetic effects induced by the
photons. This means that one can use $\vec{\nabla}_{\vec{r}}\wedge
\overrightarrow{\hat{A}}\simeq0$ on the scale of the whole nanocircuit. This
assumption is valid for all the Mesoscopic QED devices studied experimentally
so far, except Ref.\cite{Janvier:2015}. The more general case will be
discussed elsewhere. When $\vec{\nabla}_{\vec{r}}\wedge\overrightarrow{\hat
{A}}\simeq0$ it is possible to define a photonic pseudo potential $V_{\perp
}(\overrightarrow{r})$ such that%
\begin{equation}
\vec{\nabla}_{\vec{r}}.V_{\perp}(\overrightarrow{r})\simeq\omega
_{0}\mathcal{\vec{A}}(\vec{r})
\end{equation}
and
\begin{equation}
\hat{\Phi}(\vec{r})=e(\hat{a}-\hat{a}^{\dag})V_{\perp}(\overrightarrow{r}%
)/\omega_{0}%
\end{equation}
Then, one can apply to Hamiltonian (\ref{1}) the unitary transformation
$\tilde{H}_{tot}=\mathcal{U}^{\dag}H_{tot}\mathcal{U}$ with
\begin{equation}
\mathcal{U}=\exp\left(  \frac{(\hat{a}^{\dag}-\hat{a})}{\omega_{0}}%
\hat{\mathcal{V}}\right)  \label{UU}%
\end{equation}
and
\begin{equation}
\hat{\mathcal{V}}=-e\int d^{3}rV_{\perp}(\overrightarrow{r})\hat{\psi
}^{\dagger}(\vec{r})\hat{\psi}(\vec{r})
\end{equation}
This leads to the Hamiltonian%

\begin{equation}
\tilde{H}_{tot}=H_{0}+\hat{\mathcal{V}}(\hat{a}+\hat{a}^{\dag})+\omega_{0}%
\hat{a}^{\dag}\hat{a} \label{hi}%
\end{equation}
with%
\begin{align}
H_{0}  &  =\int d^{3}r\hat{\psi}^{\dagger}(r)\widetilde{h}_{\mathcal{\rho}%
}(\vec{r})\hat{\psi}(\vec{r})+\hat{H}_{Coul}\nonumber\\
&  +(\hat{\mathcal{V}}^{2}/\omega_{0})+\int d^{3}r\left(  \Delta(\vec{r}%
)\hat{\psi}_{\uparrow}^{\dagger}(\vec{r})\hat{\psi}_{\downarrow}^{\dagger
}(\vec{r})+H.c.\right)
\end{align}
and
\begin{equation}
\widetilde{h}_{\mathcal{\rho}}(\vec{r})=-\Delta_{\overrightarrow{r}}%
/2m-e\Phi_{harm}(\overrightarrow{r})-eV_{conf}(\overrightarrow{r})
\end{equation}
In the Hamiltonian of Eq.(\ref{hi}), the light-matter interaction is greatly
simplified since it involves a single linear term in $\hat{\mathcal{V}}%
(\hat{a}+\hat{a}^{\dag})$. Interestingly, this Hamiltonian bridges between
Cavity QED and Circuit QED. Indeed, the dipolar electric approximation of
Cavity QED corresponds to a photonic potential which evolves linearly in space
i.e. $V_{\perp}(\overrightarrow{r})=\overrightarrow{\hat{E}}_{0}%
.\overrightarrow{r}$, whereas Circuit QED corresponds to a constant photonic
potential inside each node of the circuit model.

\subsubsection{Anderson-like Hamiltonian for mesoscopic QED\label{Anderson}}

Since tunneling physics is at the heart of quantum transport, it is useful to
reexpress Hamiltonian (\ref{hi}) to describe tunneling explicitly. For this
purpose, one needs to decompose the field operator $\hat{\psi}^{\dagger}%
(\vec{r})$ associated to quasiparticles modes of the black circuit on the
ensemble of the creation operators $\hat{c}_{n}^{\dag}$ for electrons in an
orbital $n$ with energy $\varepsilon_{n}$ of a given circuit element
(reservoir, dot,...). At lowest order in tunneling, one can use $\hat{\psi
}^{\dag}(\vec{r})=\varphi_{n}^{\ast}\hat{c}_{n}^{\dag}$\cite{Prange}. Then,
Hamiltonian (\ref{hi}) directly gives%
\begin{equation}
\hat{H}_{tot}^{tun}=\hat{H}_{0}^{t}+\hat{h}_{int}(\hat{a}+\hat{a}^{\dag
})+\omega_{0}\hat{a}^{\dag}\hat{a} \label{Ht}%
\end{equation}
with%
\begin{equation}
\hat{H}_{0}^{t}=\sum_{n}\varepsilon_{n}\hat{c}_{n}^{\dag}\hat{c}_{n}%
+\sum_{n\neq n^{\prime}}(t_{n,n^{\prime}}\hat{c}_{n^{\prime}}^{\dag}\hat
{c}_{n}+H.c.)
\end{equation}%
\begin{equation}
\hat{h}_{int}=\sum_{n}g_{n}\hat{c}_{n}^{\dag}\hat{c}_{n}+\sum_{n\neq
n^{\prime}}(\gamma_{n,n^{\prime}}\hat{c}_{n^{\prime}}^{\dag}\hat{c}%
_{n}+H.c.)~\text{.} \label{hint2}%
\end{equation}%
\begin{equation}
g_{n}=-e%
{\textstyle\int}
dr^{3}\left\vert \varphi_{n}(\vec{r})\right\vert ^{2}V_{\perp}(\vec
{r})~\text{.} \label{jj1}%
\end{equation}
and%
\begin{equation}
\gamma_{n^{\prime},n}=-e%
{\textstyle\int}
dr^{3}\varphi_{n}^{\ast}(\vec{r})\varphi_{n^{\prime}}(\vec{r})V_{\perp}%
(\vec{r}) \label{jj2}%
\end{equation}
Above, $\hat{H}_{0}^{t}$ is the Anderson-like Hamiltonian of the nanocircuit,
with $t_{n,n^{\prime}}$ the tunnel coupling between orbitals $n$ and
$n^{\prime}$, which is finite only if $n$ and $n^{\prime}$ correspond to two
orbitals in two different circuit elements coupled through a tunnel junction.
The term $\hat{h}_{int}(\hat{a}+\hat{a}^{\dag})$ describes the interaction of
the nanocircuit with the cavity. Cavity photons can have different effects.
First, they can shift the energy of orbital $n$ due to the term in $g_{n}$. In
the limit where $V_{\perp}(\vec{r})$ can be considered as constant inside a
given circuit element, $g_{n}$ can be considered to be the same for all the
orbitals of this element (at zeroth order in tunneling). In this limit, the
coupling of the cavity to the element can be seen as a capacitive coupling due
to the finite capacitance between this element and the cavity central
conductor. This recalls the case of a superconducting charge quantum bit in
Circuit QED, where the qubit island potential is modulated due to the
capacitive coupling between the island and the cavity. In principle, cavity
photons can also produce a direct coupling between two different orbitals of
the nanocircuit, due to the term in $\gamma_{n^{\prime},n}$. This corresponds
to two physically different situations. First, there can be photo-induced
transition terms between two different orbitals of the same circuit element,
which recalls the orbital transitions inside an atom, which are used in Cavity
QED. Second, there can also be a photo-induced tunneling term between two
different circuit elements. Nevertheless, the terms $\gamma_{n^{\prime},n}$
are expected to be weak because $\varphi_{n}(\vec{r})$ and $\varphi
_{n^{\prime}}(\vec{r})$ have a small matrix element in the tunnel theory.

In this review, we will mainly discuss the effects of the $g_{n}$ elements
which are expected to be dominant in most mesoscopic QED devices and have been
sufficient so far to interpret experiments. For standard coplanar cavities
similar to that of Ref.\cite{Wallraff:2004}, the pseudo photonic potential
$V_{\perp}(\vec{r})$ typically varies by $V_{rms}\simeq1~\mathrm{\mu
eV}=240~\mathrm{MHz}$ from the cavity central conductor to the ground plane,
across a spatial gap of $5~\mathrm{\mu m}$. From one nanocircuit site to
another, $V_{\perp}(\vec{r})$ can vary by a significant fraction of $V_{rms}$,
especially if ac gates are fabricated to concentrate the cavity voltage drop
between these sites. Consequently, the value of $g_{n}$ can strongly depend on
the orbital $n$ considered. Even for a given quantum dot, the value of $g_{n}$
can strongly vary from one orbital to the other due to variations in the
spatial profile of $\varphi_{n}(\vec{r})$. For instance, in a single quantum
dot with multiple gates made in a carbon nanotube, $g_{n}/2\pi$ was found to
vary from $55~\mathrm{MHz}$ to $120~\mathrm{MHz}$\cite{Bruhat:2016a}. This
could mean that $V_{\perp}(\vec{r})$ cannot be considered as homogeneous on
the scale of this dot.

\subsection{Semiclassical cavity response in the linear coupling
regime\label{linear}}

\subsubsection{Expression of the cavity photon amplitude\label{epsin}}

So far, in Mesoscopic QED, most experiments have focused on the modification
of the cavity microwave transmission or reflection due to the presence of the
nanocircuit. To describe such a measurement, one can use the general Hamiltonian%

\begin{align}
\hat{H}  &  =\hat{H}_{0}^{t}+\hat{h}_{int}(\hat{a}+\hat{a}^{\dag})+\omega
_{0}\hat{a}^{\dag}\hat{a}+H_{diss}\label{HH}\\
&  +i\hat{a}^{\dag}\varepsilon_{in}e^{-i\omega_{RF}t}-i\hat{a}\varepsilon
_{in}^{\ast}e^{i\omega_{RF}t}\nonumber
\end{align}
which is a generalization of Eq.(\ref{Ht}). Above, we describe cavity
dissipation with a bosonic bath
\begin{equation}
H_{diss}=\int d\epsilon\eta\left(  \omega_{\epsilon}\hat{b}_{\epsilon}^{\dag
}\hat{b}_{\epsilon}+\beta\hat{b}_{\epsilon}^{\dag}\hat{a}+\beta^{\ast}\hat
{a}^{\dag}\hat{b}_{\epsilon}\right)  \label{hdiss}%
\end{equation}
with $\hat{b}_{\epsilon}^{\dag}$ the creation operator for a bosonic mode at
energy $\epsilon$, and $\eta$ the density of modes. For simplicity, we assume
that the coupling constant $\beta$ between the cavity and the bath is
energy-independent. This term was not included in Eq.(\ref{Ht}) but it can be
added by generalizing Eqs.(\ref{6}) and (\ref{UU}) \cite{multimode}. We also
use a drive term with amplitude $\varepsilon_{in}$ which describes the effect
of the continuous microwave tone which is injected at the input of the cavity.
Following the discussion in section \ref{Anderson}, we assume that the
light-matter interaction is well approximated by
\begin{equation}
\hat{h}_{int}=%
{\textstyle\sum\nolimits_{n}}
g_{n}\hat{c}_{n}^{\dag}\hat{c}_{n} \label{hint}%
\end{equation}
From Eqs.(\ref{HH}) and (\ref{hdiss}), one has
\begin{equation}
\frac{d}{dt}\hat{a}=i[\hat{H},\hat{a}]=-i\omega_{0}\hat{a}-i\hat{h}%
_{int}-\varepsilon_{in}e^{-i\omega_{RF}t}-\Lambda_{0}\hat{a} \label{eom}%
\end{equation}
with $\Lambda_{0}=\pi\eta\left\vert \beta\right\vert ^{2}$ the cavity mode
decay rate. Note that the full width at half maximum (FWHM) of the bare cavity
transmission corresponds to the parameter $\varkappa=2\Lambda_{0}$ used in
many Refs. In most experiments performed so far, a large number of cavity
photon has been used, i.e. $\left\langle \hat{a}^{\dag}\hat{a}\right\rangle
>10$. In this case, it is sufficient to treat $\hat{a}$ as a classical
quantity i.e. $\hat{a}\simeq\left\langle \hat{a}\right\rangle =a$. In the case
$\omega_{RF}\simeq\omega_{0}$, one can furthermore use the resonant approximation:%

\begin{equation}
\left\langle \hat{a}\right\rangle \simeq\bar{a}e^{-i\omega_{RF}t} \label{aaa}%
\end{equation}
Above, $\left\vert \bar{a}\right\vert ^{2}$ corresponds to the average cavity
photon number $\left\langle \hat{a}^{\dag}\hat{a}\right\rangle $. One obtains
from the linear response theory
\begin{align}
\left\langle \hat{c}_{n}^{\dag}\hat{c}_{n}\right\rangle (t)  &  =p_{n}+%
{\textstyle\sum\nolimits_{n^{\prime}}}
g_{n^{\prime}}\chi_{n,n^{\prime}}(\omega_{RF})\bar{a}e^{-i\omega_{RF}%
t}\label{nn}\\
&  +%
{\textstyle\sum\nolimits_{n^{\prime}}}
g_{n^{\prime}}\chi_{n,n^{\prime}}(-\omega_{RF})\bar{a}^{\ast}e^{i\omega_{RF}%
t}\nonumber
\end{align}
Above, $\chi_{n,n^{\prime}}(\omega_{RF})$ is by definition the charge
susceptibility expressing how the occupation of level $n$ responds at first
order to a classical modulation of the energy of level $n^{\prime}$, in
stationary conditions. We note $p_{n}$ the average occupation of state $n$ for
$\hat{h}_{int}=0$. One has in the framework of the linear response theory
\begin{equation}
\chi_{n,n^{\prime}}(t-t^{\prime})=-i\theta(t)\left\langle [\hat{c}_{n}^{\dag
}(t)\hat{c}_{n}(t),\hat{c}_{n^{\prime}}^{\dag}(t^{\prime})\hat{c}_{n^{\prime}%
}(t^{\prime})]\right\rangle _{\hat{h}_{int}=0} \label{Chi}%
\end{equation}
where $\left\langle {}\right\rangle _{\hat{h}_{int}=0}$ denotes the
statistical averaging for $g_{n}=0$ for any $n$. Inserting Eqs.(\ref{aaa}) and
(\ref{nn}) into Eq.(\ref{eom}), and keeping only resonant terms, one gets
\begin{equation}
\bar{a}=\frac{\varepsilon_{in}}{\omega_{RF}-\omega_{0}+i\Lambda_{0}-\Xi
(\omega_{RF})} \label{aa}%
\end{equation}
with~%
\begin{equation}
\Xi(\omega_{RF})=%
{\textstyle\sum\nolimits_{n,n^{\prime}}}
g_{n}g_{n^{\prime}}\chi_{n,n^{\prime}}(\omega_{RF}) \label{iiii}%
\end{equation}
the global charge susceptibility of the nanocircuit. Note that in the general
case the indices $n,n^{\prime}$ in Eq.(\ref{aa}) can belong to the
nanoconductors as well as the fermionic reservoirs. From Eq.(\ref{aa}), the
presence of the nanocircuit modifies the apparent frequency and linewidth of
the cavity. For most the experiments reported in the review, the cavity
response is measured at $\omega_{RF}=\omega_{0}$. In the rest of this section,
we will also assume that $\Xi(\omega_{RF})$ can be considered as constant for
$\omega_{RF}-\Lambda_{0}\lesssim\omega_{RF}\lesssim\omega_{RF}+\Lambda_{0}$,
which occurs for instance if the electronic relaxation rates in the
nanocircuit are much larger than the photon relaxation rate $\Lambda_{0}$. In
this case, the cavity frequency and linewidth shifts caused by the presence of
the nanocircuit are, from Eq.(\ref{aa}),%
\begin{equation}
\Delta\omega_{0}=\operatorname{Re}[\Xi(\omega_{0})] \label{j1}%
\end{equation}
and%
\begin{equation}
\Delta\Lambda_{0}=-\operatorname{Im}[\Xi(\omega_{0})] \label{j2}%
\end{equation}
Note that this type of measurement has been pioneered by works in which a
coplanar waveguide resonator\cite{Reulet:1995} or a lumped element
resonator\cite{Deblock:2000} was coupled to an array of $10^{5}$ isolated
mesoscopic rings. The resonator frequency and linewidth shifts revealed the
global electric and magnetic response of the rings to the resonator field, at
a frequency $\omega_{0}\sim350~$\textrm{MHz}. More recently, the admittance of
a single double quantum dot was measured at frequencies $\omega_{0}%
<400~$\textrm{MHz} by using a lumped element (L,C)
resonator\cite{Petersson:2010,Chorley:2012}. One important advantage of the
Circuit QED architecture is the higher frequency of the cavity. One has
typically $\omega_{0}\sim5~$\textrm{GHz}, which is higher than the cryogenic
temperatures obtained with a dilution fridge $k_{b}T\simeq25~$\textrm{mK}%
$\simeq0.5~$\textrm{GHz}. Therefore, the quantum regime with a low number of
cavity photons is accessible. In this regime, the semiclassical approximation
used in this section is not valid anymore. However, understanding the
semiclassical regime of Mesoscopic QED is an important prerequisite before
realizing quantum experiments. Most experiments realized so far correspond to
this regime and can be well understood with the semiclassical picture.

\subsubsection{Cavity signals from the input-output formalism}

\begin{figure}[h]
\includegraphics[width=1\linewidth]{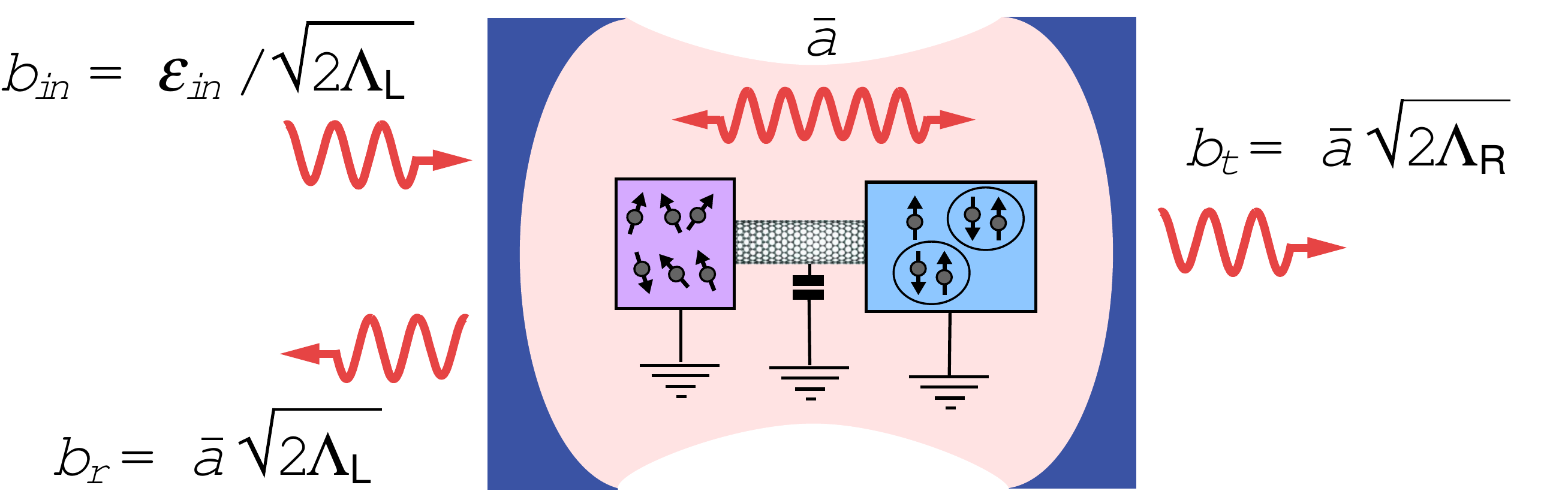}\caption{Representation of
the input-output relations of Circuit QED in the semiclassical limit.}%
\label{Figure13}%
\end{figure}

Above, we have discussed how a nanocircuit induces modifications $\Delta
\omega_{0}$ and$\ \Delta\Lambda_{0}$ of the cavity apparent frequency and
linewidth. For a full description of a Mesoscopic QED experiment, it is
necessary to describe how one can measure these quantities. For that purpose,
one has to take into account of the existence of the input and output ports of
the cavity, through which the cavity is excited and measured. These ports
correspond to the pieces of waveguide connected capacitively to the cavity, on
both sides of the cavity central conductor (see Fig.\ref{Figure0}a). In the
semiclassical limit, the incident, transmitted and reflected photon fluxes in
these ports can be characterized by complex amplitudes $b_{in}$, $b_{t}$ and
$b_{r}$. We assume that the cavity is excited through its input port, which
can be for instance the left port in Fig.\ref{Figure0}. The transmission
$b_{t}/b_{in}$ of the cavity can be obtained experimentally by measuring the
microwave amplitude $b_{t}$ going out through the output port, which can be
the right port in Fig.\ref{Figure0}. In the semiclassical limit, the
correspondence between the approach of section \ref{epsin} and the
input-output formalism of Circuit QED\cite{Walls:2008} gives the relations of
Fig.\ref{Figure13} between $b_{in}$, $b_{t(r)}$, $\varepsilon_{in}$ and
$\bar{a}$\cite{Bruhat:2016a}.

There, $\Lambda_{L(R)}$ corresponds to the contribution of the
left(right) port to the cavity linewidth, which implies $\Lambda_{L}%
+\Lambda_{R}\leq\Lambda_{0}$. From Eq.(\ref{aa}) and Fig.\ref{Figure13}, the
cavity microwave transmission can be expressed as%
\begin{equation}
\frac{b_{t}}{b_{in}}=\frac{2\sqrt{\Lambda_{L}\Lambda_{R}}}{\omega_{RF}%
-\omega_{0}+i\Lambda_{0}-\Xi(\omega_{RF})} \label{ii}%
\end{equation}
Hence, in the semiclassical linear limit, the cavity transmission amplitude is
set by the charge susceptibility of the
nanocircuit\cite{Cottet:2011,Bruhat:2016a,Viennot:2014a,Schiro:2014,Dmytruk:2015}%
. A similar result can be obtained for the cavity reflection amplitude. Note
that above, $b_{t}/b_{in}$ is expressed with the usual quantum mechanics
convention for the definition of Fourier transforms, i.e. $g(\omega)=%
{\textstyle\int\nolimits_{-\infty}^{+\infty}}
dt~g(t)e^{i\omega t}$, which is used in this whole review. In order to
interpret experiments, one has to take into account that microwave equipment
uses the electrical engineering Fourier transform convention, which is complex
conjugated to the former, so that $(b_{t}/b_{in})^{\ast}$ is obtained experimentally.

In practice, the experimental signals which are directly measured are the
transmission phase shift $\Delta\varphi$ and amplitude shift $\Delta A$
defined by%
\begin{equation}
\frac{b_{t}}{b_{in}}=(A_{0}+\Delta A)e^{i(\varphi_{0}+\Delta\varphi)}
\label{iii}%
\end{equation}
In the limit $\omega_{RF}=\omega_{0}$ and $\left\vert \Delta\omega
_{0}\right\vert ,\left\vert \Delta\Lambda_{0}\right\vert \ll\left\vert
\omega_{0}\right\vert ,\left\vert \Lambda_{0}\right\vert $, one finds that the
cavity signals are directly related to the cavity parameters' shifts, i.e.
\begin{equation}
\Delta\varphi=\frac{\Delta\omega_{0}}{\Lambda_{0}}=\operatorname{Re}%
[\Xi(\omega_{0})]\frac{1}{\Lambda_{0}}%
\end{equation}
and%
\begin{equation}
\Delta A=-\frac{\Delta\Lambda_{0}A_{0}}{\Lambda_{0}}=\operatorname{Im}%
[\Xi(\omega_{0})]\frac{A_{0}}{\Lambda_{0}} \label{deltaA}%
\end{equation}
so that $\Delta\varphi$ and $\Delta A$ correspond to the dispersive and
dissipative parts of the signal. Beyond this limit, the experimental data can
be understood by combining Eqs.(\ref{ii}) and (\ref{iii}).

Depending on the regime of parameters fulfilled by the nanocircuit, and in
particular the order of magnitude of the tunnel rates between the dots and
reservoirs, different calculation techniques can be used to calculate
$\Xi(\omega_{0})$. We will discuss several possibilities in the next sections.
In this review, we will only consider cases where the summation on indices $n$
and $n^{\prime}$ in Eq.(\ref{iiii}) can be restricted to internal sites of the
nanoconductor. This requires that the coupling between the cavity and the
nanoconductor sites is much larger than the coupling between the cavity and
the reservoirs. This is not a priori obvious since the nanoconductor is much
smaller than the reservoirs and thus tends to have a smaller capacitance
towards the cavity resonator. However, this feature can be compensated by
using for instance ac top gates which reinforce the coupling between the
nanoconductor and the cavity. In this limit, $\Xi(\omega_{0})$ corresponds to
the charge susceptibility of the nanoconductor at frequency $\omega_{0}$. In
the opposite limit where the coupling between the cavity and the source/drain
of the nanocircuit is dominant, it has been observed experimentally that the
cavity signals essentially show replicas of the conductance
signal\cite{Delbecq:2011}.

\section{Mesoscopic QED experiments in the artificial atom limit\label{Closed}%
}

\subsection{Measuring the internal degrees of freedom of a nanocircuit with
cavity photons}

In this section, we consider a nanocircuit with a very small tunnel coupling
to metallic reservoirs. Hence, the nanocircuit behaves as an artificial atom
and a direct analogy with Cavity or Circuit QED experiments can be drawn.
Assuming that the summation on index $n$ in Eq.(\ref{hint}) can be restricted
to sites which do not correspond to a reservoir, one gets\begin{figure}[th]
\includegraphics[width=1\linewidth]{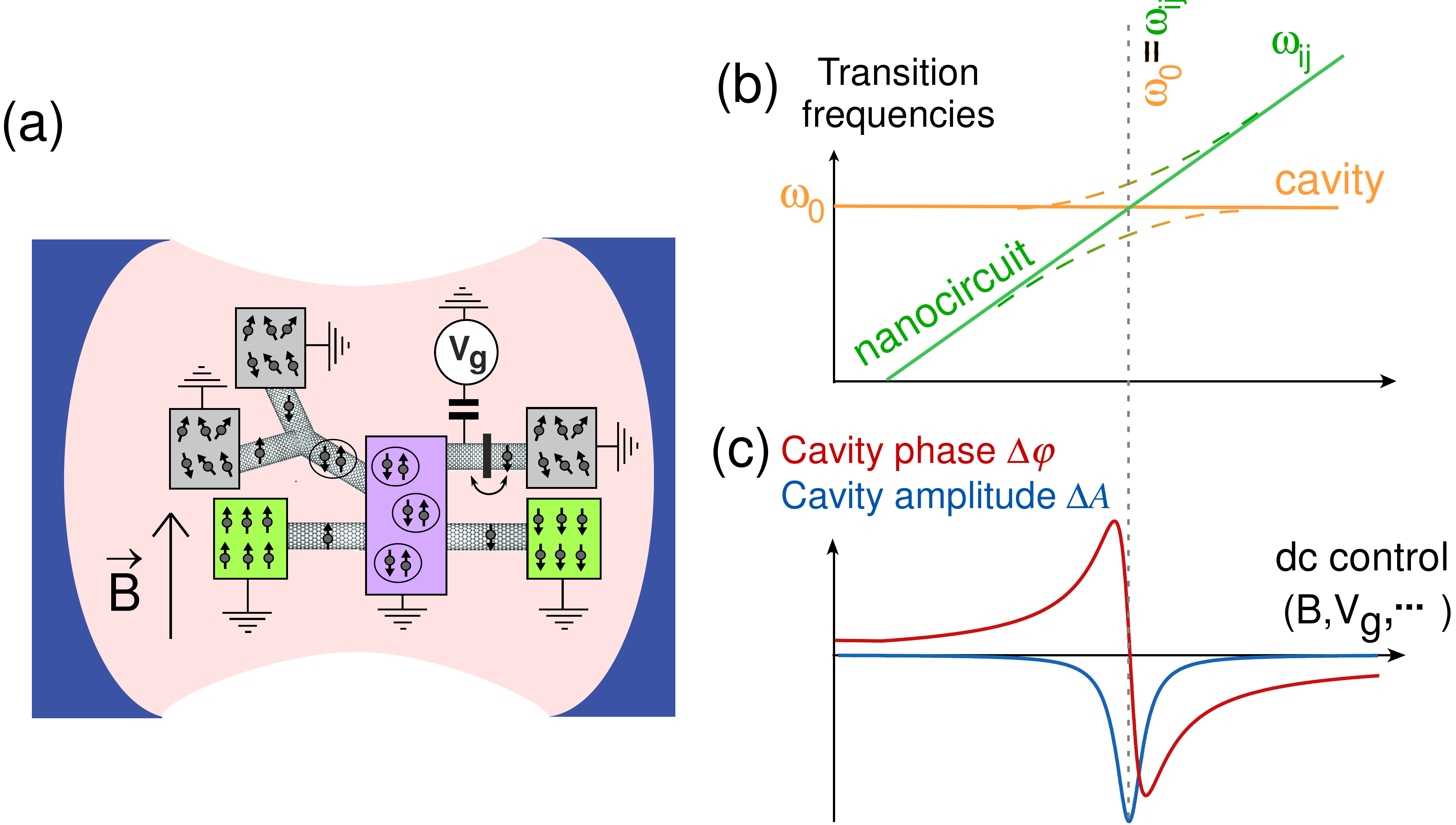}\caption{Principle of a cavity
measurement at a constant frequency $\omega_{RF}=\omega_{0}$, in the presence
of a nanocircuit with internal degrees of freedom, represented in panel (a).
When the control parameters like a magnetic field $B$ or a gate voltage
$V_{g}$, are varied, a nanocircuit transition frequency can become resonant
with the cavity as shown in panel (b). This gives variations of the cavity
transmission phase and cavity transmission amplitude represented in panel
(c).}%
\label{Figure5}%
\end{figure}%
\begin{equation}
\Xi(\omega_{RF})\simeq%
{\textstyle\sum\limits_{ij}}
\frac{g_{ij}^{2}(n_{j}-n_{i})}{\omega_{RF}-\omega_{ij}+i\Gamma_{ij}}
\label{XiInt}%
\end{equation}
Above, $\omega_{ij}$ is an internal transition frequency of the nanoconductor
between states $i$ and $j$, $\Gamma_{ij}$ is the decoherence rate associated
to this transition, and $n_{i}$ is the average occupation of state $i$. We
will see later in a particular example how to derive this expression which is
valid for $\Gamma_{ij}\ll k_{B}T$ and all transitions frequencies well
separated (see section \ref{MasterEq}).

We now discuss a measurement of $\Delta\varphi$ and $\Delta A$ made with a
constant cavity excitation frequency $\omega_{RF}=\omega_{0}$. In the absence
of external dc bias voltages, one has necessarily $\Delta A<0$, because the
nanocircuit can only damp cavity photons. By combining Eqs.(\ref{ii}),
(\ref{iii}) and (\ref{XiInt}), one can see that the cavity signals are
resonant for $\omega_{0}=\omega_{ij}$. In a typical experiment, the
transitions frequencies $\omega_{ij}$ can be tuned with nanocircuit control
parameters such as dc gate voltages or the external magnetic field. When these
parameters are swept, the cavity signals provide a cut of the nanocircuit
excitation spectrum at frequency $\omega_{0}$, as illustrated by
Fig.\ref{Figure5}. For well separated resonances, $\Delta A$ presents a
negative peak at $\omega_{0}=\omega_{ij}$ whereas $\Delta\varphi$ presents a
variation with a sign change at $\omega_{0}=\omega_{ij}$ (see
Fig.\ref{Figure5}c). Therefore, the excitation spectrum of the nanocircuit is
more straightforwardly readable in the $\Delta\Lambda_{0}$ signal, in principle.

For quantum information applications, the strong coupling regime between the
nanocircuit and the cavity is intensively sought after. This regime
corresponds to having one of the nanocircuit transitions $i\leftrightarrow j$
such that $g_{ij}>\Gamma_{ij},\Lambda_{0}$. We will also assume below that the
other nanocircuit transitions do not affect significantly the cavity. For the
most simple characterization of the nanocircuit/cavity coupling, the
nanocircuit is kept in its ground state ($p_{j}=1$, $p_{i}=0$). In these
conditions, the cavity response is set by the charge susceptibility
\begin{equation}
\Xi(\omega_{RF})\simeq\frac{g_{t}^{2}}{\omega_{RF}-\omega_{ij}+i\Gamma
_{2}^{\ast}} \label{res}%
\end{equation}
with $\Gamma_{2}^{\ast}=\Gamma_{ij}$ the decoherence rate of the resonant
nanocircuit transition, and $g_{t}$ its coupling to the cavity. In the strong
coupling limit, the cavity resonance versus $\omega_{RF}$ shows two peaks
instead of the single peak of the weakly coupled regime, due to the strong
hybridization of the cavity states with the nanocircuit states $i$ and $j$
(see Figure \ref{Figure18}d). This regime has already been reached for
instance with atomic cavity QED\cite{Thompson:1992,Brune:1996}, or isolated
quantum dots in optical cavities\cite{Reithmaier:2004,Yoshie:2004}, or
superconducting quantum bits coupled to microwave cavities\cite{Wallraff:2004}%
. It has also also been reached more recently in Mesoscopic QED, as will be
discussed in section 4.2.3

It is useful to define figures of merit to characterize the strength of a
light/matter resonance. We first define the cooperativity
\begin{equation}
C_{e-ph}=g_{t}^{2}/\Lambda_{0}\Gamma_{2}^{\ast}%
\end{equation}
From Eq.(\ref{ii}), in the resonant regime $\omega_{ij}=\omega_{0}$, this
figure of merit indicates whether the cavity dissipation is dominated by the
intrinsic cavity damping $\Lambda_{0}$ ($C_{e-ph}<1$) or by the nanocircuit
dissipation ($C_{e-ph}>1$). The cooperativity can also be used to express the
lasing threshold $C_{e-ph}\gtrsim1/2$ to obtain a lasing effect with a single
qubit in a cavity, in a situation where $\omega_{ij}=\omega_{0}$ and the qubit
dephasing $\Gamma_{\varphi}^{\ast}$ is much stronger than the qubit relaxation
$\Gamma_{1}$ ($\Gamma_{\varphi}^{\ast}\gg\Gamma_{1}$ so that $\Gamma_{2}%
^{\ast}=\Gamma_{\varphi}+(\Gamma_{1}/2)\simeq\Gamma_{\varphi}$) (see for
instance Refs.\cite{Andre:2006,Andre:2010}).

In the devices considered in the present review, $\Lambda_{0}\ll\Gamma
_{2}^{\ast}$ is always fulfilled due to the high quality of the resonators
used. In this context, the ratio\cite{defC}%

\begin{equation}
Q_{e-ph}=\sqrt{1+\sqrt{2}}g_{t}/\Gamma_{2}^{\ast} \label{Qeh}%
\end{equation}
is also instructive. For a resonant cavity/nanocircuit transition
($\omega_{ij}=\omega_{0}$) and in the limit of a negligible bare cavity
linewidth, two resonance peaks are visible in the cavity response $\left\vert
b_{t}/b_{in}\right\vert $ as soon as $Q_{e-ph}\geq1$. Below, we discuss
various circuit geometries where the transition $i\leftrightarrow j$
corresponds to charge, spin or electron/hole degrees of freedom. The values of
$C_{e-ph}$ and $Q_{e-ph}$ obtained for in these different cases are presented
in Table 1.

\subsection{Charge double quantum dots with normal metal
contacts\label{DQDcharge}}

\subsubsection{Hamiltonian of the device\label{HHdqd}}

\begin{figure}[th]
\includegraphics[width=0.5\linewidth]{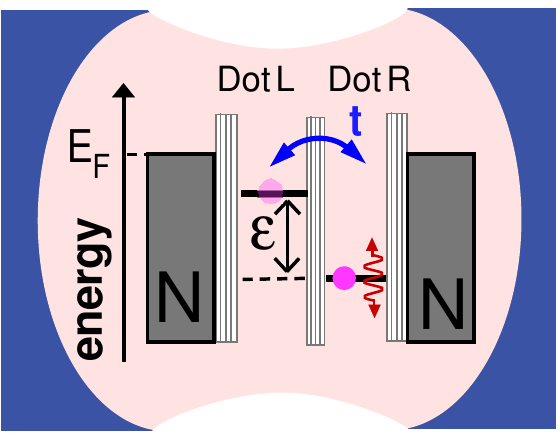}\caption{Schematic energy
representation of a double quantum dot in a microwave cavity. The double dot
orbitals in the left and right dots have an energy separation $\varepsilon$
and are coupled with a hoping constant $t$. The normal metal reservoirs states
are filled up to the Fermi energy $E_{F}$. Here, we represent a case where
only the right orbital energy level is modulated by\ the cavity electric field
(red arrow).}%
\label{Figure19}%
\end{figure}

The case of a double quantum dot embedded in a microwave cavity has received a
lot of experimental attention
\cite{Frey:2011,Frey:2012a,Petersson:2012,Schroer:2012,Frey:2012b,Toida:2012,Toida1:2013,Basset:2013,Delbecq:2013,Zhang:2014,Viennot:2014a,Basset:2014,Wang:2016,Schroer:2012,Deng:2015}%
.. The intrinsic level separation $\Delta_{o}$\ between the orbitals of one
dot (see Eq.(\ref{sep})) is usually very large in comparison with the other
energy scales of the device. Therefore, it is sufficient to consider a single
orbital with energy $\varepsilon_{L(R)}$ in dot $L(R)$. These two orbitals are
coupled with a hopping constant $t$. In practice, each dot is also contacted
to a normal metal reservoir, which enables one to control and measure the
double dot charge. One can use the energy diagram of Fig.\ref{Figure19}, where
the Fermi levels in the reservoirs are filled up to the Fermi energy $E_{F}$
and the orbital levels of the two dots have an energy separation
$\varepsilon=\varepsilon_{L}-\varepsilon_{R}$. This last parameter can be
controlled with the gate voltages $V_{g}^{L}$ and $V_{g}^{R}$ shown in
Fig.\ref{Figure6}.

It is possible to tune $V_{g}^{L}$ and $V_{g}^{R}$ such that there is a single
electron in the double dot due to Coulomb blockade. In the spin-degenerate
case, the spin degree of freedom can be disregarded to describe this situation
since the two spin species play the same role and are not present
simultaneously in the double dot. Therefore, the only internal degree of
freedom relevant to describe the internal dynamics of the double dot in this
limit is the left/right charge degree of freedom. In this framework, the
double dot Hamiltonian writes
\begin{equation}
\hat{H}_{0}^{t}=\frac{\varepsilon}{2}(\hat{c}_{L}^{\dag}\hat{c}_{L}-\hat
{c}_{R}^{\dag}\hat{c}_{R})+t\hat{c}_{L}^{\dag}\hat{c}_{R}+t^{\ast}\hat{c}%
_{R}^{\dag}\hat{c}_{L}+\hat{H}_{Coul} \label{HH1}%
\end{equation}
where $\hat{H}_{Coul}$ forbids the double occupation of the double dot. Using
the basis of bonding and antibonding states of the double dot, one gets%
\begin{equation}
\hat{H}_{0}^{t}\simeq\frac{\omega_{DQD}}{2}(\hat{c}_{-}^{\dag}\hat{c}_{-}%
-\hat{c}_{+}^{\dag}\hat{c}_{+})+\hat{H}_{Coul}%
\end{equation}
with $\omega_{DQD}=\sqrt{\varepsilon^{2}+4t^{2}}$ the double dot transition
frequency. Above, we have used the creation operators%
\begin{equation}
\hat{c}_{+}^{\dag}=\cos[\theta/2]\hat{c}_{L}^{\dag}+\sin[\theta/2]\hat{c}%
_{R}^{\dag}%
\end{equation}
and%
\begin{equation}
\hat{c}_{-}^{\dag}=-\sin[\theta/2]\hat{c}_{L}^{\dag}+\cos[\theta/2]\hat{c}%
_{R}^{\dag}%
\end{equation}
for bonding and antibonding states in the dot, and the parameter
$\theta=\arctan[2t/\varepsilon]$.

In most experiments designed so far, the samples have been designed with an
asymmetric coupling of the two dots to the cavity, in order to modulate the
parameter $\varepsilon$ with the cavity electric field (see for instance
Figure \ref{Figure6}a). In this case, following section \ref{Anderson}, the interaction
term between the double dot and the cavity can be expressed as
\begin{equation}
\hat{h}_{int}(\hat{a}+\hat{a}^{\dag})=\left(  g_{L}\hat{c}_{L}^{\dag}\hat
{c}_{L}+g_{R}\hat{c}_{R}^{\dag}\hat{c}_{R}\right)  (\hat{a}+\hat{a}^{\dag})
\end{equation}
with $g_{L}\neq g_{R}$. Using a rotating wave approximation, this term can be
expressed as%
\begin{equation}
\hat{h}_{int}(\hat{a}+\hat{a}^{\dag})\simeq g_{t}(\hat{a}^{\dag}\hat{c}%
_{-}^{\dag}\hat{c}_{+}+\hat{a}\hat{c}_{+}^{\dag}\hat{c}_{-}) \label{HH2}%
\end{equation}
with%
\begin{equation}
g_{t}=\frac{g_{R}-g_{L}}{2}\sin[\theta] \label{gt}%
\end{equation}
This coefficient depends on the differential coupling $g_{R}-g_{L}$ because
the coupling to the cavity occurs through the modulation of the dot orbital
detuning $\varepsilon$.

As we have seen in the previous section, in order to obtain the strong
coupling regime, one needs to have a small enough $\Gamma_{2}^{\ast}$. The
coherence of a charge double dot is mainly limited by charge noise due to
charge fluctuators which move in the vicinity of the double dot. This induces
fluctuations of the parameters which are electrically controlled, i.e.
$\varepsilon$ in the present case. This effect should be minimal at the charge
noise sweet spot $\varepsilon=0$ where $\partial\omega_{DQD}/\partial
\varepsilon=0$, similar to what has been done for early days charge
superconducting quantum bits which are also affected by this
problem\cite{Cottet: 2002}. Hence, it would be interesting to perform a
systematic study of the figures of merit of the cavity/double dot resonance
when the double dot parameters, and in particular $\varepsilon$, are varied.

In principle, the coupling of a double-dot dot to a microwave cavity can be
mediated by other variables than $\varepsilon$, depending on the sample
design. The first manipulations of the quantum state of a double dot were
performed by modulating the parameter $\varepsilon$ with a strong classical
drive\cite{Hayashi:2003}. Recently, similar experiments were performed by
modulating the interdot tunnel parameter $t$%
\cite{Bertrand:2015,Reed:2016,Martins:2016}, along the theory proposal of
Ref.\cite{Loss:1998}. One could push further this idea by building ac gates
connecting the double dot barrier to the cavity central conductor, in order to
modulate the interdot tunnel parameter $t$ with the cavity electric field. In
principle, this could enable one to obtain double dots with a better
coherence, since the electric (and charge noise) control of the variable
$\varepsilon$ can be shunted, in this case. In section \ref{ABS}, we will
discuss an alternative strategy to control electrically $t$\ which consists in
using a superconducting contact.

\subsubsection{Master equation description\label{MasterEq}}

This section shows how to calculate the cavity charge susceptibility
$\Xi(\omega_{RF})$ of the double dot when a microwave cavity is coupled to a
single transition $i\leftrightarrow j$ which corresponds to the left/right
charge degree of freedom of a double quantum dot. The dynamics of this
mesoscopic QED device can be described by using a master equation approach (or
Lindbladt formalism) already widely used for Cavity or Circuit
QED\cite{Cohen-Tannoudji:book1}. From Eqs. (\ref{HH}), (\ref{HH1}) and
(\ref{HH2}), one gets
\begin{equation}
\frac{d}{dt}\hat{a}=-i\omega_{r}\hat{a}-ig_{t}\hat{c}_{-}^{\dag}\hat{c}%
_{+}-\Lambda_{0}\hat{a}+\varepsilon_{in}e^{-i\omega_{RF}t} \label{v1}%
\end{equation}%
\begin{equation}
\frac{d}{dt}\hat{c}_{-}^{\dag}\hat{c}_{+}=-i\omega_{DQD}\hat{c}_{-}^{\dag}%
\hat{c}_{+}+ig_{t}(\hat{n}_{+}-\hat{n}_{-})\hat{a}-\Gamma_{2}^{\ast}\hat
{c}_{-}^{\dag}\hat{c}_{+} \label{v2}%
\end{equation}
with $\Gamma_{2}^{\ast}$ the total decoherence rate of the double dot
transition, which includes relaxation and dephasing effects. Importantly, the
above equations are valid for $\Gamma_{2}^{\ast}\ll k_{B}T$. In the
semiclassical limit with $\omega_{RF}\simeq\omega_{0}$, one can use
Eq.(\ref{aaa}) and the resonant expression%
\begin{equation}
\hat{c}_{-}^{\dag}\hat{c}_{+}\simeq\left\langle \hat{c}_{-}^{\dag}\hat{c}%
_{+}\right\rangle _{0}+\overline{\hat{c}_{-}^{\dag}\hat{c}_{+}}e^{-i\omega
_{RF}t}%
\end{equation}
Hence, Eqs. (\ref{v1}) and (\ref{v2}) give%
\begin{equation}
\bar{a}=\frac{\varepsilon_{in}}{\left(  \omega_{RF}-\omega_{0}+i\Lambda
_{0}-\frac{g_{t}^{2}(n_{-}-n_{+})}{\omega_{RF}-\omega_{DQD}+i\Gamma_{2}^{\ast
}}\right)  }%
\end{equation}
with $n_{-}$ and $n_{+}$ the average occupation numbers of the bonding and
antibonding states. A comparison with Eq.(\ref{aa}) gives
\begin{equation}
\Xi(\omega_{RF})=\frac{g_{t}^{2}(n_{-}-n_{+})}{\omega_{RF}-\omega
_{DQD}+i\Gamma_{2}^{\ast}} \label{XiDQD}%
\end{equation}
The above equation corresponds to a well known result in the dispersive regime
$\omega_{RF}-\omega_{DQD}\gg\Gamma_{2}^{\ast}(V_{g}^{L},V_{g}^{R})$ where
$\Gamma_{2}^{\ast}$ can be disregarded (see for instance Ref.\cite{Blais:2004}%
). In this limit, depending on whether the nanocircuit is in the state $+$ or
$-$, the cavity shows a frequency pull $\Delta\omega_{0}=\pm g_{t}^{2}%
/(\omega_{RF}-\omega_{DQD})$. This can be used to read out the state of the
nanocircuit in a nondestructive way, since in this limit, $\chi(\omega_{RF})$
accounts for second order processes which do not change the state of the
nanocircuit. This method is widely used to read out the state of
superconducting quantum bits\cite{Wallraff:2004}. In section \ref{DQDcharge},
we consider double dots with no voltage bias, and we also assume that the dot
levels are not resonant with the normal metal reservoirs, so that the electron
which is trapped in the double dot cannot escape. We also assume that the
power of the microwave tone applied to the cavity is too low to excite the
transition between the bonding and antibonding states. In this case, at
equilibrium, one has $n_{-}=1$ and $n_{+}=0$, which leads to Eq.(\ref{res}).

\subsubsection{Experimental results\label{JayPRB}}

\paragraph{Weak coupling limit}

\begin{figure}[th]
\includegraphics[width=1\linewidth]{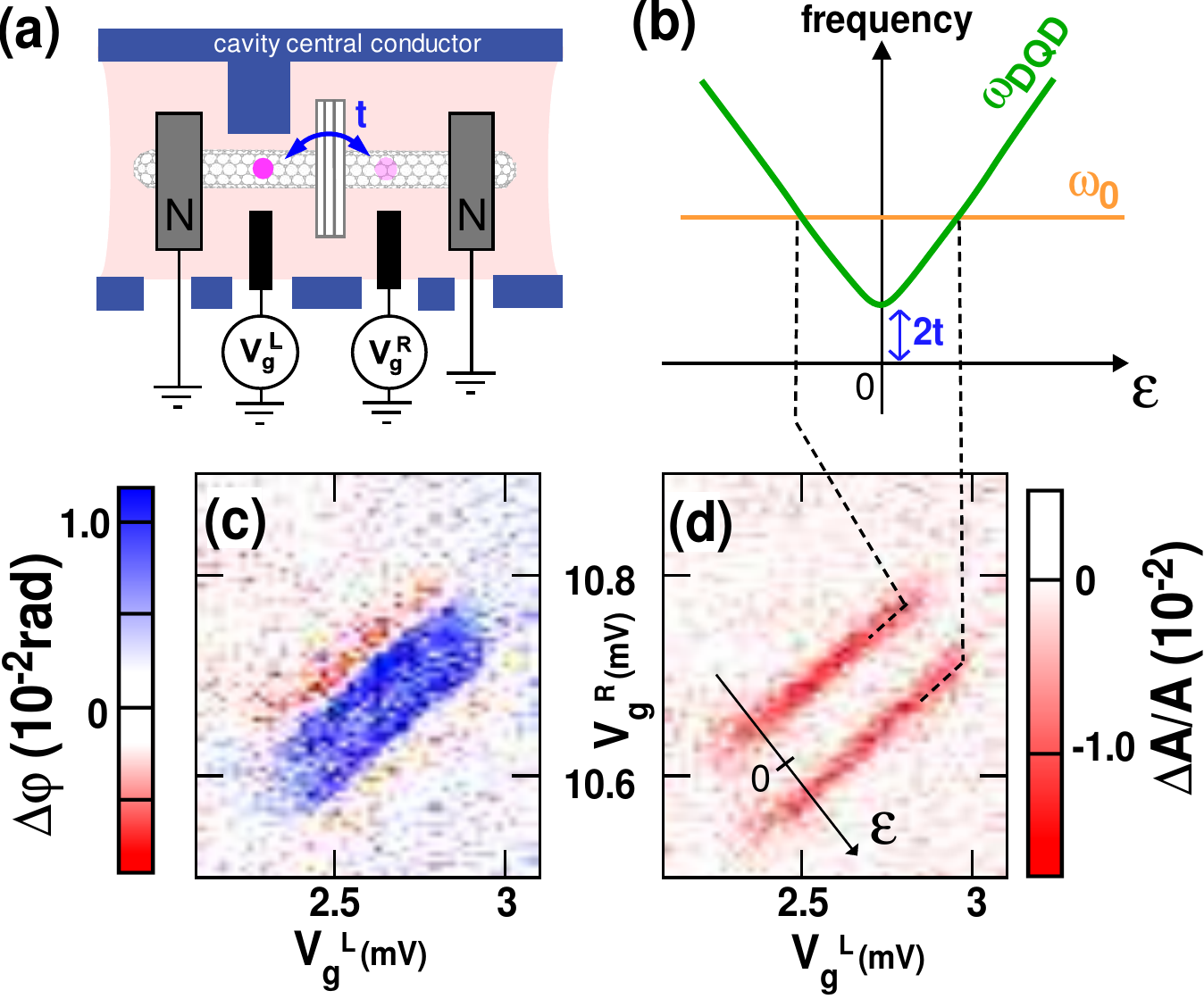}\caption{Response of a
microwave cavity coupled to a double quantum dot with normal metal contacts.
The double dot is represented schematically in panel (a). The energy levels in
the left and right dots can be shifted using the gate voltages $V_{g}^{L}$ and
$V_{g}^{R}$. Using the simplified dispersion relation $\omega_{DQD}%
=\sqrt{\varepsilon^{2}+4t^{2}}$ for the double quantum dot, the cavity an
double dot are resonant twice when increasing $\varepsilon$, as shown by the
dispersion curves of $\omega_{DQD}$ and $\omega_{0}$ in panel (b). These
resonances are visible in the cavity phase signal of panel (c) as sign
changes, and in the cavity dissipative signal of panel (d) as dissipation
peaks. Adapted from Ref.[\cite{Viennot:2014a}].}%
\label{Figure6}%
\end{figure}

The resonance between a closed charge double-dot and a microwave cavity in the
linear coupling regime has been measured by many groups, with different types
of nanoconductors and grounded normal metal reservoirs. In all these
experiments, the differential coupling $g_{R}-g_{L}$ between the double dot
and the cavity is reinforced thanks to a local ac gate connected only to one
dot. First experiments have revealed a very weak light matter coupling, i.e.
$Q_{e-ph}\ll1$ (see Table 1 for various examples). Figure \ref{Figure6} shows
an example of experimental data obtained with a double dot made in a carbon
nanotube\cite{Viennot:2014a}, with an average photon number in the cavity
$\left\langle \hat{a}^{\dag}\hat{a}\right\rangle \simeq40$. If $2t>\omega_{0}%
$, one has $\omega_{DQD}>\omega_{0}$ for any value of $\varepsilon$ so that
the double dot and cavity are always off resonant and the signals
$\Delta\varphi$ and $\Delta A$ show broad responses centered on $\varepsilon
=0$ (not shown). For $2t<\omega_{0}$, two resonances between the cavity and
the double dot are expected when $\varepsilon$ varies, for $\varepsilon
=\pm\sqrt{\omega_{0}^{2}-4t^{2}}$ (see Fig.\ref{Figure6}b). As expected from
Eq.(\ref{XiDQD}), $\Delta\varphi$ shows two sign changes along the
$\varepsilon$ axis, corresponding to these two resonances (Fig.\ref{Figure6}%
c), whereas $\Delta A$ keeps a constant negative sign and shows two simple
resonances (Fig.\ref{Figure6}d). From the cavity signals of Fig.\ref{Figure6},
one can determine $(g_{L}-g_{R})/2\pi=3.3~\mathrm{MHz}$ and $\Gamma_{2}^{\ast
}/2\pi=345~\mathrm{MHz}$ at $\varepsilon=0$. With $\Lambda_{0}%
=0.96~\mathrm{MHz}$ , this gives $Q_{e-ph}=0.015$ and $C_{e-ph}=0.033$.
Therefore, the strong coupling regime is far from being reached in this experiment.

In spite of a weak electron-photon coupling, it is possible to obtain
interesting photon emission effects by applying a strong microwave drive to a
double dot. This was shown recently with an InAs double dot. The same electron
trapped in the double dot is repeatedly driven to the excited state by a
microwave excitation with a strong amplitude which is off resonant with the
cavity and applied directly on the double dot gates. This generates a double
dot population inversion. which leads to cavity photon emission or absorption,
with a rate which depends on the double dot and cavity dynamics but also on
the dissipation caused by phonons in the InAs nanowire\cite{Stehlik:2016}%
.

\paragraph{Strong coupling limit}

\begin{figure}[th]
\includegraphics[width=1\linewidth]{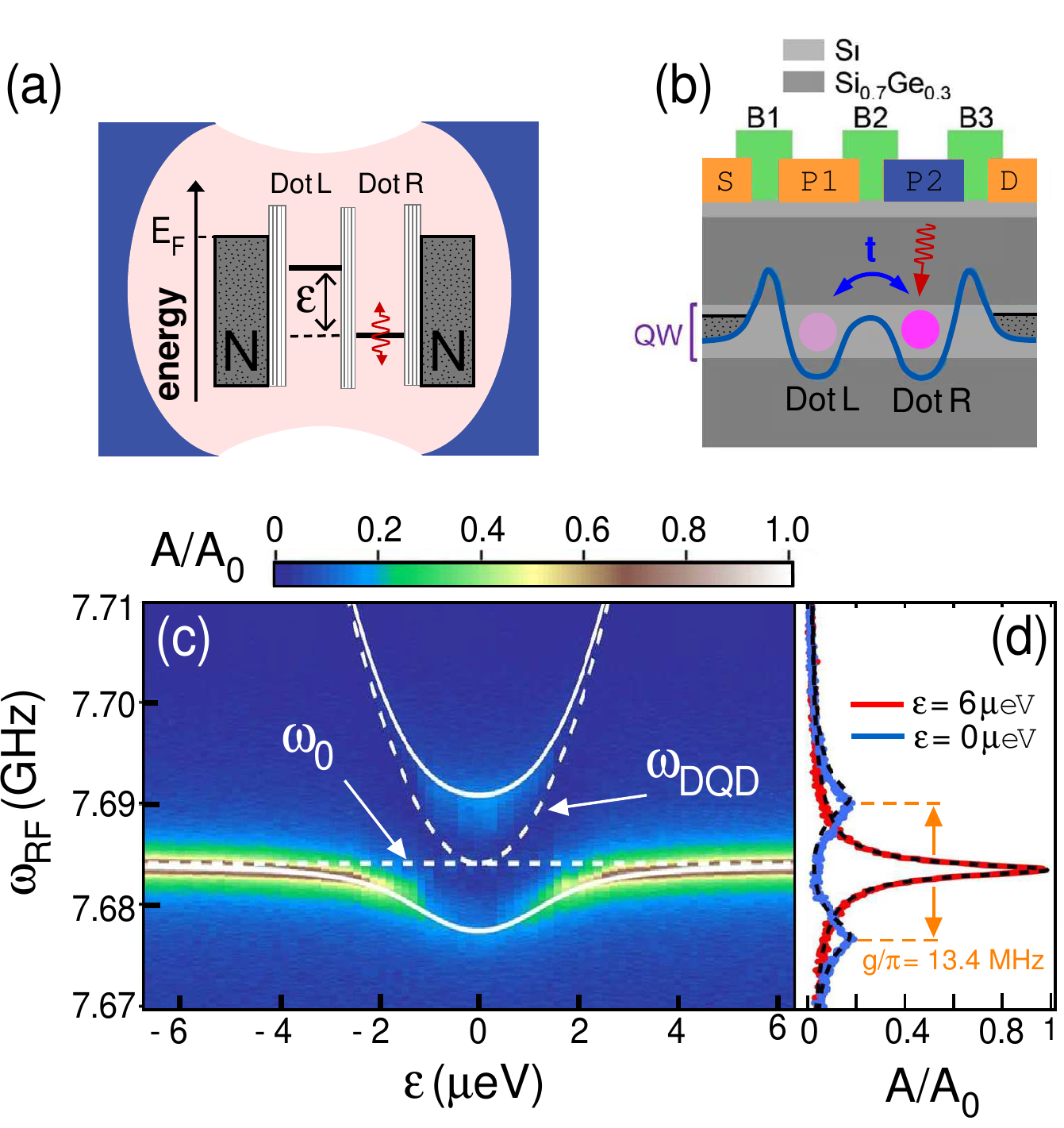}\caption{(a): Schematic
representation of a double quantum dot with an orbital energy detuning
$\varepsilon$, coupled to a microwave cavity. (b): Scheme of the cross-section
through the DQD gates and Si/SiGe heterostructure used in Ref.\cite{Mi:2017}
to implement the device of panel (a). The plunger gate P2 that is located
above the quantum dot R is electrically connected to the central conductor of
the superconducting coplanar cavity. An excess electron (magenta dot) is
confined in the quantum well (QW) within the double well potential (blue line)
created by the gate electrodes in green (c): Cavity transmission amplitude
$A/A_{0}$ as a function of $\omega_{RF}$ and $\varepsilon$ for $2t=\omega
_{0}=7.68~\mathrm{GHz}.$ The system eigenenergies for a coupling
$g=g_{L}-g_{R}=0$ are represented over the data as white dashed lines. The
calculated eigenenergies for $g/2\pi=6.7~\mathrm{MHz}$ are represented as
solid lines. (d): Cavity signal $A/A_{0}$ as a function of $\omega_{RF}$ at
$\varepsilon=6~\mathrm{\mu eV}$ (blue line) and $0~\mathrm{\mu eV}$ (red
line). The data for $\varepsilon=6~\mathrm{\mu eV}$ show a cavity resonance
splitting $2g=2\pi\times13.4~\mathrm{MHz}$ characteristic of the strong
coupling regime. Dashed lines are predictions from cavity input-output theory.
Adapted from Ref.[\cite{Mi:2017}].}%
\label{Figure18}%
\end{figure}

Very recently, the strong coupling regime was reached simultaneously in three
experiments based on different types of charge double quantum
dots\cite{Mi:2017,bruhat:2017,Stockklauser:2017}. In this regime, for a low
number of photons $\left\langle \tilde{n}\right\rangle \rightarrow0$, the
cavity transmission (or reflection) amplitude versus the frequency excitation
$\omega_{RF}$ shows a double peak, due to the strong hybridization between the
cavity and the L/R charge degree of freedom of the double dot (see
Fig.\ref{Figure18}d). To reach this regime, the differential light-matter
coupling $g_{L}-g_{R}$ must be sufficiently large in comparison with the
decoherence rate of the L/R degree of freedom, which is typically dominated by
dephasing due to charge noise. In this case, the dephasing rate $\Gamma
_{\varphi}$ of the charge double dot takes the form\cite{Viennot:2015}%
\begin{equation}
\Gamma_{\varphi}\simeq\frac{\partial\omega_{DQD}}{\partial\varepsilon}%
E_{c}A+\frac{1}{2}\frac{\partial^{2}\omega_{DQD}}{\partial\varepsilon^{2}%
}E_{c}^{2}A^{2} \label{noise}%
\end{equation}
where $E_{c}$ is the local charging energy of one dot (we disregard the mutual
charging energy between the dots). Above, $A$ is the dimensionless prefactor
in the noise spectrum $A^{2}/f$ which adds up to the reduced gate charge
$C_{g}^{L(R)}V_{g}^{L(R)}/e$ with $C_{g}^{L(R)}$ the capacitance between dot
$L(R)$ and its gate voltage source with voltage $V_{g}^{L(R)}$. From Eqs.
(\ref{Qeh}), (\ref{gt}) and (\ref{noise}), in order to have $Q_{e-ph}>1$,
three different technical strategies are possible: either use a nanoconductor
technology with an intrinsically lower charge noise (i.e. decrease $A$), or
use a double dot with. larger capacitances (i.e decrease $E_{c}$) in order to
shunt the effect of charge noise\cite{bruhat:2017}, or change the cavity
technology in order to increase the cavity electric field and thus the
$g_{L}-g_{R}$ factor\cite{Stockklauser:2017}. These three strategies have
already been implemented experimentally, as discussed in the paragraphs below.
In all cases, using a device with $\omega_{0}=2t$ should be advantageous, so
that $\Gamma_{\varphi}$ is only a second order effect in $A$ when the device
is tuned at the anticrossing between the cavity and the double dot (one has
$\varepsilon=0$ and thus $\partial\omega_{DQD}/\partial\varepsilon
=0$)~

In Ref \cite{Mi:2017}, a double quantum dot in an undoped Si/SiGe
heterostructure was used (see Fig.\ref{Figure18}b). When the double dot is far
off resonant with the cavity, a single resonance is visible along the
$\omega_{RF}$ axis, which corresponds to the bare cavity resonance (see blue
line in Fig.\ref{Figure18}c). When $\omega_{0}=2t$ is used, and when the
double dot is tuned near its sweet spot ($\varepsilon\sim0$), the vacuum Rabi
splitting is observed, i.e. a double cavity resonance is visible along the
$\omega_{RF}$ axis (see red line in Fig.\ref{Figure18}c). In this device, the
charge-photon coupling $(g_{L}-g_{R})/2\pi=13.4~\mathrm{MHz}$ is comparable to
what has been obtained with other charge double dots (see Table 1). The
microwave cavity is also similar to the ones used in previous experiments,
with $\Lambda_{0}=1.0~\mathrm{MHz}$. The vacuum Rabi splitting is achieved
thanks to an unusually small decoherence rate $\Gamma_{2}^{\ast}%
=2.6~\mathrm{MHz}$ of the left/right charge degree of freedom in the double
dot. This gives light-matter coupling ratios $Q_{e-ph}=4$ and $C_{e-ph}=34$.
In the Si/SiGe two-dimensional structure used for this experiment, the dot
charging energies are typically of the order of $E_{c}\simeq7~\mathrm{meV}%
$\cite{Mi:2016}. In comparison, the GaAs/AlGaAs structure of
Ref.\cite{Stockklauser:2015} has smaller charging energies $E_{c}%
\sim1~\mathrm{meV}$, but it is far from the strong coupling regime. This
suggests that the low $\Gamma_{2}^{\ast}$ value in Ref.\cite{Mi:2017} might be
due to a much lower intrinsic charge noise in Si/SiGe devices. In agreement
with this, in GaAs/AlGaAs devices, one of the smallest reported value of
charge noise is \cite{Petersson:2010b} is $A=2.10^{-4}$, whereas the values
$2.3~10^{-6}<A<1~10^{-5}$ have been reported\cite{Freeman:2016} for doped
Si/SiGe heterostructures. Undoped Si/SiGe heterostructures might have an even
lower charge noise\cite{Obata:2014}. The value of charge noise in Si/SiGe
deserves a thorough investigation in order to confirm this picture.

Ref.\cite{Stockklauser:2017} has used a GaAs/AlGaAs heterostructure similar to
in Ref.\cite{Stockklauser:2015}. However, the coplanar waveguide architecture
has been modified by replacing the central resonator of the cavity by an array
of 32 SQUIDs (Superconducting QUantum Interference Devices). This increases by
a factor $\sim10$ the couplings $g_{L(R)}$. One has $(g_{L}-g_{R}%
)/2\pi=238~\mathrm{MHz}$, and $\Gamma_{2}^{\ast}=93~\mathrm{MHz}$ which gives
$Q_{e-ph}=2$ and $C_{e-ph}=25$. These figures of merit are very close to those
of Ref.\cite{Mi:2017}. However, they have not been obtained at the double dot
sweet spot since $\omega_{0}=1.22\ast2t$ was used. This could suggest that the
above figure of merits are not the optimal ones for this setup. Interestingly,
with the SQUID array architecture, the cavity frequency can be tuned by using
an external magnetic field. However, the cavity decoherence rate is stronger
with this architecture ($\Lambda_{0}=6.2~\mathrm{MHz}$).

Alternatively, Ref.\cite{bruhat:2017} has reached the strong coupling regime
to the left/right charge degree of freedom of a double dot by using a device
with small charging energies. However, the double dot has a fundamentally
different architecture in this experiment, since it comprises a
superconducting contact, and since the coupling to the cavity photons seems to
occur through the variable $\varepsilon_{L}+\varepsilon_{R}$ instead of
$\varepsilon_{L}-\varepsilon_{R}$. Therefore, we will discuss this experiment
in section \ref{dSd}.

\subsection{Mesoscopic QED with spins in quantum dot circuits\label{spinQubit}%
}

\subsubsection{Spin-photon coupling due to spin-orbit coupling\label{so}}

The electronic spin degree of freedom draws a lot of interest in
nanoconductors because it could be a good means to encode quantum information.
Indeed, spins are expected to have a long coherence time in nanoconductors
because they are more weakly coupled to their environment than charges. The
counterpart of this immunity is that the natural magnetic coupling $g_{m}$
between a spin and a standard coplanar microwave cavity is only a few
$10\,$Hz, which is not sufficient for manipulation and readout operations. It
is possible to circumvent this difficulty by using a large number of spins, as
demonstrated recently with several types of crystals coupled to coplanar
microwave cavities\cite{Schuster:2010,Kubo:2010,Zhu:2011}. However, in this
case, the anharmonicity which is inherent to a two level system is lost so
that the spin ensemble can only be used as a quantum memory. To remain at the
single spin level, it has been suggested to include in the microwave cavity a
nanometric constriction to concentrate the cavity field, which would yield
$g_{m}\sim10~$\textrm{kHz }\cite{Tosi:2014,Haikka:2017}. Alternatively,
various theory Refs. have suggested to use a weak hybridization between the
spin\ and charge degrees of freedom of a quantum dot
circuit\cite{Trif:2008,Hu:2012,Kloeffel:2013,Beaudoin:2016,Cottet:2010},
provided by a real or artificial spin-orbit coupling.\ To understand this
effect, let us assume that the state of the dot circuit can be decomposed on a
basis of pure spin eigenstates states $\left\vert \varphi_{n\uparrow
}\right\rangle $ and $\left\vert \varphi_{n\downarrow}\right\rangle $ with
$n\in\mathbb{N}$ an orbital index. In the presence of a spin-orbit coupling
term the Hamiltonian of the dot circuit will write:
\begin{align}
\hat{H}_{0}^{t}  &  =%
{\displaystyle\sum\nolimits_{n,\sigma}}
\left(  E_{n}+\frac{E_{z}\sigma}{2}\right)  \left\vert \varphi_{n\sigma
}\right\rangle \left\langle \varphi_{n\sigma}\right\vert \\
&  +%
{\displaystyle\sum\nolimits_{n,n^{\prime}}}
\left(  h_{nn^{\prime}}\left\vert \varphi_{n^{\prime}\downarrow}\right\rangle
\left\langle \varphi_{n\uparrow}\right\vert +H.c.\right) \nonumber
\end{align}
Above, $E_{n}$ is the orbital energy of state $n$, $E_{z}$ is the external
Zeeman field applied to the circuit, and $h_{nn^{\prime}}$ corresponds to the
matrix elements of the spin-orbit interaction on the basis of states
$\left\vert \varphi_{n\sigma}\right\rangle =c_{n\sigma}^{\dag}\left\vert
\varnothing\right\rangle $. The photonic pseudo potential $V_{\bot}$ is
spin-conserving, so that the light-matter interaction given by Eq.(\ref{hi})
is%
\begin{equation}
\hat{h}_{int}=%
{\displaystyle\sum\nolimits_{n,\sigma}}
V_{\bot}^{n,n^{\prime}}\left\vert \varphi_{n^{\prime}\sigma}\right\rangle
\left\langle \varphi_{n\sigma}\right\vert
\end{equation}
Hence, at first order in spin-orbit coupling, the eigenstates
\begin{equation}
\left\vert \widetilde{\varphi_{n\uparrow(\downarrow)}}\right\rangle
=\left\vert \varphi_{n\uparrow(\downarrow)}\right\rangle +%
{\displaystyle\sum\nolimits_{n^{\prime}}}
\frac{h_{nn^{\prime}}^{(\ast)}}{E_{n\uparrow(\downarrow)}-E_{n\downarrow
(^{\prime}\uparrow)}}\left\vert \varphi_{n^{\prime}\downarrow(^{\prime
}\uparrow)}\right\rangle
\end{equation}
of $H_{0}$ are coupled to the cavity with the matrix element%
\begin{align}
&  \left\langle \widetilde{\varphi_{n\downarrow}}\right\vert \hat{h}%
_{int}\left\vert \widetilde{\varphi_{n\uparrow}}\right\rangle \label{couplSO}%
\\
&  =%
{\displaystyle\sum\nolimits_{n^{\prime}}}
\left(  \frac{h_{nn^{\prime}}}{E_{n}-E_{n^{\prime}}+E_{z}}-\frac
{h_{nn^{\prime}}^{\ast}}{E_{n}-E_{n^{\prime}}-E_{z}}\right)  V_{\bot
}^{n,n^{\prime}}\nonumber
\end{align}
One can imagine to build a qubit by using two states $\left\vert
\widetilde{\varphi_{n\uparrow}}\right\rangle $ and $\left\vert
\widetilde{\varphi_{n\downarrow}}\right\rangle $ which can be considered as
quasi-spin states if the spin-charge hybridization is small. Due to this
hybridization, this qubit will be sensitive to charge noise. Therefore one has
to choose a design which establishes a good compromise between having a small
enough decoherence and a high enough light-matter-coupling.

In principle, from Eq.(\ref{couplSO}), a single quantum dot with a natural
Rashba or Dresselhaus spin-orbit coupling could already offer a spin-photon
interaction. Indeed, the spin of a quantum dot was manipulated by using a
large ac drive applied directly on the dot gate and coupled to the spin
through the spin-orbit coupling\cite{Nowack:2007,Liu:2014}. One can imagine to
replace the direct ac drive by the cavity field. Then, the indices $n$,
$n^{\prime}$ in the above equations can correspond to the natural subbands in
the dot spectrum. However, for most quantum dots, the spin-orbit interaction
is too weak to enable the strong coupling regime with a single quantum dot
circuit. For instance, for GaAs quantum dots, it has been shown theoretically
that the effect of spin-orbit coupling is limited by the small spatial
extension of the quantum dot\cite{Khaetskii :2000,Khaetskii
:2001,Erlingsson:2002}. As we will see in section \ref{FdotdotF}, an
alternative approach is to engineer extrinsically an artificial spin-orbit
coupling by using a quantum dot circuit with ferromagnetic contacts, which
induce local effective Zeeman fields such as those of Eq.(\ref{ZeemanEff}).
Then, it is not necessary to invoke the existence of several levels in each
dot. For instance, in the case of a double quantum dot, the indices $n$,
$n^{\prime}$ can be restricted to a pair of bonding and antibonding states,
formed by the coherent coupling of left and right orbitals of the double dot.
In principle, this should enable one to tune the value of the spin-orbit
interaction, thanks to the electric control of the orbital energy detuning
$\varepsilon$. Another interesting possibility could be to use designs which
exploit stray fields from micromagnets\cite{Tokura :2006,Takeda:2016}.

\subsubsection{Charge readout of spin-blockaded states in a double dot}

As shown in section \ref{DQDcharge}, internal tunnel hopping of charges inside
a double quantum dot modifies the cavity signals. This property can be used to
detect with a dc current measurement the spin state of a pair of electrons
trapped in a double quantum dot, thanks to a spin-rectification effect induced
by Pauli spin blockade, which has been widely studied through current
measurements \cite{Ono: 2002}. This effect was recently exploited in a
Mesoscopic QED device, based on a singlet-triplet qubit in an InAs double
quantum dot, with one electron in each dot\cite{Petersson:2012}. The readout
of this qubit requires to discriminate the two states $\left\vert
S_{1,1}\right\rangle =(\left\vert L\uparrow,R\downarrow\right\rangle
-\left\vert L\uparrow,R\downarrow\right\rangle )/\sqrt{2}$ and $\left\vert
T_{0}\right\rangle =(\left\vert L\uparrow,R\downarrow\right\rangle +\left\vert
R\uparrow,L\downarrow\right\rangle )/\sqrt{2}$. When the orbital detuning
$\varepsilon$ between the left and right dot is modulated by the cavity
electric field, transitions to the state $\left\vert L\uparrow,L\downarrow
\right\rangle $ are possible only if the double dot initially occupies the
state $\left\vert S_{1,1}\right\rangle $, due to the spin-conserving character
of interdot tunneling. This leads to $\Xi(\omega_{RF})\neq0$. In contrast, if
the double dot is in the state $\left\vert T_{0}\right\rangle $, one has
$\Xi(\omega_{RF})=0$ due to the Pauli exclusion principle. Therefore, the spin
state of the double dot can be detected through the cavity signals. It is
nevertheless important to point out that a direct spin-photon coupling was not
implemented in the experiment of Ref.\cite{Petersson:2012}. The state of the
double dot was manipulated by applying a strong microwave drive directly to
the dots gates, to rotate the spins thanks to spin-orbit coupling. The cavity
was used only to perform the charge readout of the spin qubit. To date, no
experiment could detect a spin-cavity coupling caused by intrinsic spin-orbit
coupling in a nanocircuit.

\subsubsection{Spin-photon coupling in a double quantum dot with non-collinear
ferromagnetic contacts\label{FdotdotF}}

\begin{figure}[th]
\includegraphics[width=1\linewidth]{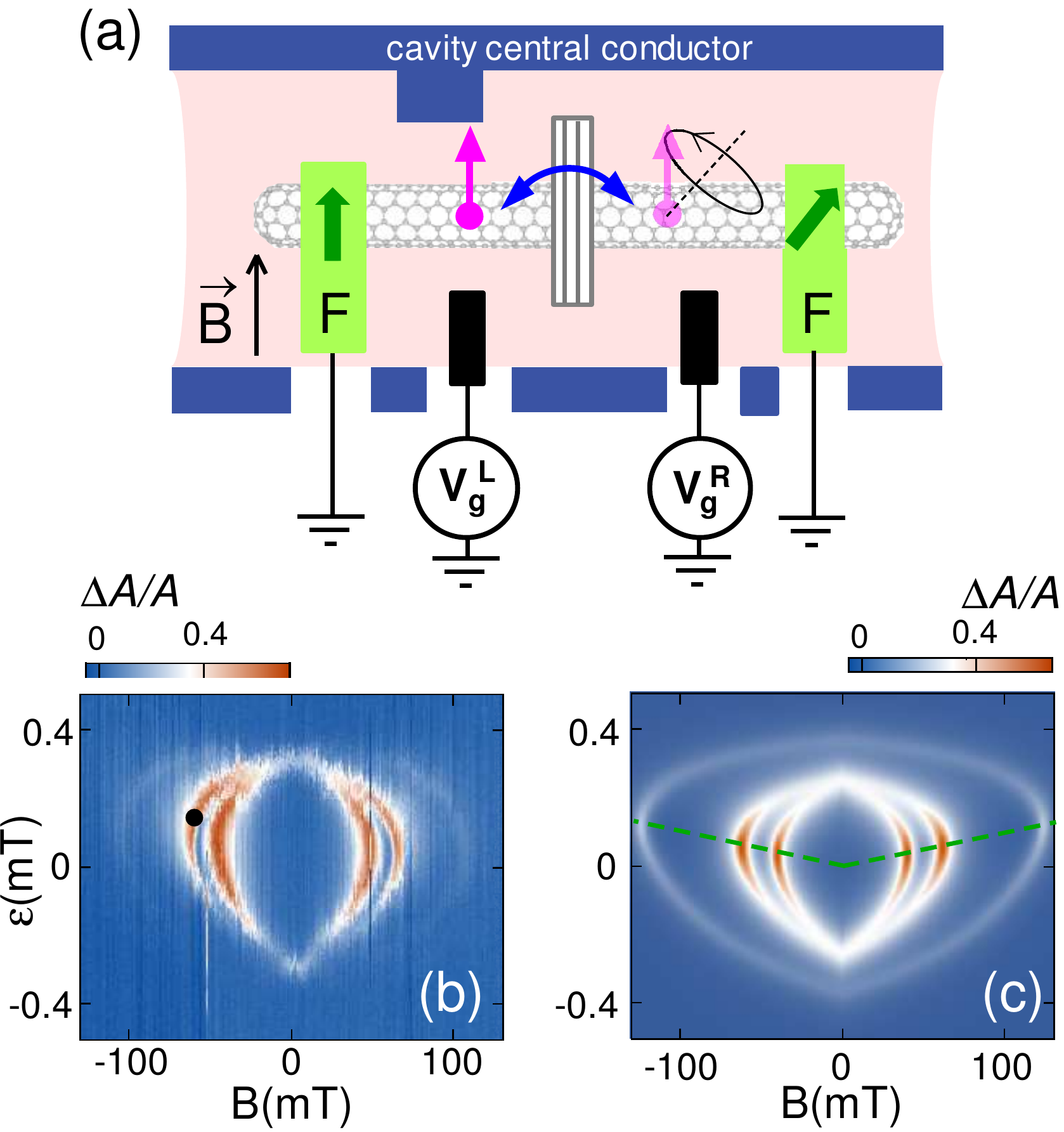}\caption{Response of a
microwave cavity coupled to a double quantum dot with non colinear
ferromagnetic contacts. The device is represented schematically in panel (a).
The cavity dissipation of panel (b) displays various resonances which
dependent on the orbital detuning $\varepsilon$ of the double dot and the
applied magnetic field $B$. This signal is reproduced theoretically in panel
(c) (see text). The green dotted line corresponds to a sweet line with respect
to charge noise. Adapted from Ref.[\cite{Viennot:2015}].}%
\label{Figure7}%
\end{figure}

It was recently suggested that the coupling of Eq.(\ref{couplSO}) could be
realized by using a double quantum dot with two ferromagnetic contacts
magnetized in non-collinear directions\cite{Cottet:2010}, represented in
Fig.\ref{Figure7}a. These contacts cause a spin-mixing of the double dot
eigenstates, which can be viewed as an artificial spin-orbit coupling. This
effect occurs due to intradot effective Zeeman fields similar to those of
Eq.(\ref{ZeemanEff}). By tuning the orbital detuning $\varepsilon$, one can in
principle control the degree of delocalization of the electron between the two
dots, in order to tune the magnitude of the artificial spin-orbit interaction.

A first version of this device has been realized recently, by
using a double quantum dot made in a single wall carbon nanotube on top of
which two ferromagnetic PdNi contacts are evaporated\cite{Viennot:2015}. When
the microwave transmission amplitude of the cavity is measured versus
$\varepsilon$ and the external magnetic field $B$ applied to the double dot,
three resonant lines appear (see Fig.\ref{Figure7}b). Various features suggest
that the spin degree of freedom is an important ingredient in this pattern.
First, the resonances split and strongly move with the external magnetic field
$B$, with a maximum of contrast/coherence for a finite value of $B$. Second,
the black point of Fig.\ref{Figure7}b corresponds to a coupling $\ g_{s}%
=1.3~\mathrm{MHz}$ and a double dot decoherence rate $\Gamma_{2}^{\ast}%
/2\pi=2.5~\mathrm{MHz}$. This last number is about 200 smaller than the charge
decoherence rate determined for a similar carbon nanotube device (see section
\ref{JayPRB}). One has $Q_{e-ph}=0.81$ which means that this device is almost
in the strong coupling regime.

To understand better the contribution of the spin degree of freedom to the
cavity signals, one can use Eq.(\ref{XiInt}) which is a generalization of
Eq.(\ref{XiDQD}), valid if the different transition frequencies $\omega_{ij}$
of the nanocircuit are well separated. To calculate $\omega_{ij}$ and the
couplings $g_{ij}$, one has to use a double dot Hamiltonian which takes into
account the existence of the left/right and spin degree of freedoms of the
double dot, but also the K/K' local orbital degree in each dot (or valley
degree of freedom), which is due to the fact that electrons can rotate
clockwise or anticlockwise around the carbon nanotube. The linewidth of the
resonances can be modeled by taking into account the effect or charge noise.
This gives Fig.\ref{Figure7}c, which reproduces well the behavior of
Fig.\ref{Figure7}b. The two strongest resonances mainly correspond to spin
transitions with a conserved K/K' index. These two resonances are slightly
split due to a small lifting of the K/K' degeneracy. The third weaker
resonance mainly corresponds to a transition where both the spin and the K/K'
index are reversed. In Fig.\ref{Figure7}c, this transition is less visible
than the two others because the K/K' degree of freedom is only weakly coupled
to cavity photons, probably due to weak microscopic disorder in the carbon
nanotube structure. However, this resonance is very interesting in the light
of recent works which investigate the coupling between the valley degree of
freedom of a silicon dot and a microwave cavity \cite{Burkard:2016, Mi:2017b}.

Remarkably, the coherence (or, visually, the contrast) of the three
transitions is maximum along the green dashed line in Fig.\ref{Figure7}c. This
is because the derivative of the transition frequencies $\omega_{ij}$ with
respect to $\varepsilon$ vanishes along this line, which is a charge noise
sweet line. This behavior also occurs in the data, which confirms that charge
noise is an important source of decoherence in this device. It may be possible
to enhance these performances by reducing the spin-charge hybridization to
decrease decoherence due to charge noise. It is expected that $\Gamma
_{2}^{\ast}$ will decrease more quickly than $g_{s}$ with $\varepsilon$, so
that the strong coupling regime is accessible with this geometry, in
principle\cite{Cottet:2010}.

\subsubsection{Spin-photon coupling in multiple particle devices with
collinear fields}

Various theory Refs. have suggested to couple electrically the spin degree of
freedom to the cavity electric field by using two or three electron states in
a quantum dot circuit.\ For that purpose, one can use a multi-quantum dot
circuit with proper spin-symmetry breaking ingredients, in order to transduce
the charge-photon into a spin-photon coupling. For instance, in a double dot
with a finite interdot hopping, the transition between the singlet and triplet
spin states $\left\vert S_{1,1}\right\rangle $ and $\left\vert T_{0}%
\right\rangle $ is coupled to cavity photons due to the presence of a Zeeman
field with constant direction but a different amplitude in the two
dots\cite{PeiQing:2012b,Pei-Qing:2016,Burkard:2006,Wang:2015,Han:2016}. This
field can correspond to an Overhauser field due to nuclear spins in a
two-dimensional electron gas, or to stray fields from a ferromagnet. In the
case of a triple quantum dot, it is possible to use three electron states from
the $(S=1,S_{z}=1/2)$ subspace, with $S$ the total spin of the dots, to define
the resonant exchange qubit\cite{Russ:2015,Srinivasa:2016,Russ:2016}. In this
case, the spin-photon coupling can be obtained with a homogeneous Zeeman field
if the spatial symmetry of the triple dot is adequately broken. Note that the
above setups do not involve any real or effective spin-orbit interaction. On
the contrary, they consider devices where the individual spin of electrons
would be conserved in the single electron regime. At present, the
multiparticle spin-photon coupling of
Refs.\cite{PeiQing:2012b,Pei-Qing:2016,Burkard:2006,Wang:2015,Han:2016,Russ:2015,Srinivasa:2016,Russ:2016}
is awaiting an experimental realization.

\subsection{Probing Andreev states with cavity photons\label{ABS}}

When superconducting elements are included in a nanocircuit, the electron and
hole excitations become coupled by Andreev reflections, so that Andreev bound
states appear inside the nanoconductors (see Fig.\ref{Figure1}c). This
superconducting proximity effect raises a strong attention presently because
it is at the heart of phenomena such as Majorana bound states in hybrid
structures or Cooper pair splitting. Furthermore, Andreev bound states can
appear on interfaces such as an atomic contact, for which the charge orbital
confinement is not a relevant concept. One could hope that such states are
weakly sensitive to charge noise and could be a good support of quantum
information. It is therefore very interesting to investigate the properties of
this degree of freedom with a microwave cavity, as suggested by
Ref.\cite{Skoldberg:2008}. In the presence of superconductivity, the
Hamiltonian of the hybrid\ nanocircuit can be written as
\begin{equation}
\hat{H}_{0}^{t}=%
{\textstyle\sum\nolimits_{\alpha}}
E_{\alpha}\gamma_{\alpha}^{\dag}\gamma_{\alpha} \label{HoAndreev}%
\end{equation}
with $E_{\alpha}>0$ and $\gamma_{\alpha}^{\dag}$ a Bogoliubov-De Gennes
excitation creation operator which is a superposition of $\hat{c}_{n}^{\dag}$
and $\hat{c}_{n}$ operators. Hence, the interaction term with the cavity takes
the general form\cite{Cottet:2015}
\begin{equation}
\hat{h}_{int}=%
{\textstyle\sum\nolimits_{\alpha}}
M_{\alpha\beta}\hat{\gamma}_{\alpha}^{\dag}\hat{\gamma}_{\beta}+N_{\alpha
\beta}\hat{\gamma}_{\alpha}^{\dag}\hat{\gamma}_{\beta}^{\dag}+N_{\alpha\beta
}^{\dag}\hat{\gamma}_{\alpha}\hat{\gamma}_{\beta} \label{HintAndreev}%
\end{equation}
In the absence of magnetic coupling between the device and the cavity, the
elements $M_{\alpha\beta}$ and $N_{\alpha\beta}$ can be expressed as matrix
elements induced by the cavity photonic pseudopotential between the
wavefunctions associated to $\hat{\gamma}_{\alpha(\beta)}^{\dag}$ and
$\hat{\gamma}_{\alpha(\beta)}$ (see Ref.\cite{Cottet:2015} for details). At
zero temperature ($T=0$), Eq.(\ref{HintAndreev}) gives\cite{Dartiailh:2016}%

\begin{equation}
\chi(\omega_{0})\simeq\frac{1}{2}%
{\textstyle\sum\nolimits_{\alpha\neq\beta}}
\frac{\left\vert N_{\alpha\beta}\right\vert ^{2}}{\omega_{0}-E_{\alpha
}-E_{\beta}+i0^{+}} \label{XiAndreev}%
\end{equation}
Importantly, due to the Pauli exclusion principle, one has $N_{\alpha\alpha
}=0$ since a term in $\hat{\gamma}_{\alpha}^{\dag}\hat{\gamma}_{\alpha}^{\dag
}$ cannot occur in $\hat{h}_{int}$. Hence, from Eq.(\ref{XiAndreev}),
$\chi(\omega_{0})$ does not involve transitions between electron and holes
states associated to conjugated operators $\hat{\gamma}_{\alpha}^{\dag}$ and
$\hat{\gamma}_{\alpha}$\cite{Dartiailh:2016,Vayrynen}. This selection rule can
be extended to a finite temperature ($T\neq0$) or a level broadening smaller
than the inter-level separation\cite{Dartiailh:2016}. Nevertheless, having a
nanocircuit response at $\omega_{0}=2E_{\alpha}$ is possible provided there
exists a state degeneracy $E_{\alpha}=E_{\alpha^{\prime}}$ in the nanocircuit
so that a coefficient $N_{\alpha\alpha^{\prime}}$ comes into
play\cite{Skoldberg:2008}, as observed in spin-degenerate superconducting
atomic contacts\cite{Janvier:2015}. In this experiment, an atomic contact
between two superconductors was coupled to a microwave resonator through a
superconducting loop. A light matter coupling $g=74~\mathrm{MHz}$ and a
decoherence rate $\Gamma_{2}^{\ast}/2\pi=26~\mathrm{MHz}$ were estimated
inside a spin-degenerate Andreev doublet, which corresponds to $Q_{e-ph}=4.4$
and $C_{e-ph}=90$. These good performances are probably related to a smaller
sensitivity of atomic point contacts to charge noise.

Interestingly, devices have been built, where a microwave cavity is coupled to
a superconducting circuit which includes a Josephson junction made out of a
semiconducting nanowire quantum dot\cite{Larsen:2015, Lange:2015}. The
superconducting current in the Josephson junction is mediated by Andreev bound
states inside the quantum dot. Since the spectrum of Andreev states is tunable
with the dc gate of the dot, the critical current of the Josephson junction is
electrically controllable. This can represent a technical advantage in
comparison with usual magnetically tunable Josephson junctions made out of a
SQUID. The nanowire junction is used to form the "Gatemon" superconducting
quantum bit which involves a coupling between a microwave cavity and the
superconducting phase difference between two metallic
islands\cite{Larsen:2015, Lange:2015}. This variable is a macroscopic
collective degree of freedom of the superconducting circuit. Therefore, these
devices belong more to the family of Circuit QED devices than to the family of
Mesoscopic QED devices. This is why we will not discuss them further in this review.

\subsection{Double quantum dot with a central superconducting
contact\label{dSd}}

\begin{figure}[th]
\includegraphics[width=0.7\linewidth]{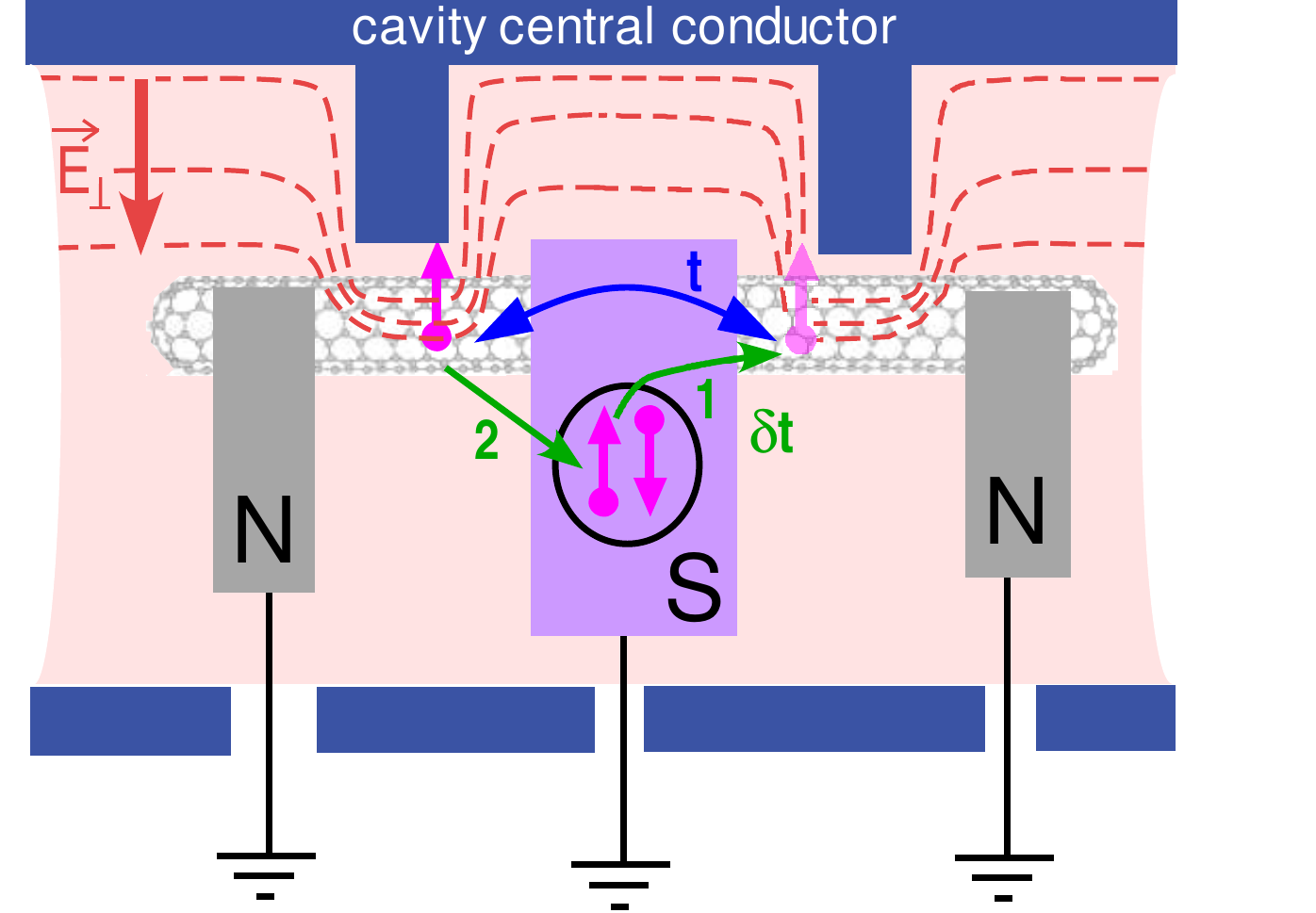}\caption{Scheme of a double
quantum dot made in a carbon nanotube, with a central superconducting contact.
Electrons can go from one dot to the other by direct tunneling through the
nanotube section below the S contact (blue arrow), or by second order
tunneling through the S contact (green arrows)}%
\label{Figure16}%
\end{figure}

Recently, the coupling scheme between a double quantum dot and a cavity was
drastically modified by placing a superconducting contact between the two
quantum dots instead of an insulating barrier, and two identical ac top gates
on the two quantum dots\cite{bruhat:2017} (see Figure \ref{Figure16}), instead
of coupling asymmetrically the two dots to the cavity as done usually (see
Fig.\ref{Figure0}). In the symmetric coupling case of Fig. \ref{Figure16}, the
differential coupling $g_{L}-g_{R}$ between the two dots and the cavity is
expected to be small. However, anticrossings were observed in the cavity
response for a low number of photons in the cavity. These anticrossings can be
switched on/off with the double dot gate voltages, and they vanish when the
photon number is large so that the double dot transitions are saturated. This
suggests that the cavity anticrossings are due to a strong coupling between
the cavity and the double dot. A fitting of these anticrossings yields a
coupling $\ g_{t}=10~\mathrm{MHz}$ and a decoherence rate $\Gamma_{2}^{\ast
}/2\pi\sim2~\mathrm{MHz}$ which corresponds to $Q_{e-ph}=3.9$ and
$C_{e-ph}=60$. \begin{figure*}[t]%
\begin{tabular}
[c]{|c|c|c|c|c|c|c|c|c|c|c|c|}\hline
Geometry & double dot material &
\begin{tabular}
[c]{l}%
degree of\\
freedom
\end{tabular}
& cavity design & Refs. & Fig. &
\begin{tabular}
[c]{l}%
$\omega_{0}/2\pi$\\
(GHz)
\end{tabular}
& $Q_{0}$ &
\begin{tabular}
[c]{l}%
$2g_{t}/2\pi$\\
(MHz)
\end{tabular}
&
\begin{tabular}
[c]{l}%
$\Gamma_{2}^{\ast}/2\pi$\\
(MHz)
\end{tabular}
& $Q_{e-ph}$ & $C_{e-ph}$\\\hline
N/dot/dot/N & graphene & charge & Al stripe & \cite{Deng:2015,NoteDeng} &
$\times$ & $6.24$ & $1600$ & $12.4$ & $430$ & $0.022$ & $0.046$\\\hline
N/dot/dot/N & InAs nanowire & charge & Nb stripe & \cite{Liu:2014,NoteLiu} &
$\times$ & $7.86$ & $3930$ & $32$ & $1500$ & $0.016$ & $0.17$\\\hline
N/dot/dot/N & carbon nanotube & charge & Al stripe & \cite{Viennot:2014a} &
\ref{Figure6} & $6.72$ & $3500$ & $6.6$ & $345$ & $0.015$ & $0.033$\\\hline
F/dot/dot/F & carbon nanotube & quasi-spin & Nb stripe & \cite{Viennot:2015} &
\ref{Figure7} & $6.75$ & $11200$ & $2.6$ & $2.5$ & $0.81$ & $2.3$\\\hline
N/dot/S/dot/N & carbon nanotube & charge & Nb stripe & \cite{bruhat:2017} &
\ref{Figure16} & $6.64$ & $16000$ & $10$ & $\sim2$ & $3.9$ & $60$\\\hline
N/dot/dot/N & GaAs/AlGaAs 2DEG & charge & Al stripe & \cite{Stockklauser:2015}
& \ref{Figure12} & $6.85$ & $2060$ & $22$ & $250$ & $0.068$ & $0.29$\\\hline
N/dot/dot/N & GaAs/AlGaAs 2DEG & charge & Al SQUID array &
\cite{Stockklauser:2017} & $\times$ & $5.02$ & $\sim400$ & $238$ & $93$ & $2$
& $25$\\\hline
N/dot/dot/N & Si/SiGe 2DEG & charge & Nb stripe & \cite{Mi:2017,NoteMi} &
$\times$ & $7.68$ & $7460$ & $13.4$ & $2.6$ & $4.0$ & $34$\\\hline
S/atom/S & Al atomic contact & Andreev state & Nb stripe & \cite{Janvier:2015}
& \ref{Figure18} & $10.1$ & $2200$ & $74$ & $26$ & $4.4$ & $90$\\\hline
\end{tabular}
\caption{TABLE 1: \textit{Measured performances for various Mesoscopic QED
setups in the artificial atom limit. From left to right, we give the geometry
considered, the nature of the nanoconductor used, the degree of freedom placed
in resonance with the cavity, the cavity design and material, the Ref. in
which the experiment is reported, the corresponding Figure in the review, the
cavity frequency }$\omega_{0}$,\textit{ the quality factor }$Q_{0}=\omega
_{0}/2\Lambda_{0}$\textit{ of the cavity, the light/matter coupling }$g_{t}%
$\textit{ between the electronic transition considered and the cavity, the
decoherence rate }$\Gamma_{2}^{\ast}$\textit{ of the transition coupled to the
cavity, and the ratios }$Q_{e-ph}=\sqrt{1+\sqrt{2}}g/\Gamma_{2}^{\ast}%
$\textit{ and }$C_{e-ph}=g^{2}/\Gamma_{2}^{\ast}\Lambda_{0}$\textit{. We use
the abbreviations N=normal metal, F=ferromagnet, S=superconductor. Note that
the parameters }$\Gamma_{2}^{\ast}$\textit{ and }$\Lambda_{0}$\textit{ are not
FWHM parameters. They are rather defined such that no factor 1/2 occurs in the
damping terms of Eqs.(\ref{ii}) and (\ref{res}). Hence, the full width at half
maximum (FWHM) of the bare cavity transmission corresponds to the parameter
}$\varkappa=2\Lambda_{0}$\textit{. With other conventions, the numerical
factors in the definition of }$Q_{e-ph}$\textit{ and }$C_{e-ph}$\textit{ can
differ.}}%
\end{figure*}

This result may seem surprising since a standard coupling to the left/right
charge degree of freedom through the $\varepsilon$ variable is unlikely for a
small $g_{L}-g_{R}$. However, two ingredients can help to understand the
behavior of the setup. First, due to the superconducting gap, direct
dissipative tunneling between the dot and the superconductors is forbidden,
but second order tunnel processes from one dot to the other are allowed. This
gives a renormalization $\delta t$ of the tunnel coupling between the two dots
which depends on the average orbital energy $(\varepsilon_{L}+\varepsilon
_{R})/2$. This last parameter is well coupled to cavity photons even when
$g_{L}=g_{R}$. A coupling between the left/right degree of freedom of the
double dot and the cavity photons can thus be restored. Second, there is a
large capacitance between the dots and the ac top gates, so that the dot
charging energies are decreased by a factor $\sim10$ (determined from
conductance measurements) in comparison with the experiment of
Fig.\ref{Figure6}. This strongly reduces the sensitivity of the device to
charge noise. This strategy is reminiscent from the strategy developed for the
Transmon superconducting quantum bit. In this device, a very large charging
energy is used to flatten the dispersion of the energy bands with gate
voltages and reduce the sensitivity to charge noise\cite{Koch:2007}.

\subsection{Comparison between the different systems and conclusion}

Table I presents a comparison of the performances of the different Mesoscopic
devices used to far to implement artificial atoms in a cavity. Impressive
progresses are already visible since the publication of the first experiments
in Refs.\cite{Delbecq:2011,Frey:2011}. To summarize, charge and spin states in
a double quantum dot, and Andreev bound states in atomic contacts have been
strongly coupled to cavity photons. First experimental results are also
available regarding the local orbital degree of freedom in a quantum dot, for
Si/SiGe heterostructures\cite{Mi:2017b}, and carbon nanotube
devices\cite{Viennot:2015}. Other configurations have been proposed
theoretically which lack experimental realization, at present. For instance,
it has been suggested to use multiparticle spin states in devices with no
spin-orbit
coupling\cite{PeiQing:2012b,Pei-Qing:2016,Burkard:2006,Wang:2015,Han:2016,Russ:2015,Srinivasa:2016,Russ:2016}%
. It could also be interesting to use Shiba states which are Andreev states in
the presence of a magnetic impurity or strong Coulomb interaction in a quantum
dot\cite{Chirla:2016}. Interestingly, several theoretical works have proposed
manipulation protocols which circumvent at least partially the imperfections
of Mesoscopic QED devices, and in particular a limited
coherence\cite{Wang:2015,Beaudoin:2016,Schuetz:2016}.

\section{Mesoscopic QED experiments beyond the artificial atom
limit\label{Open}}

In the previous sections, we have shown that cavity photons are a powerful
probe for the internal dynamics of a nanoconductor. The present section will
show that a microwave cavity is also a very interesting tool to study the
dynamics of tunneling between a nanoconductor and a metallic reservoir. The
interplay between electron tunneling to reservoirs and the light matter
coupling leads to a very rich phenomenology
\cite{Xu:2013a,Schiro:2014,Hartle:2015,Wong:2016,Cirio:2016,Dmytruk:2016,Agarwalla:2016,Kumar:2016}
whose experimental investigation with microwave cavities has recently started.

\subsection{Keldysh expression of the charge susceptibility of a
nanocircuit\label{Keldysh}}

In section \ref{Closed}, the nanocircuit charge susceptibility $\Xi,$ which
sets the cavity signals in the linear semiclassical limit, was evaluated
without taking into account explicitly tunnel processes towards the fermionic
reservoirs of the nanocircuit. Hence, it is useful to introduce a more general
calculation method for $\Xi$. For simplicity, we will assume that the coupling
of the cavity to the fermionic reservoirs of the nanocircuit is negligible. In
case of a quantum dot circuit, the cavity electric field simply modulates the
potential of some quantum dots $n$ with a coupling constant $g_{n}$, like in
Eqs.(\ref{HH}) and (\ref{hint}). Alternatively, in case of a one-dimensional
conductor, a coarse grain description into sites with an index $n$ can be
used\cite{Dartiailh:2016}. Then, in the semiclassical limit, the cavity
signals (\ref{j1}) and (\ref{j2}) are set by the nanocircuit charge
susceptibility $\Xi(\omega_{0})$, defined by Eqs. (\ref{Chi}) and
(\ref{iiii}). Disregarding Coulomb interactions in the dots/sites, this
quantity can be expressed by using the Keldysh formalism as%
\begin{equation}
\Xi(\omega_{0})=-iTr[%
{\textstyle\int}
\frac{d\omega}{2\pi}\mathcal{C}(\omega)\mathcal{G}^{r}(\omega)\Sigma
^{<}(\omega)\mathcal{G}^{a}(\omega)] \label{xi}%
\end{equation}
with%
\begin{equation}
\mathcal{C}(\omega)=\hat{T}\left(  \mathcal{G}^{r}(\omega+\omega
_{RF})+\mathcal{G}^{a}(\omega-\omega_{RF})\right)  \hat{T}%
\end{equation}
Above, $\mathcal{G}^{r}(\omega)$ and $\mathcal{G}^{a}(\omega)=\left(
\mathcal{G}^{r}(\omega)\right)  ^{\dag}$ are the retarded and advanced Greens'
functions of the dots or nanocircuit internal sites. These Green's functions
have a matrix structure with elements $\mathcal{G}_{B,A}^{r}(\omega)=%
{\textstyle\int\nolimits_{-\infty}^{+\infty}}
dt\mathcal{G}_{B,A}^{r}(t)e^{i\omega t}$ with
\begin{equation}
\mathcal{G}_{B,A}^{r}(t)=-i\theta(t)\left\langle \{B(t),A(t=0)\}\right\rangle
_{\hat{h}_{int}=0} \label{defG}%
\end{equation}
Above, $A$, $B$ are quasiparticle creation and annihilation operators inside a
site $n$\ of the nanoconductor and $\left\langle {}\right\rangle _{\hat
{h}_{int}=0}$ denotes an average calculated with the light-nanocircuit
interaction turned off. The matrix $\hat{T}$ is a diagonal matrix which
corresponds to $diag(g_{n},-g_{n})$ in the orbital block $(n,n)$. The
derivation of Eq.(\ref{xi}) is given in appendix \ref{KeldyshXi}. The matrix
structure of $\mathcal{G}^{r(a)}$ takes into account the sites and spin
degrees of freedom of electrons and also the electron/hole degree of freedom
if the nanocircuit includes superconducting elements (a few examples of
$\mathcal{G}^{r(a)}$ will be presented in the next sections). The self energy
$\Sigma^{<}(\omega)$ has a matrix structure similar to $\mathcal{G}^{r(a)}$,
and it involves Fermi occupation factors of the fermionic reservoirs. The
tunnel rates to the reservoirs affect the values of both $\mathcal{G}^{r(a)}$
and $\Sigma^{<}(\omega)$, as we will see in various examples in the next
sections. One advantage of the Keldysh approach is that it is very general. It
can be used to describe the presence of many different types of reservoirs,
such as normal metals, superconductors or ferromagnets, with possibly finite
bias voltages leading to non-equilibrium transport effects. It also takes into
account properly internal hopping between neighboring dots or sites, which
leads to the internal transitions discussed in section \ref{Closed}. For
instance, Appendix B shows how to recover an expression of the charge
susceptibility of a closed non-interacting double quantum dot similar to
Eq.(\ref{XiDQD}), by using Eq. (\ref{xi}). Other examples of use of
Eq.(\ref{xi}) are given in sections \ref{relax}, \ref{emitDsot} and
\ref{Majos}.

\subsection{Effective admittance of a single quantum dot with normal metal
contacts\label{relax}}

\begin{figure}[th]
\includegraphics[width=1\linewidth]{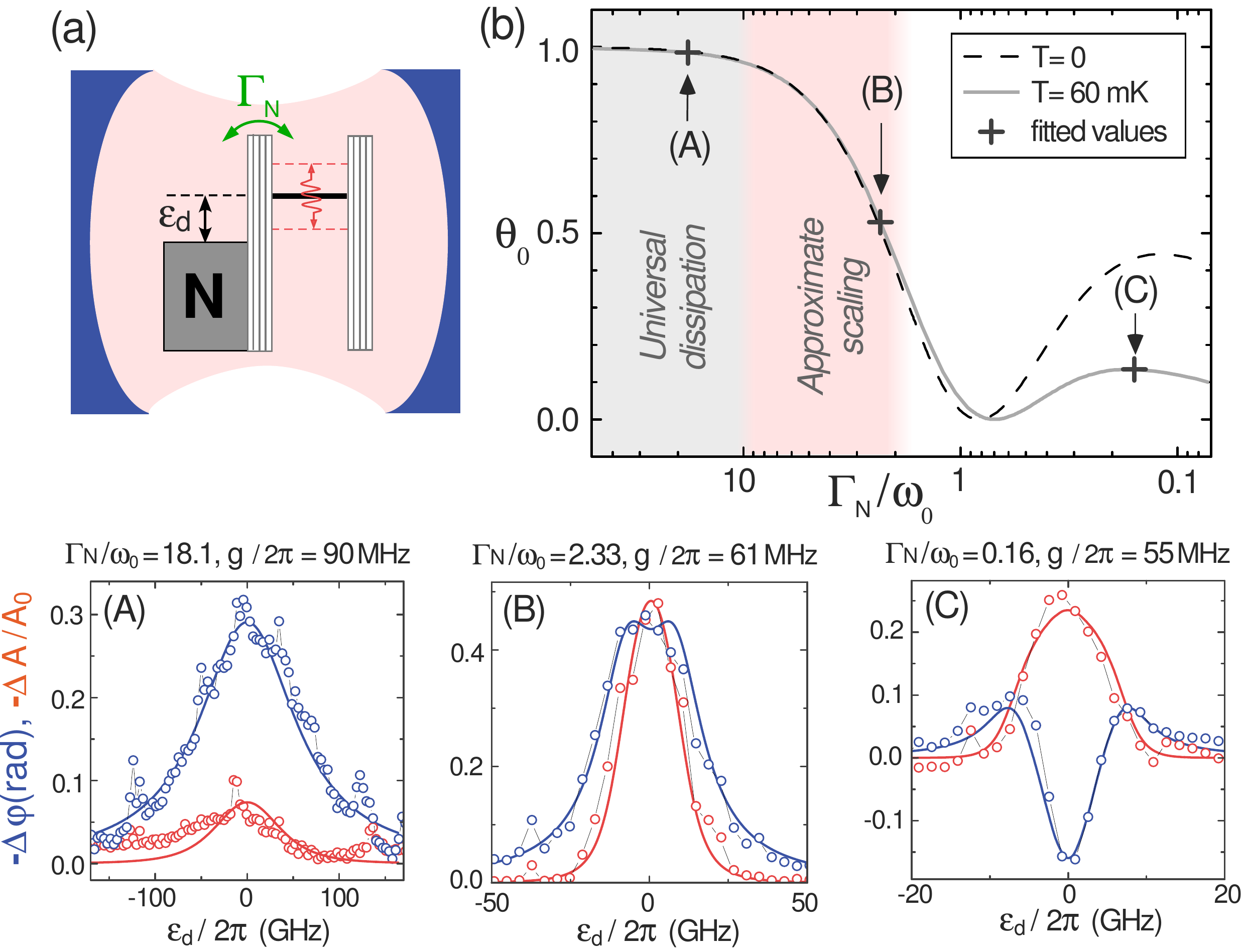}\caption{(a) Scheme of a
microwave cavity coupled to a single quantum dot with a normal metal
reservoir. (A), (B) and (C): microwave response of the cavity versus the
energy detuning $\varepsilon_{d}$ between the dot orbital and the Fermi level
of the reservoir, for various values of the tunnel rate $\Gamma_{N}$ to the
reservoir. Note that we represent the opposites of $\Delta A$ and
$\Delta\varphi$. The symboles are data and the lines is the theory discussed
in the main text (b) Ratio $\theta_{0}$ corresponding to the theory curves in
panels (A), (B), and (C). Adapted from Ref.[\cite{Bruhat:2016a}].}%
\label{Figure8.pdf}%
\end{figure}

The simplest possible example of open quantum dot circuit which can be coupled
to a microwave cavity is a single level quantum dot with a normal metal
reservoir. This situation has been studied experimentally with carbon-nanotube
quantum dots\cite{Delbecq:2011,Bruhat:2016a}. It is possible to control
electrically the energy detuning $\varepsilon_{d}$ between the dot level and
the Fermi energy of the reservoir (see Fig. \ref{Figure8.pdf}a). The signals
$\Delta\varphi$ and $\Delta A$ were measured as a function of $\varepsilon
_{d}$ for several dot levels with a different tunnel rate $\Gamma_{N}$ between
the dot and the normal metal reservoir (see Figure \ref{Figure8.pdf}, bottom
panels, which represents the opposites of $\Delta A$ and $\Delta\varphi$). For
$\omega_{0}\ll\Gamma_{N}$, the $\Delta A$ signal is very small while
$\Delta\varphi$ is always negative with a minimum at $\varepsilon_{d}=0$ (see
Fig.\ref{Figure8.pdf}A). For $\omega_{0}\gg\Gamma_{N}$, $\Delta A$ seems
globally larger but negative in any case, whereas $\Delta\varphi$ changes sign
and becomes positive around $\varepsilon_{d}=0$ (see Fig.\ref{Figure8.pdf}C).

To understand this behavior, one can model the device with Hamiltonian
(\ref{HH}) with
\begin{equation}
\hat{H}_{0}^{t}=\varepsilon_{d}\hat{c}_{d\sigma}^{\dag}\hat{c}_{d\sigma}+%
{\textstyle\sum\nolimits_{k,\sigma}}
\left(  t\hat{c}_{k\sigma}^{\dag}\hat{c}_{d\sigma}+t^{\ast}\hat{c}_{d\sigma
}^{\dag}\hat{c}_{k\sigma}\right)  \label{ou}%
\end{equation}
and%
\begin{equation}
\hat{h}_{int}=g\hat{c}_{d\sigma}^{\dag}\hat{c}_{d\sigma}\text{.} \label{hintt}%
\end{equation}
Above, $t$ is the tunnel hopping parameter between the dot and the reservoir,
and $g$ the coupling of the dot level to cavity photons. The spin index is
$\sigma\in\{\uparrow,\downarrow\}$. The tunnel rate between the dot and the
reservoir can be expressed as $\Gamma_{N}=2\pi\left\vert t\right\vert ^{2}$.
One can use the non-interacting Keldysh formalism of section
\ref{Keldysh} to calculate the charge susceptibility $\Xi(\omega_{0})$ which
corresponds to Eqs.(\ref{ou}) and (\ref{hintt}). Here, we define
\begin{equation}
\mathcal{G}^{r}(\omega)=\left[
\begin{tabular}
[c]{ll}%
$\mathcal{G}_{\hat{c}_{d\uparrow},\hat{c}_{d\uparrow}^{\dag}}^{r}$ &
$\mathcal{G}_{\hat{c}_{d\uparrow},\hat{c}_{d\downarrow}^{\dag}}^{r}$\\
$\mathcal{G}_{\hat{c}_{d\downarrow},\hat{c}_{d\uparrow}^{\dag}}^{r}$ &
$\mathcal{G}_{\hat{c}_{d\downarrow}^{\dag},\hat{c}_{d\downarrow}^{c}}^{r}$%
\end{tabular}
\ \ \ \ \right]  \label{INV}%
\end{equation}
Equation (\ref{ou}) then gives%
\begin{equation}
\mathcal{G}^{r}(\omega)=\frac{1}{\omega-\varepsilon_{d}+i\Gamma_{N}}\check{1}
\label{ggr}%
\end{equation}
and
\begin{equation}
\check{\Sigma}^{<}(\omega)=i\Gamma_{N}f(\omega)\check{1}%
\end{equation}
with $f(\omega)=1/(1+\exp[\omega/k_{B}T])$ and $\check{1}$ the identity matrix
in spin space. In this diagonal case, it is possible to reduce the number of
Green's function in expression (\ref{xi}) by using the identity%
\begin{equation}
\mathcal{G}^{r}(\omega)\mathcal{G}^{r}(\omega-\omega_{0})=\frac{\mathcal{G}%
^{r}(\omega-\omega_{0})-\mathcal{G}^{r}(\omega)}{\omega_{0}}%
\end{equation}
This leads to the result%
\begin{equation}
\Xi(\omega_{0})=\frac{G_{ac}(\omega_{0})}{i\omega_{0}} \label{XiGac}%
\end{equation}
with
\begin{equation}
G_{ac}(\omega_{0})=-ig^{2}\Gamma_{N}%
{\textstyle\int}
\frac{d\omega}{2\pi}\mathcal{G}^{a}(\omega-\omega_{0})\mathcal{G}^{r}%
(\omega)(f(\omega)-f(\omega-\omega_{0})) \label{Gac}%
\end{equation}
\begin{figure}[h]
\includegraphics[width=0.7\linewidth]{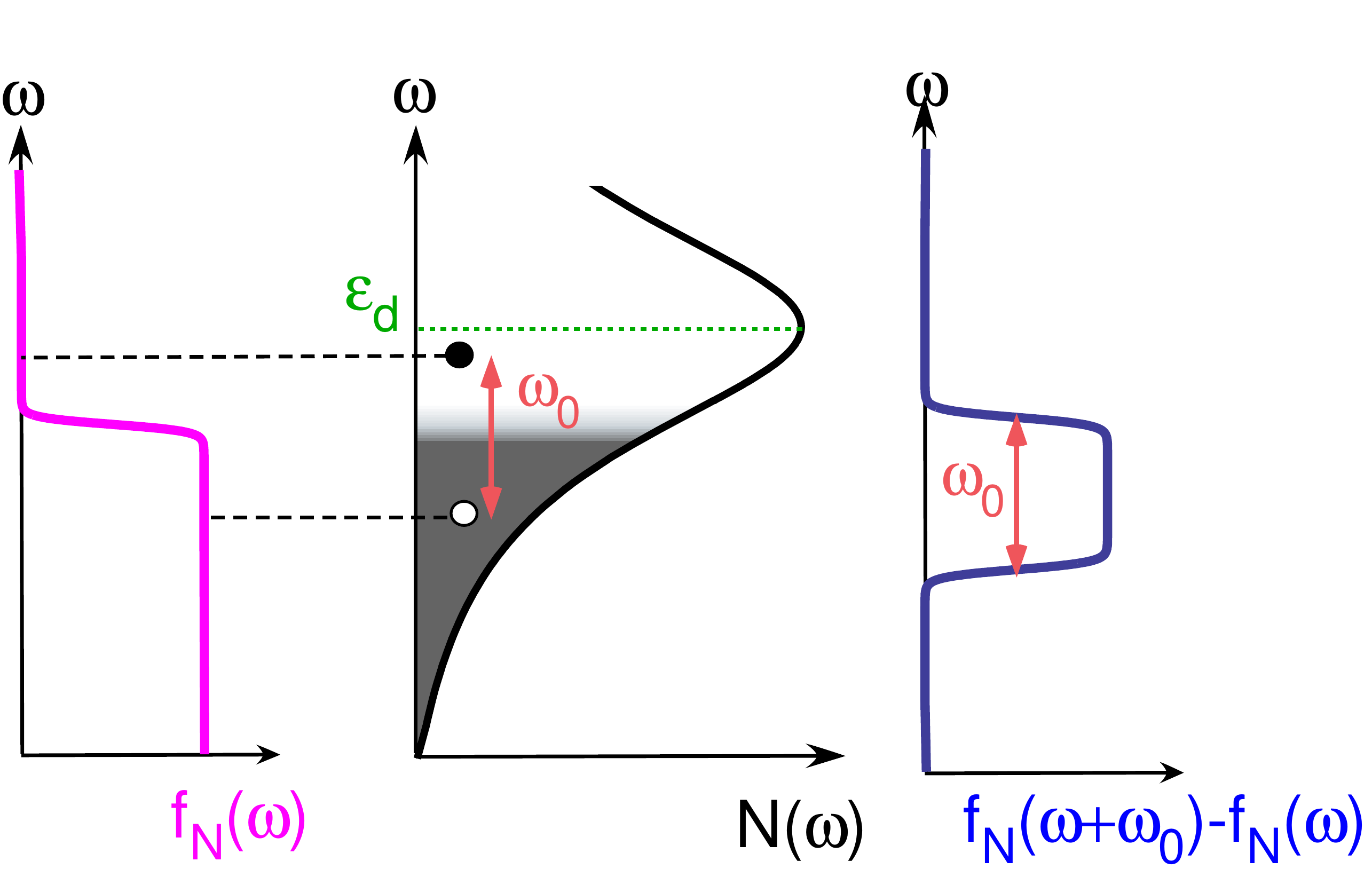}\caption{Plot of the Fermi
factor $f_{N}(\omega)$, the dot density of states $N(\omega)=-Im[\mathcal{G}%
^{r}(\omega)]/\pi$ dressed by tunneling to the reservoir, and the fermi factor
difference $f_{N}(\omega)-f_{N}(\omega-\omega_{0})$. The cavity with frequency
$\omega_{0}$ can induce the formation of electron-hole pairs in $N(\omega)$ if
the factor $f_{N}(\omega)-f_{N}(\omega-\omega_{0})$ is finite.}%
\label{Figure14}%
\end{figure}Since the current through the tunnel junction corresponds to the
time derivative of the charge on the dot, $G_{ac}(\omega_{0})$ naturally
corresponds to the admittance of the dot. Interestingly, the interpretation of
the expression (\ref{Gac}) is straightforward: $G_{ac}$ is due to the creation
of electron hole pairs at energies $\omega$ and $\omega+\omega_{0}$
respectively, in the density of states (DOS) of the dot broadened by
$\Gamma_{N}$. This creation is possible if the state at frequency
$\omega+\omega_{0}$ has a low occupation probability whereas the other state
at $\omega$ has a high occupation probability. This is taken into account by
the difference of fermi factors in Eq.(\ref{Gac}) (see Fig.\ref{Figure14}).

Using Eq.(\ref{XiGac}), it is possible to reproduce quantitatively, the data
of Fig.\ref{Figure8.pdf}A, B, and C. In particular, the sign change in
$\Delta\omega_{0}$ when $\Gamma_{N}$ decreases can be reproduced. Physically,
for large tunnel rates $\Gamma_{N}\gg\omega_{0}$, the dot is able to absorb or
emit electrons very quickly in response to the modulation of its potential by
the cavity field. The dot thus behaves as an effective capacitor. Considering
the cavity as a (L,C) resonator in parallel with this effective capacitor,
this gives $\Delta\omega_{0}<0$. This behavior can be reproduced with a simple
quasi-static model. In the semiclassical and resonant limit with $\omega
_{0}\ll\Gamma_{N}$, one has $\left\langle \hat{c}_{d}^{\dag}\hat{c}%
_{d}\right\rangle =n_{d}+(\partial n_{d}/\partial\varepsilon_{d})\bar
{a}e^{-i\omega_{RF}t}$, with $n_{d}$ is the average equilibrium value of
$\hat{c}_{d}^{\dag}\hat{c}_{d}$ for $g=0$, because the dot charge has the time
to reach its equilibrium value at each time. Therefore, one has%
\begin{equation}
\Xi(\omega_{0}\ll\Gamma_{N})=g\partial n_{d}/\partial\varepsilon_{d}%
\end{equation}
Hence, from Eqs. (\ref{j1}) and (\ref{j2}), one gets a cavity frequency shift
$\Delta\omega_{0}=g^{2}\partial n_{d}/\partial\varepsilon_{d}<0$ which is
minimum for $\varepsilon_{d}=0$, in full agreement with Fig. \ref{Figure8.pdf}%
A. In contrast, for small tunnel rates $\Gamma_{N}<\omega_{0}$, there is a
lagging effect in the charge current because the charge of the dot cannot
follow instantaneously the level oscillation caused by the cavity field.
Hence, the dot has an inductive behavior which naturally leads to
$\Delta\omega_{0}>0$ (see \ref{Figure8.pdf}C). Note that this capacitive to
inductive cross-over has also been observed in Ref.\cite{Frey:2012a}.

The dissipative part of the cavity response in Fig.\ref{Figure8.pdf} might
seem less interesting since it displays no sign change. However, many
theoretical studies on the charge relaxation resistance in single quantum dots
reveal a surprising universal charge relaxation resistance. This effect was
first predicted by Markus B\"{u}ttiker two decades
ago\cite{Buttiker:1993,Pretre:1996} for a non-interacting quantum dot tunnel
contacted to a normal metal reservoir and capacitively coupled to an ac
voltage source with frequency $\omega_{0}$. In this case, for low frequencies
$\omega_{0}\ll\Gamma_{N}$, the quantum dot circuit is expected to behave like
a $(R,C)$ circuit with a constant resistance $R$, independently of the circuit
parameters. One can recover this result from Eq.(\ref{Gac}). More precisely,
in the limit $\omega_{0}\ll\Gamma_{N}$, one finds that $G_{ac}(\omega_{0})$
can be written in the form $G_{ac}(\omega_{0})=i\omega_{0}C+RC^{2}\omega
_{0}^{2}+o(\omega_{0}^{3})$ which corresponds to the expansion of the
admittance of a $(R,C)$ circuit at low frequency
with\cite{Buttiker:1993,Pretre:1996}%
\begin{equation}
R=\frac{\pi}{4}\frac{\hbar}{e^{2}} \label{R}%
\end{equation}
independently of the circuit parameters $\varepsilon_{d}$ and $\Gamma_{N}$.
From Eqs. (\ref{j1}) and (\ref{j2}) and (\ref{XiGac}), for the device of
Fig.\ref{Figure8.pdf}a, the universality of the charge relaxation resistance
is equivalent to having a ratio
\begin{equation}
\theta=\frac{\pi}{4e^{2}R}=\frac{\pi}{2}\frac{\omega_{0}}{g^{2}}\frac
{(\Delta\omega_{0})^{2}}{\Delta\Lambda_{0}}%
\end{equation}
equal to $1$ for any value of $\varepsilon_{d}.$ Remarkably, this implies that
the variations of $(\Delta\omega_{0})^{2}$ and $\Delta\Lambda_{0}$ with
$\varepsilon_{d}$ should be similar since $\theta$ and $R$ should not depend
on $\varepsilon_{d}$.

The value $R=\pi\hbar/2e^{2}$ was experimentally confirmed for a two
dimensional electron gas structure subject to a high magnetic field measured
without a cavity, through a direct ac conductance measurement\cite{Gabelli}.
In this case, a similar phenomenology is expected, with $\Xi$ divided by a
factor 2 and $R$ multiplied by a factor $2$ due to the lifting of the spin
degeneracy. However, the independence of $R$ from $\varepsilon_{d}$ could not
be checked in this experiment. The data of Figs \ref{Figure8.pdf} are
consistent with this behavior since they can be fitted by the Eqs.(\ref{ii}),
(\ref{xi}) and (\ref{ggr})-(\ref{Gac}) (see full lines), which give
Eq.(\ref{R}) in the limit $\omega_{0}\ll\Gamma_{N}$. Figure \ref{Figure8.pdf}b
shows the values of $\theta$ calculated from this theory at $\varepsilon
_{d}=0$ for the three resonances of the bottom panels. This shows that a broad
range of frequency regimes could be addressed with this sample. The
universality of $R$ is expected in the gray area of Fig.\ref{Figure8.pdf}b
where $\omega_{0}\ll\Gamma_{N}$. In the pink area of Fig.\ref{Figure8.pdf}b,
one expects $R<1$. The absolute value of $R$ could not be measured in
Ref.\cite{Bruhat:2016a}, but it was found that, in the pink area, the curves
$(\Delta\omega_{0})^{2}$ and $\Delta\Lambda_{0}$ versus $\varepsilon_{d}$ show
similar variations (see Ref.\cite{Bruhat:2016a} for details). The scaling of
the $(\Delta\omega_{0})^{2}$ and $\Delta\Lambda_{0}$ curves in the gray area
could not be studied accurately due to experimental noise. A further
investigation of this regime would be very interesting.

Surprisingly, the data of Fig.\ref{Figure8.pdf} could be interpreted with a
non-interacting theory, although Coulomb blockade was visible in finite bias
voltage measurements (not shown). Hence, it would be interesting to understand
to which extent the universality of the charge relaxation resistance is robust
to
interactions\cite{Nigg:2006,Mora:2010,Hamamoto:2010,Filippone:2011,Alomar:2016}%
. There is not yet a full consensus on this question and the answer could
depend on the regime of parameter considered. Mesoscopic QED experiments
provide a new tool to study this question.

\subsection{Photon emission by a superconductor/quantum dot
interface\label{emitDsot}}

\begin{figure}[th]
\includegraphics[width=1\linewidth]{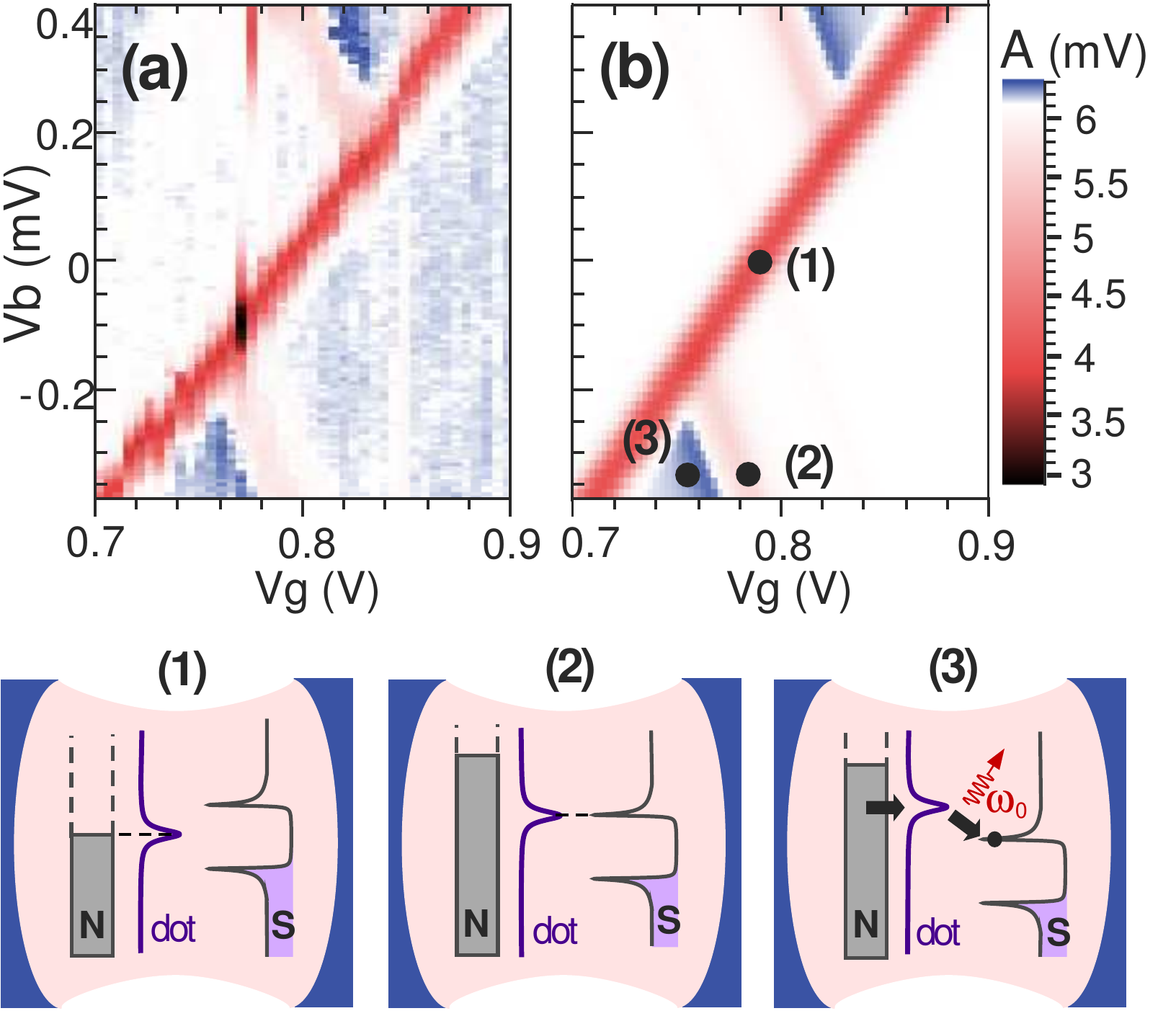}\caption{(a) Microwave
transmission amplitude measured for a cavity coupled to a
superconductor/quantum dot/normal metal bijonction. The white shade
corresponds to the reference amplitude $A_{0}=6.1~\mathrm{mV}$ and the blue
and red shades to $\Delta A>0$ and $\Delta A<0$ respectively (b) Similar
quantity, calculated with the Keldysh approach (see text). Bottom panels:
Scheme of the bijonction in different configurations: in (1), the dot level is
resonant with the Fermi level of the normal metal, in (2), the dot level is
resonant with a BCS peak of the superconductor, and in (3), the dot level has
an energy higher than a BCS peak by $\omega_{0}$, which enables photo-assisted
tunneling. Adapted from Ref.[\cite{Bruhat:2016a}].}%
\label{Figure10}%
\end{figure}

So far, we have discussed configurations in which the nanocircuit embedded in
the microwave cavity is not voltage biased. However, the combination of the
light-matter interaction and the out-of equilibrium dynamics in a
voltage-biased nanocircuit can lead to unique features. This is well
illustrated by the case of a superconductor/quantum dot/ normal metal
bijunction embedded in a microwave cavity. Fig.\ref{Figure10}a shows the
cavity dissipative signal $\Delta A$ versus the gate voltage $V_{g}$ of the
dot, which shifts the energy level $\varepsilon_{d}$, and versus the finite
bias voltage $V_{b}$ applied to the normal metal contact. An area with $\Delta
A>0$ appears (see blue area in Fig.\ref{Figure10}a). This suggests that the
bijunction emits photons for some regimes of parameters.

In order to understand this behavior, one can use again expression (\ref{xi}).
To take into account the presence of superconducting correlations in the
device, one can use the dot Green's function
\begin{equation}
\mathcal{\check{G}}^{r(a)}=\left[
\begin{tabular}
[c]{ll}%
$\mathcal{G}_{\hat{c}_{d\uparrow},\hat{c}_{d\uparrow}^{\dag}}^{r(a)}$ &
$\mathcal{G}_{\hat{c}_{d\uparrow},\hat{c}_{d\downarrow}}^{r(a)}$\\
$\mathcal{G}_{\hat{c}_{d\downarrow}^{\dag},\hat{c}_{d\uparrow}^{\dag}}^{r(a)}$
& $\mathcal{G}_{\hat{c}_{d\downarrow}^{\dag},\hat{c}_{d\downarrow}^{c}}%
^{r(a)}$%
\end{tabular}
\ \ \ \ \right]  \label{green}%
\end{equation}
where $\hat{c}_{d\sigma}^{\dag}$ creates an electron with spin $\sigma$ on the
dot. One finds%
\begin{align}
\mathcal{\check{G}}^{r}(\omega)  &  =\left[
\begin{tabular}
[c]{ll}%
$\omega-\varepsilon_{d}+i\frac{\Gamma_{N}}{2}$ & $0$\\
$0$ & $\omega+\varepsilon_{d}+i\frac{\Gamma_{N}}{2}$%
\end{tabular}
\ \ \ \ \right] \label{ggg}\\
&  +i\frac{\Gamma_{S}}{2}\left[
\begin{tabular}
[c]{ll}%
$G_{\omega}$ & $F_{\omega}$\\
$F_{\omega}$ & $G_{\omega}$%
\end{tabular}
\ \ \ \ \right] \nonumber
\end{align}
Above, $\Gamma_{N(S)}$ is the bare tunnel rate between the dot and the
normal(superconducting) reservoir. The presence of superconducting
correlations in the superconducting reservoir is described by the functions
$G_{\omega}=-i(\omega+i\frac{\Gamma_{n}}{2})/D_{\omega}$, and $F_{\omega
}=i\Delta/D_{\omega}$ with $D_{\omega}=\sqrt{\Delta^{2}-(\omega+i\frac
{\Gamma_{n}}{2})^{2}}$ and $\Delta$ is the gap of the superconductor. The
parameter $\Gamma_{n}$ accounts for a broadening of the BCS peaks in the DOS
of the superconductor, which is observed experimentally. In the absence of
superconductivity ($\Delta=0$), one has $G_{\omega}=1$ and $F_{\omega}=0$ so
that the off-diagonal part of the Green's function which describes the
presence of superconducting correlations in the dot vanishes.

The self-energy $\check{\Sigma}^{<}(\omega)$ corresponding to Eq.(\ref{ggg})
is given in Ref.\cite{Bruhat:2016a}. Using the above theory, Equation
(\ref{xi}) gives the results of Fig.\ref{Figure10}b, which reproduce closely
the data of Fig.\ref{Figure10}a. The line marked with the circle (1)
corresponds to a resonance between the dot orbital and the Fermi level in the
normal metal, whereas the circle (2) corresponds to a resonance between the
dot level and a BCS peak in the superconducting reservoir. The $\Delta A>0$
effect is due to inelastic tunneling from the dot level to one BCS peak, which
triggers the emission of a cavity photon, as represented in Fig.\ref{Figure10}.3

It is interesting to evaluate the performance of the device of
Fig.\ref{Figure10} as a photon emitter. In order to evaluate the emission rate
$\Gamma_{e}$ of photons in the cavity by the quantum dot circuit, one needs to
\ calculate the time evolution of the average number of cavity photons
$\left\langle \hat{a}^{\dag}\hat{a}\right\rangle =\left\vert \bar
{a}\right\vert ^{2}$. By combining Eqs.(\ref{eom}), (\ref{aaa}) and the top
left Eq. of Fig.(\ref{Figure13}), one gets
\begin{equation}
\frac{\partial\left\vert \bar{a}\right\vert ^{2}}{\partial t}=\Gamma
_{e}-2\Lambda_{0}\left\vert \bar{a}\right\vert ^{2}-\sqrt{2\Lambda_{L}}\left(
\bar{a}b_{in}^{\ast}+b_{in}^{\ast}\bar{a}\right)
\end{equation}
with%
\begin{equation}
\Gamma_{e}=ig\left(  \left\langle \hat{c}_{d\uparrow}^{\dag}\hat{c}%
_{d\uparrow}\right\rangle +\left\langle \hat{c}_{d\downarrow}^{\dag}\hat
{c}_{d\downarrow}\right\rangle \right)  (\bar{a}e^{-i\omega_{RF}t}-\bar
{a}^{\ast}e^{i\omega_{RF}t})
\end{equation}
the emission rate of photons inside the cavity by the nanocircuit. This rate
can be simplified by eliminating fast oscillating terms which do not
contribute significantly to photon emission. Using the linear response
expression
\begin{equation}%
{\displaystyle\sum\limits_{\sigma}}
\left\langle \hat{c}_{d\sigma}^{\dag}\hat{c}_{d\sigma}\right\rangle
(t)=\Xi(\omega_{RF})\bar{a}e^{-i\omega_{RF}t}+\Xi(-\omega_{RF})\bar{a}^{\ast
}e^{i\omega_{RF}t}%
\end{equation}
with $\Xi(-\omega_{RF})=\Xi(\omega_{RF})^{\ast}$, we get%
\begin{equation}
\Gamma_{e}=2\operatorname{Im}[\Xi(\omega_{RF})])\left\vert \bar{a}\right\vert
^{2}%
\end{equation}
From Eq.(\ref{j2}) and (\ref{deltaA}), this can be expressed in terms of the
experimental signals as:%
\begin{equation}
\Gamma_{e}=2\frac{\Delta A\Lambda_{0}}{A_{0}}\left\langle \hat{n}\right\rangle
\label{ge}%
\end{equation}
The data of Fig.\ref{Figure10} were measured with an average photon number
$\left\langle \hat{a}^{\dag}\hat{a}\right\rangle \sim120$ in the cavity. The
calibration of the experiment also gives $\Lambda_{0}\sim2\pi\times
0.26~\mathrm{MHz}$ and $A_{0}\sim6.1~\mathrm{mV}$. The area with $\Delta A>0$
in Fig.\ref{Figure10} corresponds to $\Delta A/A_{0}\simeq0.03$. From
Eq.(\ref{ge}), this gives $\Gamma_{e}\sim2\pi\times2MHz$, which corresponds to
an inelastic current of the order of $0.3~\mathrm{pA}$. This current was not
detectable in the experiment because of current noise.

\subsection{Voltage-biased double quantum dot in a cavity\label{emitDQD}}

\subsubsection{Probing out-of-equilibrium double dot populations with a
microwave cavity\label{DQDJay}}

\begin{figure}[th]
\includegraphics[width=1\linewidth]{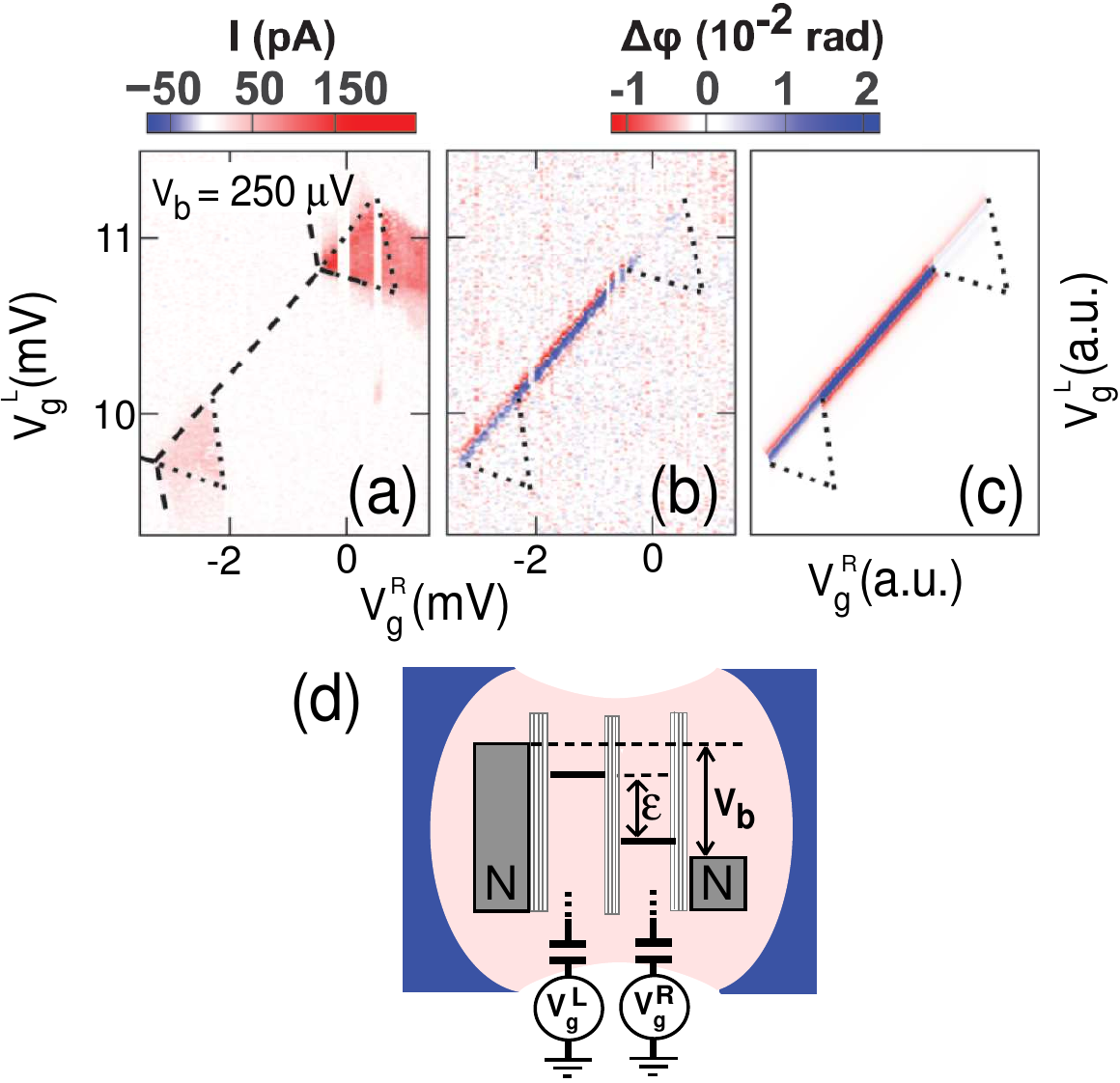}\caption{(a) dc current
through a carbon-nanotube double quantum dot versus the gate voltages
$V_{g}^{L}$ and $V_{g}^{R}$ for a finite bias voltage $V_{b}=250\mathrm{\mu
V}$. (b) Corresponding cavity signal $\Delta\varphi$ (c) Theoretical
predictions for $\Delta\varphi$, obtained with Eq.(\ref{XiDQD}) and a master
equation calculation of $n_{+}$ and $n_{-}$ at zeroth order in $g_{t}$.
Adapted from Ref.[\cite{Viennot:2014a}].}%
\label{Figure17}%
\end{figure}

From Eq.(\ref{XiDQD}), current transport in the double dot can modify the
cavity signals by modifying the populations $n_{-}$ and $n_{+}$ of the bonding
and antibonding states. The transport configuration can be tuned electrically
through the double dot gate voltages $V_{g}^{L}$ and $V_{g}^{R}$ and
source-drain bias voltage $V_{b}$. Figure \ref{Figure17} presents results
obtained with a carbon nanotube double quantum dot with a finite Coulomb
interaction \cite{Viennot:2014a}. As observed usually, the current $I$ through
the double dot is finite only inside some triangles in the $V_{g}^{L}%
-V_{g}^{R}$ plane where the bonding or antibonding states are inside the
transport window opened by the source-drain voltage $V_{b}$
(Fig.\ref{Figure17}a). The cavity signal $\Delta\varphi$ is maximum along the
line $\varepsilon=0$ where $\omega_{DQD}-\omega_{0}$ is minimum
(Figs.\ref{Figure17}b), and it takes a different value along the transport
triangles, because the populations of $n_{-}$ or $\ n_{+}$ are modified by
transport. In Fig.\ref{Figure17}c, this behavior is well reproduced by using
Eq.(\ref{XiDQD}), with $n_{-}$ and $n_{+}$ calculated with a master equation
approach at lowest order in the light matter coupling ($g_{t}=0$) (see details
in Ref.\cite{Viennot:2014a}). In the regime $\Gamma_{L(R)}\ll k_{b}T$, this
approach is sufficient to take into account Coulomb blockade, which
essentially affects the structure of the nanocircuit state space and tunnel
rates. The inclusion of Coulomb interactions in the Keldysh formalism is a
more complex task\cite{Ng:1996}. Note that however, in the non-interacting
case, the Keldysh approach also reproduces well the fact that $\Delta\varphi$
is affected by current transport through a modification of $n_{-}-n_{+}$ (see
Appendix \ref{DQDkeldysh}). Remarkably, the dc current through the double dot
and the cavity signals are qualitatively different, since the cavity signals
directly probe $n_{-}-n_{+}$ whereas the current $I$ is a more complex
combination of $n_{-}$, $n_{+}$ and the double dot parameters. Therefore, the
simultaneous study of the two signals can again enable a more accurate
characterization of the double dot parameters.

\subsubsection{Photon emission below the lasing threshold}

\begin{figure}[th]
\includegraphics[width=1\linewidth]{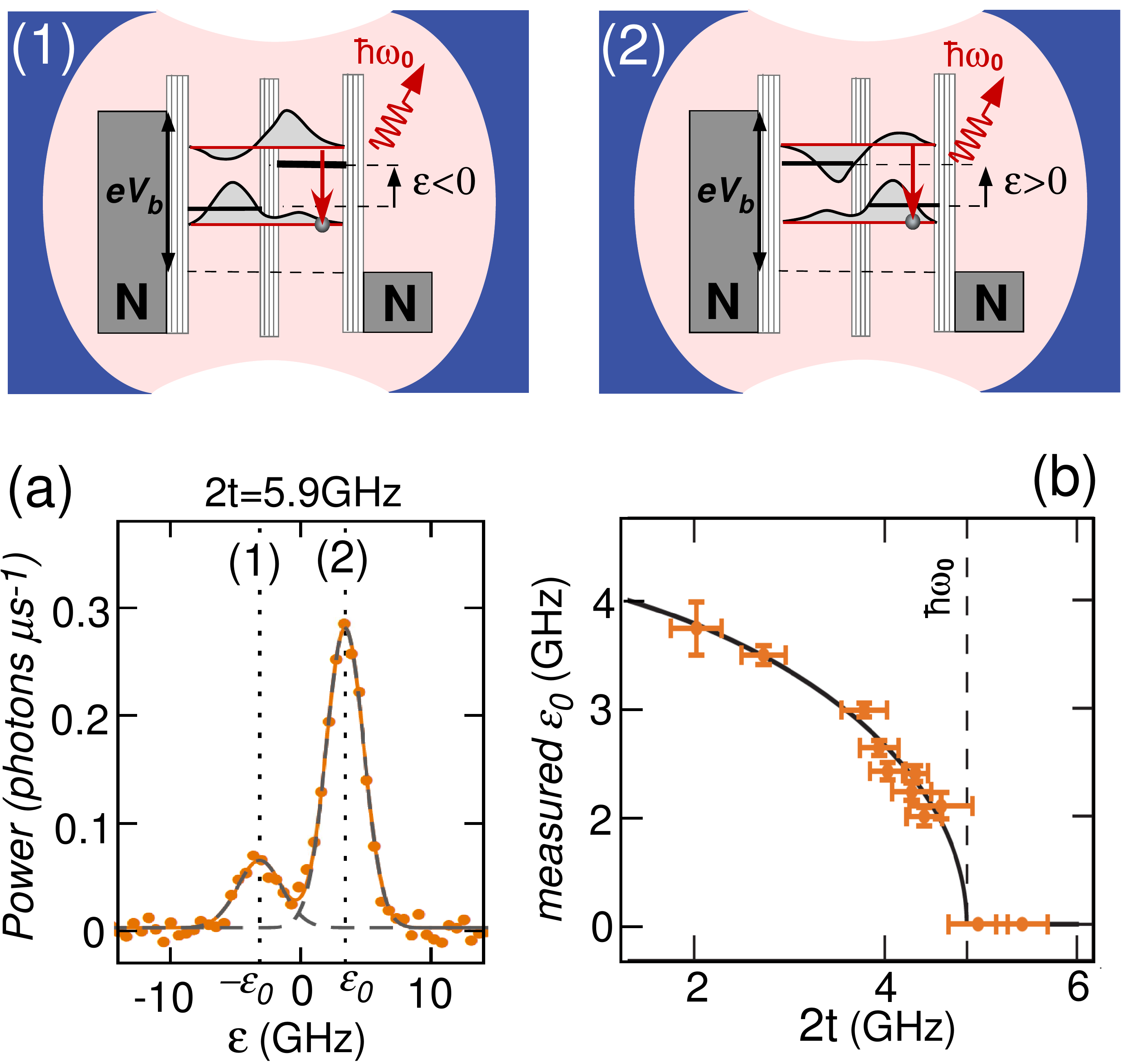}\caption{Top panels: Scheme
of a double dot subject to a finite bias voltage $V_{b}$, with a negative
orbital detuning ($\varepsilon<0$, panel (1)) or a positive orbital detuning
($\varepsilon>0$, panel (2)). Bottom panels: measurements performed with a
GaAs/AlGaAs heterostructure. (a) Number of photons $P$ emitted by the cavity
per unit time versus the orbital detuning $\varepsilon$. The dotted lines
corresponds to the situations depicted in panels (1) and (2). (b) Resonant
detuning $\varepsilon_{0}$ measured for different values of interdot coupling
$t$. The dots/squares are data points and the full lines a theory based on a
master equation approach. Adapted from Ref.[\cite{Stockklauser:2015}].}%
\label{Figure12}%
\end{figure}

The splitting between the bonding and antibonding states of the double dot can
become resonant with the cavity if $\varepsilon$ is tuned properly. In this
regime, it is possible to obtain photon emission due to the dc voltage bias.
This phenomenon has been investigated in two dimensional electron gas
structures and semiconducting nanowires\cite{Stockklauser:2015,Liu:2014}. In
Ref.\cite{Stockklauser:2015}, the number $P$ of photons emitted by the cavity
per unit time was measured as a function of the double dot detuning
$\varepsilon$ for a constant interdot detuning $t$ (see Figure \ref{Figure12}%
a). For $2t>\omega_{0}$, two resonances appear in $P$, for orbital detunings
$\varepsilon=\pm\varepsilon_{0}$ with $\varepsilon_{0}=\sqrt{\omega_{0}%
^{2}-4t^{2}}$. The resonance for $\varepsilon<0$ is less pronounced than the
resonance for $\varepsilon>0$ because it requires that the electrons tunnel
from the left reservoir to an antibonding state which has little extension on
the left dot (see Figure \ref{Figure12}.1). Figure \ref{Figure12}b shows the
measured value of $\varepsilon_{0}$ versus $t$, which can be tuned with gate
voltages in two dimensional electron gas structures. As expected,
$\varepsilon_{0}$ vanishes for $2t>\omega_{0}$, because it is not possible to
satisfy the resonant condition $\omega_{DQD}=\omega_{0}$ in this case. In this
limit, the cavity shows a single resonance centered on $\varepsilon=0$ (not
shown). The light/matter coupling in this experiment can be estimated from the
parameters $g/2\pi=11~\mathrm{MHz}$, $\Gamma_{2}^{\ast}/2\pi=250~\mathrm{MHz}$
and $\Lambda_{0}=1.7~\mathrm{MHz}$, which corresponds to $Q_{e-ph}=0.068$ and
$C_{e-ph}=0.29$. The data were interpreted with a master equation approach
similar to the one of Ref.\cite{Viennot:2014a} (see section \ref{DQDJay}). In
the out-of-equilibrium regime, the rate of photon emission in the cavity by
the double dot is $\Gamma_{e}\sim2\pi\times0.3~\mathrm{MHz}$. This number is
similar to the performances obtained with the normal metal/dot/superconductor
bijunction of section \ref{emitDsot}.

\subsubsection{Photon emission above the lasing threshold}

\begin{figure}[th]
\includegraphics[width=1\linewidth]{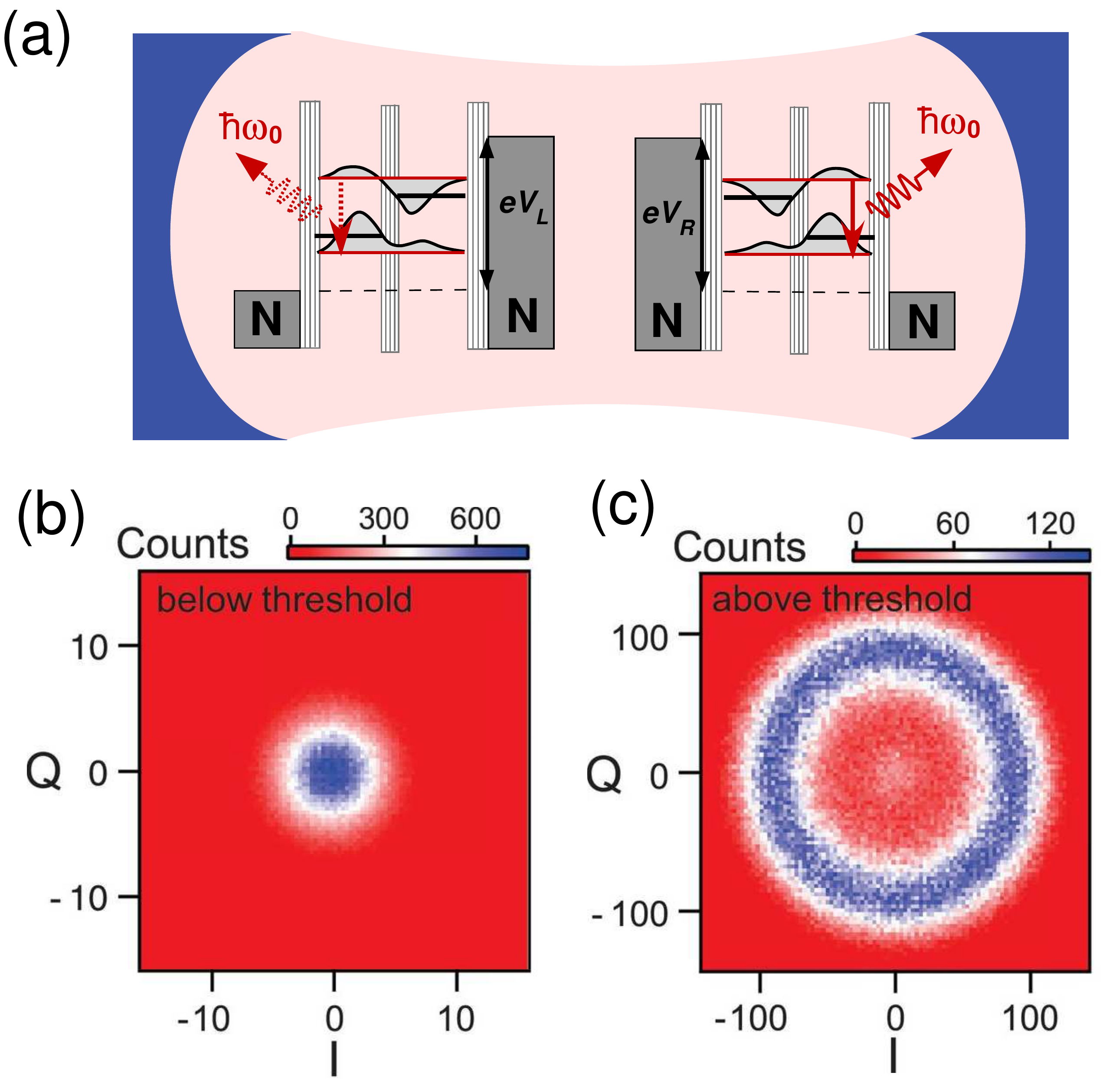}\caption{(a): Scheme of a
microwave cavity coupled to two double quantum dots (b) Measured I-Q
tomography of a cavity output field, in the presence of two InAs nanowire
double dots, when one double dot is resonant with the cavity an the other off
resonant (b) Measured I-Q tomography of the cavity output field when the two
double dots are resonant with the cavity. Adapted from Ref.[\cite{Liu:2015a}%
].}%
\label{Figure11}%
\end{figure}

In principle, when a double dot with $\Gamma_{\varphi}^{\ast}\gg\Gamma_{1}$ is
resonant with a cavity ($\omega_{ij}=\omega_{0}$), it is possible to obtain a
lasing effect, which corresponds to an emission of a coherent microwave
radiation by the double dot, if $C_{e-ph}\gtrsim1/2$%
\cite{PeiQing2011,PeiQing2012a,PeiQing2013,Kulkarni:2014,Lambert:2015,Karlewski:2016,Marthaler:2015,Gullans:2016}%
. When $C_{e-ph}<1/2$ , it is possible to reach the lasing regime by coupling
several double quantum dots to the cavity. This was recently realized with two
double dots made in InAs nanowires\cite{Liu:2015a,Liu:2015}. Figure
\ref{Figure11} shows the in-phase and quadrature phase components I and Q of
the output field of the cavity, measured when only one of the dots fulfills
$\omega_{DQD}=\omega_{0}$ (panel (b)) or when the two dots satisfy
$\omega_{DQD}=\omega_{0}$ (panel (c)). In the first case, the cavity photons
show a thermal distribution, because the device is below the lasing threshold.
In the second case, the "ring" in the tomography reveals a coherent photonic
emission, because the device is above the lasing threshold. An average photon
number in the cavity $\left\langle \hat{a}^{\dag}\hat{a}\right\rangle
\simeq8000$ is estimated in this last case. The rate of photon emission by the
double dot can be estimated as $\Gamma_{e}\simeq\left\langle \hat
{n}\right\rangle \omega_{0}/2Q_{0}=10~\mathrm{GHz}$, which is significantly
larger than in Refs.\cite{Stockklauser:2015,Bruhat:2016a}. Charge noise limits
the maser linewidth. However, a linewidth narrowing by more than a factor of
10 can be obtained by using a microwave input tone that stabilizes the
frequency of laser emission by triggering stimulated emission\cite{Liu:2015}.
This technique is known as injection locking in the field of conventional
lasers\cite{Stover}. Note that in the lasing regime, the linear theory of
section \ref{linear} fails. We refer interested readers to
Ref.\cite{Andre:2006} for a simple theoretical description of this limit.
Interestingly, there is a close analogy between the lasing effect produced in
a microwave cavity by an out-of-equilibrium quantum dot, and the one obtained
with a Josephson superconducting transistor with a finite voltage
bias\cite{Astafiev:2007}. In this second case, the lasing transition
corresponds to a change in the number of Cooper pairs inside a superconducting
island. In Ref.\cite{Astafiev:2007}, a photon emission rate inside the cavity
$\Gamma_{e}\simeq2~\mathrm{GHz}$ has been obtained, which is comparable to the
performance of Ref.\cite{Liu:2015}.

\subsection{Characterizing exotic condensed matter systems with a microwave
cavity}

In the previous sections, we have considered relatively simple situations with
nanocircuits made out of only one or two dots and fermionic reservoirs, in the
non-interacting regime or in the Coulomb blockade regime. However, this
section shows that the use of more complex circuits with possibly many body
correlations enables one to study non-trivial quantum transport effects or
condensed matter problems in a new way. Cavity photons appear as a new tool
which can give qualitatively different information on exotic excitations in a
nanocircuit, in comparison with traditional dc transport measurements. With
mesoscopic QED devices, it is possible to measure simultaneously dc currents
through the nanostructure and the microwave cavity response. The joint study
of these two non-equivalent signals is particularly informative as we will see below.

\subsubsection{Kondo quantum dots}

The Kondo effect was observed experimentally since the 30's [see for instance
Ref.\cite{Franck:1961}]. Although the resistivity of bulk metals is expected
to decrease with temperature, for some metals containing a small amount of
magnetic impurities an increase was observed below the Kondo temperature
$T_{K}$. Thirty years later, Kondo suggested that this behavior is due to
spin-flip scattering processes between the itinerant electrons of the metal
and the magnetic impurities\cite{Kondo:1964}. These processes give birth to a
Kondo cloud which screens the spins of the impurities and reduces the
conductivity of the metal.

More recently, it was shown that quantum dots can also be used to study the
Kondo effect if they are subject to a strong Coulomb
interaction\cite{Cronenwett:1998,Goldhaber:1998,Nygard:2000}. In this case,
the spin-flip scattering processes give rise to an increase of the dot
zero-bias conductance. This effect depends on the value of the charging energy
$U$ to add a second electron in a dot level. If $U$ is large and that the dot
orbital energy $\varepsilon_{d}$ fulfills $-U<\varepsilon_{d}<0$ , there can
only be a single electron in the dot orbital, whose electronic spin simulates
a local magnetic impurity. Then, if the dot is coupled to normal metal
contacts with a high enough tunnel rate $\Gamma_{N}$, the local spin in the
dot can fluctuate due to even order tunnel processes, which change the dot
spin but not its charge. These processes involve intermediate virtual dot
states with a different charge (dot orbital empty or doubly occupied), which
are energetically forbidden but quantum mechanically allowed for a very short
amount of time $\sim\hbar/U$. Therefore, the Kondo effect should remain
transparent to a microwave cavity which is only sensitive to charge
fluctuations at frequencies $\omega\lesssim\omega_{0}$. Importantly, this test
requires that the cavity is mainly coupled to the dot level so that the cavity
signals are set by the dot charge susceptibility. On the contrary, if the
cavity modulates asymmetrically the potentials of the source and drain
reservoirs, the cavity signals can reveal resonances similar to the Kondo
conductance peak, as found experimentally in
Refs.\cite{Delbecq:2011,Deng:2016}

The charge susceptibility of a Kondo dot has been studied recently in a carbon
nanotube quantum dot which is coupled to the cavity through the dot energy
level only\cite{Desjardins:2017}. Both the Kondo regime and the Coulomb
blockade regime have been studied with the same sample. In the Coulomb
blockade regime, the conductance peaks through the dot are also visible in the
cavity signals because they correspond to real charge fluctuations which are
visible by cavity photons. However, in the Kondo regime, the low energy Kondo
peaks are visible in the dot conductance but not in the cavity signals. Hence,
the Kondo effect corresponds to conduction through essentially frozen charges
in the dot (for a detector with a frequency cutoff $\omega_{0}\ll T_{K}$).
This illustrates the decoupling of the spin and charge dynamics in the Kondo effect.

\subsubsection{The Cooper pair splitter}

The spatial separation of spin-entangled electrons from a Cooper pair is an
interesting goal in the context of quantum computation and communication. In
principle, a Cooper pair beam splitter (CPS) connected to a central
superconducting contact and two outer normal metal (N) contacts could
facilitate this process\cite{Recher:2001}. The spatial splitting of Cooper
pairs has been demonstrated experimentally from an analysis of the CPS average
currents, current noise, and current cross-correlations, in devices made out
of a carbon
nanotube\cite{Hofstetter:2009,Herrmann:2010,Hofstetter:2011,Schindele:2012,Herrmann:2012,Schindele:2015}
or a semiconducting InAs quantum wire\cite{Das:2012}. However, new tools
appear to be necessary to investigate further the Cooper pair splitting
dynamics, and in particular its coherence, which has not been demonstrated
experimentally so far in the N/dot/S/dot/N geometry. This coherence has two
intimately related aspects: the coherence of Cooper pair injection and the
conservation of spin entanglement. The first aspect is due to the fact that
Cooper pair injection into the CPS is a coherent crossed Andreev process,
which produces a coherent coupling between the initial and final states of the
Cooper pair in the superconducting contact and the double dot. In this
context, coupling the CPS to a microwave cavity would be very interesting
because it would enable one to perform the spectroscopy of the CPS and
identify anticrossings in the CPS spectrum, which are due to the coherence of
the injection process\cite{Cottet:2014}. Detecting the conservation of spin
entanglement represents an even greater challenge. In principle, microwaves
couple to transitions between the states of the CPS with matrix elements which
keep signatures of the coherent superposition of spins states displayed by a
singlet state. Therefore, a microwave cavity could help to characterize split
singlet states\cite{Cottet:2012}. Interestingly, a supercurrent was recently
observed in Josephson junction made out of two self-assembled quantum dots
coupled in parallel to two superconducting contacts\cite{Deacon:2015}. The
observation of a supercurrent necessarily implies a non-dissipative and thus
coherent pair injection process. However, even in this case, the use of a
microwave cavity would be very interesting to characterize further the device,
i.e. perform its spectroscopy and check whether the coherent Cooper pair
injection really goes together with a spin-entangled double dot state.

\subsubsection{Majorana nanocircuits\label{Majos}}

\begin{figure}[th]
\includegraphics[width=1\linewidth]{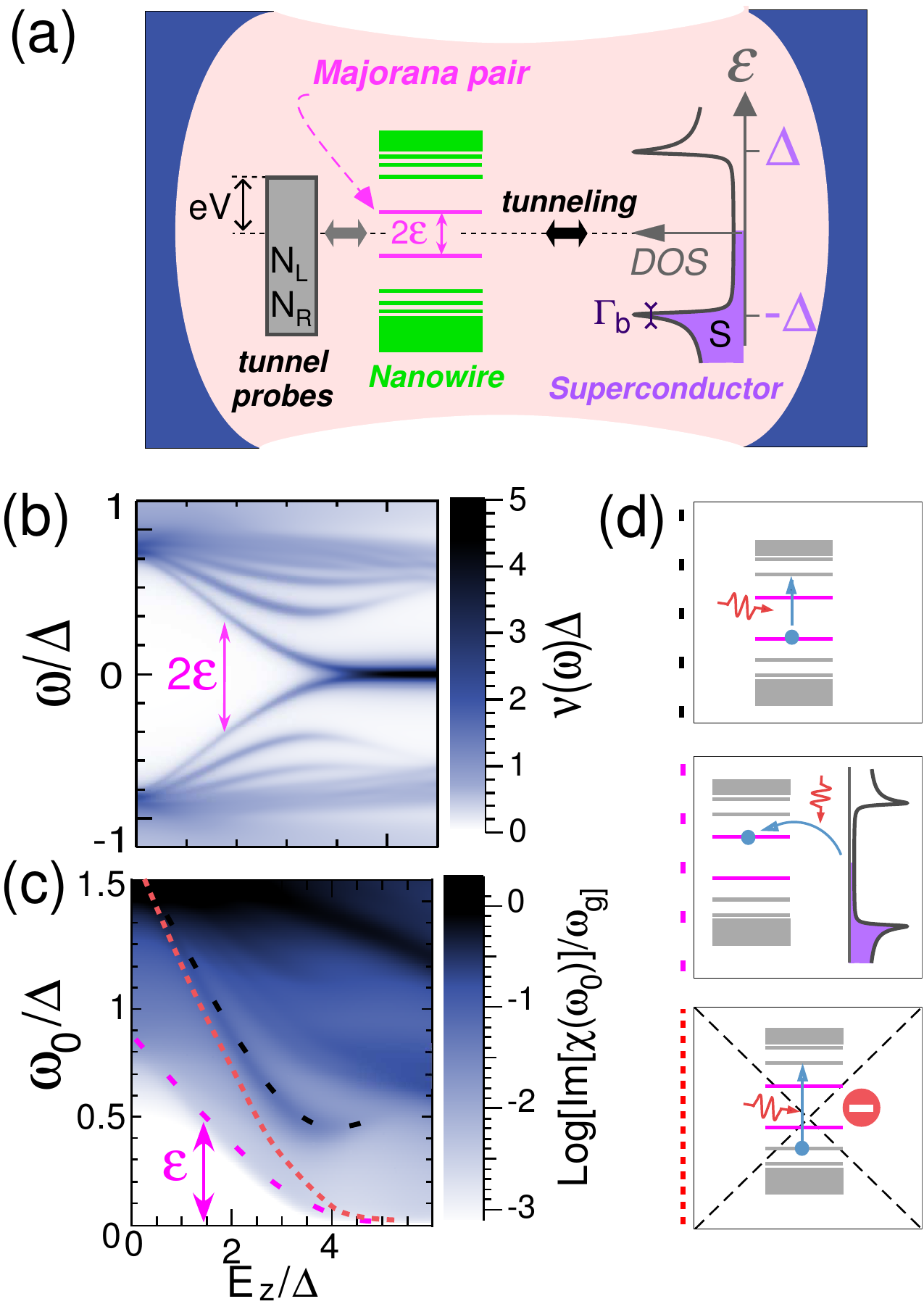}\caption{(a) Energetic scheme
of a Majorana nanowire coupled to a microwave cavity. The energy levels in the
Majorana nanowire (green and pink lines) are coupled to a superconducting
contact (DOS in purple) and two normal metal contacts (DOS in gray). The
Majorana doublet in pink has an energy splitting $2\varepsilon$ (b) Calculated
density of states in the nanowire (c) Calculated microwave response of the
cavity coupled to the nanowire (d) Schemes of the processes contributing to
the resonances highlighted with the dashed lines in panel (c) (see main text).
Adapted from Ref.[\cite{Dartiailh:2016}].}%
\label{Figure9}%
\end{figure}Majorana quasiparticles are among the most intriguing excitations
predicted in condensed matter physics\cite{Kitaev}. By definition, the
creation operator $m^{\dag}$ of a Majorana quasiparticle is self adjoint, i.e.
$m^{\dag}=m$. This property offers possibilities of non-abelian
statistics\cite{Ivanov} and topologically protected quantum
computation\cite{Nayak} in condensed matter systems. It has been found that
different types of hybrid electronic circuits could enclose Majorana
quasiparticles. In particular, hybrid structures combining a semiconducting
nanowire in contact with a superconductor raise a lot of
attention\cite{Mourik,Das,DengDeng:2012,Churchill,Albrecht,Deng:2016b,Zhang:2016}%
. It has been predicted that in some situations, a single pair of overlapping
Majorana bound states $(\hat{m}_{L},\hat{m}_{R})$ could appear inside a
semiconducting nanowire, with an overlap which can be switched off with an
external magnetic field or a gate voltage, in order to obtain two isolated
Majorana bound states\cite{Lutchyn,Oreg}. Recently, pairs of conductance peaks
with a splitting oscillating and decaying with the magnetic field were
observed, in striking agreement with these
predictions\cite{Albrecht,Deng:2016b}. However, so far, mainly dc conductance
measurements have been used, which reveal essentially the DOS of the nanowire.
This gives only a very indirect access to the property $m^{\dag}=m$. A
microwave cavity could represent an interesting tool to test more directly
this property since the self-adjoint property affects the structure of the
light-matter coupling. Here, we will mainly focus on the proposal of
Ref.\cite{Dartiailh:2016}, which considers a nanowire contacted to a
superconducting contact and two normal metal tunnel probes (Figure
\ref{Figure9}a). In practice, it is possible to measure simultaneously
$\Delta\Lambda_{0}$ from the cavity response, and the DOS of the nanowire by
using the tunnel probes. The Keldysh theory was used to calculate these two
quantities, by using a coarse grained description of the nanowire (see
Eq.(\ref{xi}) and Figures \ref{Figure9}b and c). To understand how a Majorana
pair affects physical signals it is convenient to reexpress the degree of
freedom associated to the Majorana pair by defining an ordinary fermion
operator $\hat{\gamma}_{1}^{\dag}=(\hat{m}_{L}-i\hat{m}_{R})/\sqrt{2}$ which
fulfills $\{\hat{\gamma}_{1}^{\dag},\hat{\gamma}_{1}^{\dag}\}=1$. At low
energies, in the subspace spanned by $\hat{\gamma}_{1}^{\dag}$, one gets the
nanocircuit Hamiltonian (\ref{Ht}) with $\hat{H}_{0}=\varepsilon\hat{\gamma
}_{1}^{\dag}\hat{\gamma}_{1}$ and $\hat{h}_{int}=\beta\hat{\gamma}_{1}^{\dag
}\hat{\gamma}_{1}$ (with the conventions of Eqs.(\ref{HoAndreev}) and
(\ref{HintAndreev}), one has $\varepsilon=E_{1}$ and $\beta=M_{11}$). This
means that the Majorana pair corresponds to a fermionic state which can be
split into two fully independent parts for $\varepsilon=0$. In the simplest
situation, when the nanowire is driven to its topological phase, $\varepsilon$
tends to zero, by showing or not an oscillatory behavior, depending on the
length of the nanowire. The electron-hole conjugated pair $(\hat{\gamma}%
_{1}^{\dag},\hat{\gamma}_{1})$ appears in the DOS of the nanowire as a pair of
resonances at $\varepsilon$ and $-\varepsilon$ (see Figure \ref{Figure9}b).
However, no transition should be visible in the cavity signals at $\omega
_{0}=2\varepsilon$ because from Eq.(\ref{XiAndreev}), the cavity photons
cannot induce transitions between a pair of electron-hole conjugated states
(see Figure \ref{Figure9}c). Nevertheless, the light-Majorana coupling has
physical consequences, since it can induce a step at $\omega_{0}=\varepsilon$
in the $\Delta\Lambda_{0}$ signal. This feature is due to photo-assisted
tunneling between the Majorana pair and the residual zero-energy DOS in the
imperfect superconducting reservoir. It can be used to check that the
low-energy doublet is well coupled to cavity photons, so that the absence of a
cavity resonance at $\omega_{0}=2\varepsilon$ is not due to $\beta=0$. Then,
the simultaneous presence[absence] of the step at $\omega_{0}=\varepsilon
\lbrack2\varepsilon]$ would represent a good indication that the low energy
doublet in the nanowire results from the combination of a non-degenerate
electron-hole pair, which is a natural precursor for a Majorana pair.
Importantly, this information cannot be directly obtained from the
DC\ current. Importantly, this protocol requires to measure simultaneously the
cavity response and the dc current through the tunnel probes. Although the
spectroscopic measurement described above is probably the most straightforward
measurement to perform with a cavity, many other effects are expected by
combining Majorana fermions and cavities. Hence, the direct electric coupling
between Majorana bound states and cavity photons has raised a lot of attention
recently\cite{Trif:2012,Schmidt:2013a,Schmidt:2013b,Cottet:2013,Dmytruk:2015}.
The indirect coupling of Majorana fermions to cavities through a
superconducting quantum bit also raises interest for quantum computation
purposes\cite{Vayrynen,Hassler,Hyart,Muller,Xue,Pekker,Ginossar,Ohm,Yavilberg,Yavilberg2}%
.

\section{Conclusion and perspectives}

In this review, we have shown that a microwave cavity represents a powerful
tool to investigate the properties and dynamics of electrons in a hybrid
nanocircuit. In the linear-coupling regime, the microwave cavity gives an
access to the charge susceptibility of the nanocircuit, which can be used to
understand most Mesoscopic QED experiments realized so far. First, many
different types of electronic degrees of freedom can be coupled coherently or
strongly to cavity photons. In particular, we have reviewed several promising
experiments where cavity photons are strongly or coherently coupled to the
charge\cite{Mi:2017,bruhat:2017,Stockklauser:2017} or spin\cite{Viennot:2015}
of a double dot or Andreev states\cite{Janvier:2015} in an atomic contact. The
investigation of the coherent dynamics of these degrees of freedom now seems
accessible\cite{Janvier:2015}. Second microwave cavities give a new access to
the tunneling dynamics of electrons between a dot and fermionic reservoirs. In
equilibrium conditions, the charge relaxation dynamics caused by a fermionic
reservoir can be studied with a high sensitivity. A first experiment seems
consistent with a non-interacting theory which suggests the universality of
the charge relaxation resistance in the adiabatic regime\cite{Bruhat:2016a}.
However, further study is required in the interacting case where a rich
phenomenology is expected. In out-of equilibrium conditions, there exists
several means to obtain photon emission, by using inelastic tunneling between
a quantum dot and a reservoir with an energy-dependent DOS\cite{Bruhat:2016a},
or by using inelastic tunneling between two different quantum
dots\cite{Stockklauser:2015,Liu:2014,Liu:2015a,Liu:2015}. It was possible to
obtain the lasing emission of a coherent photon field by coupling several
voltage-biased double dots to a microwave cavity\cite{Liu:2015a}. Finally,
Mesoscopic QED represents a new tool to study\ exotic condensed matter states,
as shown by a recent experiment for a quantum dot in the Kondo
regime\cite{Desjardins:2017}. Recent theory proposals also suggest to study
split Cooper pairs\cite{Cottet:2012,Cottet:2014} and Majorana bound states
with a microwave
cavity\cite{Trif:2012,Schmidt:2013a,Schmidt:2013b,Cottet:2013,Dmytruk:2015,Dartiailh:2016}%
.\ 

One interesting feature of Mesoscopic QED is that many geometries can be
realized thanks to the versatility of nanofabrication techniques. Furthermore,
many different experimental protocols are possible thanks to the control on
the nanocircuit dc bias and the microwave supply and detection. Therefore many
situations can be investigated. There are many possible research directions in
continuation of the works mentioned in this review:

$\bullet$ Most of the experiments performed so far have been realized with a
large number of photons, i.e. $\left\langle \hat{a}^{\dag}\hat{a}\right\rangle
>10$, such that a semiclassical description of the cavity response is
sufficient to understand the measurements. The $\left\langle \hat{a}^{\dag
}\hat{a}\right\rangle \sim1$ regime has been used very recently for the study
of the strong nanocircuit/cavity coupling, in the stationnary regime where the
cavity response versus the excitation frequency shows a characteristic
splitting\cite{Mi:2017,bruhat:2017,Stockklauser:2017}. Many other quantum
phenomena are expected for $\left\langle \hat{a}^{\dag}\hat{a}\right\rangle
\sim1$, especially in the time domain, by analogy with Circuit QED experiments
with superconducting quantum bits, and beyond if quantum transport to the
reservoirs is involved. It is one of the main goals of Mesoscopic QED to
explore this possibility.

$\bullet$ The nonlinear regime, where multi-photon emission or absorption by
the nanocircuit is possible, is a particularly interesting
regime\cite{Gullans:2016,Leppakangas:2017,Oliver:2005,Jooya:2016}. This regime
can be obtained by increasing the amplitude of the cavity microwave drive or
the intensity of the light/matter interaction.

$\bullet$ In this review, we have essentially discussed how to use a microwave
cavity to characterize the properties of a nanocircuit. One could also study
how to use a nanocircuit to prepare non-classical photonic states other than
the coherent field already obtained through the lasing effect
\cite{Schiro:2014}. For instance, squeezed cavity states can be prepared using
a non-linear light/matter interaction\cite{Cottet:2013,Mendes:2015}. In this
context, cavity state tomography and photon statistics would represent
important quantities to explore\cite{Xu:2013b}. So far, the cavity state
tomography was performed only in the presence of two double quantum dots in
the lasing regime\cite{Liu:2015a}.

$\bullet$ Placing several nanocircuits in a microwave cavity would enable one
to study \ a large variety of effects which involve the interaction between
nanocircuits mediated by cavity photons
\cite{Contreras-Pulido:2013,Lambert,Bergenfeldt:2013,Bergenfeldt:2014,Han:2015}%
. First experiments have already been realized with carbon nanotubes and
graphene\cite{Delbecq:2011,Deng:2015a,Han:2016}.

$\bullet$ One could go further in the hybridization of devices by coupling
both a quantum dot circuit and a superconducting quantum bit to a microwave
cavity\cite{Berg:2014,Feng:2012}. The use of the superconducting quantum bit
could for instance give further access to the electronic dynamics in the nanocircuit.

$\bullet$ Finally, it would be interesting to transpose Mesoscopic QED to
other types of cavities. First, one could imagine to couple nanocircuits to
teraherz cavities. This would give a photonic access to other energy scales
such as the charging energy of a Coulomb blockaded quantum dot. Second, the
behavior of quantum dot circuits coupled to optical cavities is discussed
theoretically in
Refs.\cite{Gudmundsson:2012,Jonasson:2012,Abdullah:2013,Abdullah:2014,Arnold:2014,Gudmundsson:2015b,Gudmundsson:2016a,Abdullah:2015,Gudmundsson:2016b}%
. The fabrication of such devices is extremely challenging, but this could
reveal effects related to the polarization of light. With the coplanar
cavities considered in this review, these effects are irrelevant because the
microwave fields profile is set by the shape of the microwave cavity.

To conclude, mesoscopic QED experiments open a new avenue to investigate the
light matter interaction under a different perspective. Many research
directions are possible for the future development of this field.

\textit{Acknowledgements: We would like to thank Fran\c{c}ois Mallet and Zaki
Leghtas for fruitful discussions.} \textit{This work was financed by the ERC
Starting grant CirQys and the EU FP7 Project No. SE2ND[271554].}

\section{Appendix A: Nanocircuit charge susceptibility with the Keldysh
formalism\label{KeldyshXi}}

In this appendix, we show how to predict the charge dynamics of a generic
nanocircuit which encloses a nanoconductor, assuming that the coupling of the
nanocircuit reservoirs to the cavity electric field can be disregarded. The
nanoconductor can be decomposed into $N$ sites with an index $n$ or $m$. For
the sake of generality, we do not specify the Hamiltonian $\hat{H}_{0}^{t}$ of
the nanocircuit, which is expressed in terms of electronic creation and
annihilation operators $c_{n\sigma}^{\dag}$ and $c_{n\sigma}$ for an electron
with spin $\sigma$ in site $n$. In the semiclassical limit and in the
framework of the resonant approximation of section \ref{epsin}, one can use an
Hamiltonian ac excitation term%

\begin{equation}
H_{ac}(t)=%
{\textstyle\sum_{n,\sigma}}
g_{n}c_{n\sigma}^{\dag}c_{n\sigma}\bar{a}e^{-i\omega_{RF}t}%
\end{equation}
to account for the semiclassical drive of the nanocircuit by cavity photons.
In the Keldysh formalism, the dynamics of the nanocircuit can be described by
using retarded, advanced, and lesser Greens functions $G^{c}(\omega)$ with
($\alpha\in\{r,a,<\}$)\cite{Jauho:1994}. These Green's function have a matrix
structure which encloses $N\times N$ site subblocks $G^{nm,c}(\omega)$ whose
structure depends on the problem considered. For instance, in the presence of
superconductivity and a single spin quantization axis, one can use Green's
functions with the Nambu structure:%
\begin{equation}
G^{nm,\alpha}=\left[
\begin{tabular}
[c]{ll}%
$G_{c_{n\uparrow},c_{m\uparrow}^{\dag}}^{\alpha}$ & $G_{c_{n\uparrow
},c_{m\downarrow}}^{\alpha}$\\
$G_{c_{n\downarrow}^{\dag},c_{m\uparrow}^{\dag}}^{\alpha}$ &
$G_{c_{n\downarrow}^{\dag},c_{m\downarrow}^{c}}^{\alpha}$%
\end{tabular}
\ \ \right]  \label{Gdef}%
\end{equation}
with
\begin{equation}
G_{B,A}^{r}(t)=-i\theta(t)\left\langle \{B(t),A(t=0)\}\right\rangle
\end{equation}%
\begin{equation}
G_{B,A}^{a}(t)=i\theta(-t)\left\langle \{B(t),A(t=0)\}\right\rangle
\end{equation}%
\begin{equation}
G_{B,A}^{<}(t)=i\left\langle B(t=0)A(t)\right\rangle
\end{equation}
Above, one must calculate the statistical average $\left\langle {}%
\right\rangle $ using the full time-dependent Hamiltonian $\hat{H}_{0}%
^{t}+H_{ac}(t)$. Since $\left\langle c_{n\sigma}^{\dag}(t)c_{n\sigma
}(t)\right\rangle =-iG_{c_{n\sigma},c_{n\sigma}^{\dag}}^{<}(t,t)$, one can
obtain the nanocircuit charge response by calculating $G^{<}(t,t)$. Below, we
perform this calculation at first order in $\bar{a}$ in order to obtain the
linear charge susceptibility of the nanocircuit. The first order perturbation
theory in $\bar{a}$ gives%
\begin{align}
G^{r(a)}(t,t^{\prime})  &  =\mathcal{G}^{r(a)}(t,t^{\prime})\label{tt}\\
&  +%
{\textstyle\int}
dt_{2}\mathcal{G}^{r(a)}(t,t_{2})\hat{E}_{ac}(t_{2})\mathcal{G}^{r(a)}%
(t_{2},t^{\prime})\nonumber
\end{align}
with%
\begin{equation}
\hat{E}_{ac}(t)=\bar{a}%
{\textstyle\sum_{n}}
g_{n}\hat{\tau}_{n}e^{-i\omega_{RF}t} \label{Eac}%
\end{equation}
Above, $\mathcal{G}^{r(a)}(t,t^{\prime})$ is the Green's function solution of
the problem for $\bar{a}=0$. Examples of expressions of $\mathcal{G}^{r(a)}$
are given in sections \ref{relax} and \ref{emitDsot}. The matrix $\hat{\tau
}_{n}$ is a diagonal matrix which corresponds to $diag(g_{n},-g_{n})$ in the
orbital block $(n,n)$ and is zero otherwise. The combination of Eqs.(\ref{tt})
and (\ref{Eac}) gives
\begin{align}
G^{r(a)}(t,t^{\prime})  &  =%
{\textstyle\int}
\frac{d\omega}{2\pi}e^{-i\omega(t-t^{\prime})}\\
&  \mathcal{G}^{r(a)}(\omega)\left(  1+\hat{\tau}\mathcal{G}^{r(a)}%
(\omega-\omega_{RF})e^{-i\omega_{RF}t}\right) \nonumber
\end{align}
From the Langreth theorem\cite{Jauho:1994} one has%
\begin{equation}
G^{<}(t,t^{\prime})=%
{\textstyle\iint}
dt_{1}dt_{2}G^{r}(t,t_{1})\Sigma^{<}(t_{1},t_{2})G^{a}(t_{2},t^{\prime})
\label{hh}%
\end{equation}
with%
\begin{equation}
\Sigma^{<}(t_{1},t_{2})=%
{\textstyle\int}
\frac{d\omega}{2\pi}e^{-i\omega^{\prime}(t_{1}-t_{2})}\Sigma^{<}(\omega)
\end{equation}
The lesser self energy $\Sigma^{<}(\omega)$ of the discrete levels due to the
presence of the reservoirs can be expressed by following standard Keldysh
rules (see e.g. \ section \ref{relax} and Ref.\cite{Bruhat:2016a} for
examples). By combining Eqs.(\ref{tt}) and (\ref{hh}), in the non-interacting
case, one gets at first order in $\bar{a}$:%

\begin{align}
G^{<}(t,t)  &  =%
{\textstyle\int}
\frac{d\omega}{2\pi}\mathcal{D}(\omega)\\
&  +\bar{a}e^{-i\omega_{RF}t}%
{\textstyle\sum_{n}}
{\textstyle\int}
\frac{d\omega}{2\pi}\mathcal{G}^{r}(\omega)g_{n}\hat{\tau}_{n}\mathcal{D}%
(\omega+\omega_{RF})\nonumber\\
&  +\bar{a}e^{-i\omega_{RF}t}%
{\textstyle\sum_{n}}
{\textstyle\int}
\frac{d\omega}{2\pi}\mathcal{D}(\omega)g_{n}\hat{\tau}_{n}\mathcal{G}%
^{a}(\omega+\omega_{RF})\nonumber
\end{align}
with%
\begin{equation}
\mathcal{D}(\omega)=\mathcal{G}^{r}(\omega)\Sigma^{<}(\omega)\mathcal{G}%
^{a}(\omega)
\end{equation}
One can identify the charge susceptibility $\chi_{nm}(\omega_{0})$ of site $n$
in response to an excitation at site $m$ by comparing the equation
\begin{equation}
Q_{n}(t)=ieTr[\hat{\tau}_{n}G^{<}(t,t)]-e \label{neu}%
\end{equation}
which is due to the definition of $G^{<}$ and the equation
\begin{equation}
Q_{n}(t)=\bar{Q}_{n}-e%
{\textstyle\sum_{m}}
\bar{a}g_{m}e^{-i\omega_{RF}t}\left(  \chi_{nm}(\omega_{RF})\right)
\end{equation}
which defines $\chi_{nm}$, with $\bar{Q}_{n}$ the average charge in dot $n$.
In Eq.(\ref{neu}), the $-e$ term is due to the fact that the Greens functions
are defined in the Nambu space in the particular case of Eq.(\ref{Gdef}).
Using the cyclic invariance of the trace, one finds%
\begin{align}
\chi_{nm}(\omega_{RF})  &  =-i%
{\textstyle\int}
\frac{d\omega}{2\pi}Tr[\hat{\tau}_{n}\mathcal{G}^{r}(\omega+\omega_{RF}%
)\hat{\tau}_{m}\mathcal{D}(\omega)]\\
&  -i%
{\textstyle\int}
\frac{d\omega}{2\pi}Tr[\hat{\tau}_{m}\mathcal{G}^{a}(\omega-\omega_{RF}%
)\hat{\tau}_{n}\mathcal{D}(\omega)]\nonumber
\end{align}
This leads to Eq.(\ref{xi}) of the main text with $\hat{T}=%
{\textstyle\sum_{n}}
g_{n}\hat{\tau}_{n}$. Note that a similar derivation can be performed without
using the Nambu space in the case of a problem without superconductivity, or
by using an extended Nambu space if the nanocircuit includes non-homogeneous
magnetic fields which induce spin rotations (see Ref.\cite{Dartiailh:2016}).
In this case, Eq.(\ref{xi}) still holds provided the matrix $\hat{T}$ is
defined consistently with the structure of the Green's functions.

\section{Appendix B: Non-interacting double dot with the Keldysh
non-interacting description\label{DQDkeldysh}}

In this appendix, we show in the simple example of a non-interacting double
quantum dot with two grounded normal metal reservoirs (see Fig.(\ref{Figure19}%
)) that the Keldysh approach of section \ref{Keldysh} leads to a nanocircuit
charge susceptibility $\Xi(\omega)$ which can account simultaneously for
internal transitions inside a nanoconductor, and tunneling between the
nanoconductor and fermionic reservoirs. For simplicity, we disregard the spin
degree of freedom. The double dot Hamiltonian is%
\begin{align}
\hat{H}_{0}^{t}  &  =\varepsilon_{L}\hat{c}_{L}^{\dag}\hat{c}_{L}%
+\varepsilon_{R}\hat{c}_{R}^{\dag}\hat{c}_{R}+t\hat{c}_{L}^{\dag}\hat{c}%
_{R}+t^{\ast}\hat{c}_{R}^{\dag}\hat{c}_{L}\label{Hdqd}\\
&  +%
{\textstyle\sum\nolimits_{\alpha\in\{L,R\}}}
\varepsilon_{\alpha k}\hat{c}_{\alpha k}^{\dag}\hat{c}_{\alpha k}+t_{N}\left(
\hat{c}_{\alpha k}^{\dag}\hat{c}_{\alpha}+\hat{c}_{\alpha}^{\dag}\hat
{c}_{\alpha k}\right) \nonumber
\end{align}
Above, $c_{\alpha}^{\dag}$ is the creation operator for an electron in the dot
$\alpha\in\{L,R\}$. The operator $\hat{c}_{\alpha k}^{\dag}$ creates an
electron in the state $k$ of the normal metal reservoir attached to dot
$\alpha$. We note $t$ the interdot hopping, and $t_{N}$ the tunnel hopping
constant between dot $\alpha$ and its reservoir. The double dot Green's
function can be defined as%
\begin{equation}
G^{c}=\left[
\begin{tabular}
[c]{ll}%
$G_{c_{L},c_{L}^{\dag}}^{c}$ & $G_{c_{L},c_{R}^{\dag}}^{c}$\\
$G_{c_{R},c_{L}^{\dag}}^{c}$ & $G_{c_{R},c_{R}^{\dag}}^{c}$%
\end{tabular}
\ \ \ \right]
\end{equation}
Therefore, using the Keldysh description of nanocircuits\cite{Jauho:1994}, one
gets from Eq. (\ref{Hdqd})
\begin{equation}
\mathcal{G}^{r}(\omega)=\left[
\begin{tabular}
[c]{ll}%
$\omega-\varepsilon_{L}+i\frac{\Gamma_{N}}{2}$ & $t$\\
$t$ & $\omega-\varepsilon_{R}-i\frac{\Gamma_{N}}{2}$%
\end{tabular}
\ \ \ \right]  ^{-1}%
\end{equation}
and
\begin{equation}
\check{\Sigma}^{<}(\omega)=i\Gamma_{N}f(\omega)\check{1}%
\end{equation}
with $\Gamma_{N}=2\pi\left\vert t_{N}\right\vert ^{2}$ and $f(\omega
)=1/(1+\exp[\omega/k_{B}T])$. Then, Eq.(\ref{xi}) gives%
\begin{equation}
\Xi(\omega)=\sum\limits_{s}\int\nolimits_{-\infty}^{+\infty}d\varepsilon
(\frac{2\Gamma_{N}f(\varepsilon)\left(  \mathcal{F}_{t,s}(\omega
)+\mathcal{F}_{l,s}(\omega)\right)  }{\pi\lbrack\left(  \varepsilon-E_{s}%
)^{2}+\left(  \frac{\Gamma_{N}}{2}\right)  ^{2}\right)  ]} \label{piDQD}%
\end{equation}
with%
\begin{equation}
E_{\pm}=\frac{1}{2}\left[  \varepsilon_{L}+\varepsilon_{R}\pm\sqrt
{(\varepsilon_{L}-\varepsilon_{R})^{2}+4t^{2}}\right]
\end{equation}%
\begin{equation}
\mathcal{F}_{t,s}(\omega)=\frac{g_{t}^{2}(\varepsilon-E_{-s})}{\left(
\varepsilon-E_{-s})^{2}-(\omega+i\frac{\Gamma_{N}}{2}\right)  ^{2}}%
\end{equation}
and%
\begin{equation}
\mathcal{F}_{l,s}(\omega)=\frac{g_{l,s}^{2}(\varepsilon-E_{s})}{\left(
\varepsilon-E_{s})^{2}-(\omega+i\frac{\Gamma_{N}}{2}\right)  ^{2}}%
\end{equation}
Above, $g_{t}$ is defined by Eq.(\ref{gt}) of the main text and one has%
\begin{equation}
g_{l,s}=\frac{\lambda_{L}+\lambda_{R}}{2}+s\frac{\lambda_{L}-\lambda_{R}}%
{2}\frac{\varepsilon_{L}-\varepsilon_{R}}{E_{+}-E_{-}}%
\end{equation}
Equation (\ref{piDQD}) is similar to Eq.(\ref{6}) of Ref.\cite{Cottet:2011},
obtained with a non-interacting diagrammatic approach. An inspection of the
double dot Hamiltonian in the bonding/antibonding state basis reveals that
$g_{t}$ corresponds to the transverse coupling between the bonding/antibonding
transition and the cavity, whereas $g_{l,+}$ and $g_{l,-}$ correspond to the
lever arms for the modulation of the bonding and antibonding energy levels by
the cavity electric field. The complex expression (\ref{piDQD}) can be
simplified in some particular limits to get a better physical insight. First,
in the limit $\omega<\Gamma_{N}\ll k_{B}T\ll t$, one gets, from
Eq.(\ref{piDQD})%
\begin{equation}
\Xi(\omega=0)\simeq-\frac{\beta}{4}%
{\displaystyle\sum\nolimits_{s}}
\left(  \frac{g_{l,s}^{2}}{\cosh^{2}[\frac{\beta E_{s}}{2}]}-2g_{t}^{2}%
\frac{f(E_{-})-f(E_{+})}{\Delta_{c}}\right)  \label{pi1therm}%
\end{equation}
Hence, the terms in $g_{l,s}$ describe a quantum capacitance contribution
which is due to the thermal equilibration of the bonding and antibonding
states with the reservoirs. The term in $g_{t}$ describes another quantum
capacitance contribution which is due to a transfer of electrons between the
bonding and antibonding states of the double dot. This last term is maximally
visible when the double dot has an occupation close to one electron, i.e.
$f(E_{-})=1$ and $f(E_{+})=0$. Second, in the limit $\omega\simeq E_{+}-E_{-}$
and $2t\gg\Gamma_{N}$, one gets%
\begin{equation}
\Xi(\omega)=g_{t}^{2}\frac{f(E_{-})-f(E_{+})}{\omega-E_{+}-E_{-}}%
\end{equation}
which is similar to Eq.(\ref{XiDQD}) of the main text, with $p_{-}=f(E_{-})$
and $p_{+}=f(E_{+})$. These values of $p_{+}$ and $p_{\_}$ stem from the fact
that in the above lines, for simplicity, we have assumed that the normal metal
reservoirs are grounded. Therefore, the population of the double dot states is
thermal. It would be possible to generalize straightforwardly the above
approach to the non-equilibrium case by using a term $\check{\Sigma}%
^{<}(\omega)$ which would depend on the bias voltage. Note that in
Ref.\cite{Dartiailh:2016} discussed in section \ref{Majos}, it is also found
that the charge susceptibility $\Xi(\omega)$ for a Majorana nanowire accounts
for both tunneling to normal metal reservoirs (see Fig.\ref{Figure9}c, step
highlighted with magenta dots) and transitions internal to the nanowire (see
Fig.\ref{Figure9}c, resonance highlighted with black dots).


\begin{thebibliography}{999}
\bibitem {multimode}In principle, in the presence of a cavity bosonic bath,
one has to include the dissipation term $H_{diss}$ of Eq.(\ref{hdiss}) in the
Hamiltonian (\ref{1}). Then, in order to obtain the photonic pseudo-potential
picture, it is still sufficient to use the transformation (\ref{UU}) if there
is a very weak spatial overlap between the nanocircuit and the modes of the
bosonic bath. The exact form for Hamiltonian (\ref{HH}) is then slightly more
complicated than what is indicated in the main text, because the
transformation of the term $H_{diss}$ by (\ref{UU}) leads to an extra coupling
term $\int d\epsilon\eta\hat{\mathcal{V}}\left(  \beta\hat{b}_{\epsilon}%
^{\dag}+\beta^{\ast}\hat{b}_{\epsilon}\right)  /\omega_{0}$ between the
electrons of the nanocircuit and the bosonic bath. One could expect an extra
widening of the nanocircuit electronic levels by this term, with an order of
magnitude $\Lambda_{0}(g_{n}/\omega_{0})^{2}$. However, in the experiments,
one has $g_{n}/\omega_{0}\ll1$ and $\Lambda_{0}$ is very small in comparison
with the intrinsic damping of the dot levels by for instance by the metallic
contacts. Therefore, equations (\ref{HH}) and (\ref{hdiss}) seem to be a good
approximation for the description of cavity dissipation in the photonic
pseudo-potential picture.
\bibitem {drive}The drive term in $\varepsilon_{in}$ can be obtained by
treating explicitly the classical part $\left\langle \hat{b}_{in}^{\dag
}\right\rangle $ of the bosonic bath mode $\hat{b}_{in}^{\dag}$ excited by a
microwave source.
\bibitem {defC}To define $Q_{e-ph}$ mathematically, we consider the resonant
limit ($\omega_{ij}=\omega_{0}$) with a negligible bare cavity linewidth
($C_{e-ph}\gg1$). Then, the threashold $Q_{e-ph}=1$ corresponds to having
$\partial^{2}(\left\vert b_{t}/b_{in}\right\vert )/\partial\omega_{RF}^{2}=0$
at $\omega_{RF}=\omega_{0}$, with $\left\vert b_{t}/b_{in}\right\vert $ given
by Eq.(\ref{ii}) with $\Lambda_{0}=0$.
\end{thebibliography}
\end{document}